\documentclass[final,nopageno,nowidows,noorphans]{jpaper}

\usepackage[normalem]{ulem}
\usepackage{algorithm}
\usepackage[T1]{fontenc}
\usepackage[scaled=0.94]{XCharter}
\usepackage{pifont}
\usepackage{inconsolata}

\usepackage{xspace}
\usepackage{mystyle}
\usepackage{algpseudocode}
\usepackage{pifont}
\usepackage{color}
\usepackage{xcolor}
\usepackage{color,soul}
\definecolor{pastelyellow}{rgb}{0.99, 0.99, 0.59}
\usepackage{setspace}
\usepackage{indentfirst}
\usepackage{fancyhdr}
\usepackage[keeplastbox]{flushend}
\usepackage{natbib}

\def\cameraready{}

\newcommand{\X}[0]{Zorua\xspace}

\newcommand{\thoughts}[1]{{\leavevmode\color{blue}{\emph{ #1}}}}

\newcommand{\revision}[1]{{\leavevmode\color{black}{ #1}}}

\newcommand{\new}[1]{{{#1}}}
\newcommand{\One}{\emph{(i)}~}
\newcommand{\two}{\emph{(ii)}~}
\newcommand{\three}{\emph{(iii)}~}
\newcommand{\PrintTodo}{true}
\newcommand{\todo}[2]{
\ifx\PrintTodo\undefined
\textcolor{red}{#2\xspace}
\else
\textcolor{red}{TODO[#1]:} \textcolor{blue}{\textit{#2}\xspace}
\fi
}

\ifx\cameraready\undefined
\fancypagestyle{firstpage}{
  \fancyhf{}
\setlength{\headheight}{50pt}

  \fancyhead[C]{\normalsize{--- Version 9, Feb.\ 4, 2018, 5:12PM EST ---}} 
}  
  \newcommand{\newII}[1]{{\leavevmode\color{BrickRed}{#1}}}
\else
  \newcommand{\newII}[1]{{{#1}}}
\fi

\usepackage[font={small,bf}]{caption}
\usepackage{caption}
\usepackage{subcaption}
\newcommand{\ignore}[1]{}
\newcommand{\squishlist}{
   \begin{list}{$\bullet$}
    { \setlength{\itemsep}{0pt}      \setlength{\parsep}{3pt}
      \setlength{\topsep}{3pt}       \setlength{\partopsep}{0pt}
      \setlength{\leftmargin}{1.5em} \setlength{\labelwidth}{1em}
      \setlength{\labelsep}{0.5em} } }

\newcommand{\squishlisttwo}{
   \begin{list}{$\numbered$}
    { \setlength{\itemsep}{0pt}    \setlength{\parsep}{0pt}
      \setlength{\topsep}{0pt}     \setlength{\partopsep}{0pt}
      \setlength{\leftmargin}{2em} \setlength{\labelwidth}{1.5em}
      \setlength{\labelsep}{0.5em} } }

\newcommand{\squishend}{
    \end{list}  }

\begin{document}
\vspace{0.5cm}
\title{\Large{Zorua: Enhancing Programming Ease, Portability, and Performance in
GPUs \\by Decoupling  Programming Models from Resource Management}}
\author{Nandita Vijaykumar$^\dag$\quad Kevin Hsieh$^\dag$\quad Gennady
Pekhimenko$^\ddagger$$^\dag$\quad Samira
Khan$^\aleph$\quad \\Ashish Shrestha$^\dag$\quad Saugata Ghose$^\dag$\quad Adwait
Jog$^\dotplus$\quad
Phillip B. Gibbons$^\dag$\quad Onur Mutlu$^\S$$^\dag$
\vspace{3mm}\\
\emph{$^\dag$Carnegie Mellon University}
\quad\quad\quad\emph{$^\ddagger$Microsoft Research}\quad\quad\quad
\emph{$^\aleph$University of Virginia}\\
\emph{$^\dotplus$College of William and Mary}
\quad\quad\quad\quad\quad\emph{$^\S$ETH
Z{\"u}rich}
\vspace{2mm}\\
}

\date{}
\maketitle
 
\pagestyle{plain}

\begin{abstract}
This paper introduces a new resource virtualization framework,
\emph{Zorua}, that \emph{decouples} the GPU programming model from the
management of key on-chip resources in hardware to enhance programming ease, portability, and
performance.
The application resource specification{\textemdash}a static specification of several parameters
such as the number of threads and the scratchpad memory usage per thread
block{\textemdash}forms a critical component of the existing GPU programming models.
This specification determines the parallelism, and hence performance, of the
application during execution because the corresponding on-chip hardware
resources are allocated and managed purely based
on this specification. This \emph{tight coupling} between the software-provided
resource specification and resource management in hardware leads to
significant challenges in programming ease, portability, and performance, as we
demonstrate in this work using real data obtained on state-of-the-art GPU
systems. 

Our goal in this work is to reduce the dependence of performance on the
software-provided static resource specification to simultaneously alleviate the above challenges. To this end, we introduce Zorua, a new resource virtualization framework,
that \emph{decouples} the programmer-specified resource usage of a
GPU application from the actual allocation in the on-chip hardware
resources. 
\X enables this decoupling by \emph{virtualizing} each
resource transparently to the programmer.  

The virtualization provided
by \X builds on two key concepts{\textemdash}\emph{dynamic allocation}
of the on-chip resources, and their \emph{oversubscription} using a
swap space in memory. \X provides a holistic GPU resource virtualization strategy designed
to {\One}adaptively \emph{control the extent} of oversubscription, and
\two \emph{coordinate} the dynamic management of multiple on-chip resources to maximize the
effectiveness of virtualization. We demonstrate that by providing the illusion of more resources than
physically available via controlled and coordinated virtualization, \X
offers several important benefits: \One \textbf{Programming Ease.} \X
eases the burden on the programmer to provide code that is tuned to
efficiently utilize the physically available on-chip resources. \two
\textbf{Portability.} \X alleviates the necessity of re-tuning an
application's resource usage when porting the application across GPU
generations. \three \textbf{Performance.} By dynamically allocating
resources and carefully oversubscribing them when necessary, \X
improves or retains the performance of applications that are already
highly tuned to best utilize the resources. The holistic
virtualization provided by \X has many other potential uses, e.g.,
fine-grained resource sharing among multiple kernels, low-latency 
preemption of GPU programs, and support for dynamic parallelism, which we
describe in this paper.

\end{abstract}

\section{Introduction}

Modern Graphics Processing Units (GPUs) have evolved into powerful
programmable machines over the last decade, offering high performance
and energy efficiency for many classes of applications
by concurrently executing thousands of threads. In order to execute,
each thread requires several major on-chip resources: \One registers,
\two scratchpad memory (if used in the program), and
\three a thread slot in the thread scheduler that keeps all the
bookkeeping information required for execution.

Today, these hardware resources are {\em statically} allocated to threads based on several parameters{\textemdash}the number of threads
per thread block, register usage per thread, and scratchpad usage per
block. We refer to these static application parameters as the
\emph{resource specification} of the application. This resource specification
forms a critical component of modern GPU programming models (e.g.,
CUDA~\cite{cuda},
OpenCL~\cite{opencl}). The static
allocation over a fixed set of hardware resources based on the
software-specified resource specification creates a
\emph{tight coupling} between the program (and the programming model) 
and the physical hardware resources. As a result of this tight coupling, for
each application, there are only a few optimized resource specifications that
maximize resource utilization. Picking a suboptimal specification
leads to underutilization of resources and hence, very often,
performance degradation. This leads to three key
difficulties related to obtaining good performance on modern GPUs: programming
ease, portability, and resource inefficiency (performance). 

\textbf{Programming Ease.} First, the burden falls upon the
programmer to optimize the resource specification.
For a naive programmer, this is a very challenging
task~\cite{OptGPU1,OptGPU2,OptGPU3,toward-autotuning,OptGPU4,jpeg-occ, asplos-sree}.
This is because, in addition to selecting a specification suited to an
algorithm, the programmer needs to be aware of the details of the GPU architecture
to 
fit the specification to the underlying hardware resources. This
\emph{tuning} is easy to get wrong because there are \emph{many}
highly suboptimal performance points in the specification space, and
even a minor deviation from an optimized specification can lead to a
drastic drop in performance due to lost parallelism. We refer to such
drops as \emph{performance cliffs}. We analyze the effect of
suboptimal specifications on real systems for 20 workloads 
(Section~\ref{sec:perf-cliffs}), and experimentally demonstrate that
changing resource specifications can produce as much as a 5$\times$
difference in performance due to the change in parallelism. Even a
minimal change in the specification (and hence, the resulting allocation) of one resource can result in a significant
performance cliff, degrading performance by as much as 50\%
(Section~\ref{sec:perf-cliffs}).

\textbf{Portability.} Second, different GPUs have varying quantities
of each of the resources. Hence, an optimized specification on one GPU
may be highly suboptimal on another. In order to determine the extent
of this portability problem, we run 20 applications on three
generations of NVIDIA GPUs: Fermi, Kepler, and Maxwell
(Section~\ref{sec:motivation:port}). An example result demonstrates
that highly-tuned code for Maxwell or Kepler loses as much as 69\%
of its performance on Fermi. This lack of \emph{portability} necessitates
that the programmer \emph{re-tune} the resource specification of the application for
\emph{every} new GPU generation. This
problem is especially significant in virtualized environments, such as cloud or
cluster computing, where the same program may run on a wide range of GPU
architectures, depending on data center composition and hardware availability.

\textbf{Performance.} Third, for the programmer who chooses to employ software
optimization tools (e.g., auto-tuners) or manually tailor the program to fit the
hardware,
performance is still constrained by the \emph{fixed, static} resource
specification. It is well
known~\cite{virtual-register,gebhart-hierarchical,compiler-register,shmem-multiplexing,caba,
caba-bc, virtual-thread}
that the on-chip resource requirements of a GPU application vary throughout
execution. Since the program (even after auto-tuning) has to {\em statically}
specify its {\em worst-case} resource requirements, severe
\emph{dynamic underutilization} of several GPU
resources~\cite{virtual-register,compiler-register,gebhart-hierarchical,kayiran-pact16,caba,caba-bc} ensues,
leading to suboptimal performance (Section~\ref{sec:underutilized_resources}).

\textbf{Our Goal.} To address these three challenges at the same time, we propose to
\emph{decouple} an application's resource specification from the available hardware
resources by \emph{virtualizing} all three major resources in a holistic manner. This
virtualization provides the illusion of \emph{more} resources to the GPU programmer and
software than physically available, and enables the runtime system and the
hardware to {\em dynamically} manage multiple physical resources in a manner that is transparent
to the programmer, thereby alleviating dynamic underutilization.

Virtualization is a concept that has been applied to the management of hardware
resources in many contexts (e.g.,~\cite{virtual-memory1,
virtual-memory2,
virtualization-1,virtualization-2,ibm-360,pdp-10,how-to-fake,vmware-osdi02}),
providing various benefits. We believe that applying the general principle of
virtualization to the management of \emph{multiple} on-chip resources in GPUs offers
the opportunity to alleviate several important challenges in modern GPU
programming, which are described above. However, at the same time, effectively adding a new level
of indirection to the management of multiple latency-critical GPU resources
introduces several new challenges\ignore{, as we demonstrate in this paper}
(see Section~\ref{sec:virt-challenges}). This necessitates the design of a new
mechanism to effectively address the new challenges and enable the benefits of
virtualization. In this work, we introduce a new framework,
\emph{\X},\footnote{Named after a Pok\'{e}mon~\cite{pokemon} with the power of illusion, able to
take different shapes to adapt to different circumstances (not unlike our proposed
framework).} to
decouple the programmer-specified resource specification of an application from
its physical on-chip hardware resource allocation by effectively virtualizing
the multiple
on-chip resources in GPUs. 

\textbf{Key Concepts.} The virtualization strategy used by \X is
built upon two key concepts. First, to mitigate performance cliffs when we do
not have enough physical resources, we \emph{oversubscribe} resources by a small
amount at runtime, by leveraging their dynamic underutilization and maintaining a
swap space (in main memory) for the extra resources required. Second, \X
improves utilization by determining the runtime resource requirements of an
application. It then allocates and deallocates resources dynamically, managing
them \One \emph{independently} of each other to maximize their utilization; and \two in a \emph{coordinated} manner, to enable efficient execution of each
thread with all its required resources available.

\textbf{Challenges in Virtualization.} Unfortunately, oversubscription means
that latency-critical resources, such as registers and scratchpad, may be
swapped to memory at the time of access, resulting in high overheads in performance
and energy. This leads to two critical challenges in designing a framework to
enable virtualization. The first challenge is to effectively determine the
\emph{extent} of virtualization, i.e., by how much each resource appears to be
larger than its physical amount, such that we can
minimize oversubscription while still reaping its benefits. This is difficult as
the resource requirements continually vary during runtime. The second challenge
is to minimize accesses to the swap space. This requires \emph{coordination} in
the virtualized management of \emph{multiple resources}, so that enough of each
resource is available
on-chip when needed.

\textbf{\X}. In order to address these challenges, \X employs a
hardware-software codesign that comprises three components: \One
\textbf{\emph{the compiler}} annotates the program to specify the resource needs
of {\em each phase} of the application; \two \textbf{\emph{a runtime system}},
which we refer to as the \textbf\emph{coordinator}, 
uses the compiler annotations to dynamically manage the virtualization of the
different on-chip resources; and \three \textbf{\emph{the hardware}}
employs mapping tables to locate a virtual resource in the physically available
resources or in the swap space in main memory. The coordinator plays the key role
of scheduling threads {\em only when} the expected gain in thread-level
parallelism outweighs the cost of transferring oversubscribed resources from the
swap space in memory, and coordinates the oversubscription and allocation of
multiple on-chip resources.

\textbf{Key Results.} We evaluate \X with many resource specifications
for eight applications across three GPU architectures (Section~\ref{sec:eval}).
Our experimental results show that \X \One reduces the range in performance for
different resource specifications by 50\% on average (up to 69\%), by alleviating
performance cliffs, and hence eases the burden on the programmer to provide
optimized resource specifications, \two improves performance for code with optimized
specification by 13\%  on average (up to 28\%), and \three enhances portability by reducing
the maximum porting performance loss by 55\% on average (up to 73\%) for three different
GPU architectures. We conclude that decoupling the resource specification and
resource management via virtualization significantly eases programmer burden,
by alleviating the need to provide optimized specifications and enhancing
portability, while still improving or retaining performance for programs that already have
optimized specifications.
 
\textbf{Other Uses.} We believe that \X offers the opportunity to address
several other key challenges in GPUs today, for example: (i) By providing an new level of
indirection, \X provides a natural way to enable dynamic and fine-grained
control over resource partitioning among {\em multiple GPU kernels and
applications}. (ii) \X can be utilized for {\em low-latency preemption} of GPU
applications, by leveraging the ability to swap in/out resources from/to
memory in a transparent manner. (iv) \X provides a simple mechanism to provide
dynamic resources to support other programming paradigms such as nested
parallelism, helper threads, etc. and even system-level tasks. (v) The dynamic
resource management scheme in \X improves the energy efficiency and scalability
of expensive on-chip resources (Section~\ref{sec:applications}).

The main \textbf{contributions} of this work are: \squishlist

\item{This is the first work that takes a holistic approach to
decoupling a GPU application's resource specification from its physical on-chip
resource allocation via the use of virtualization. We develop a comprehensive
virtualization framework that provides \emph{controlled} and \emph{coordinated}
virtualization of \emph{multiple} on-chip GPU resources to maximize the efficacy
of virtualization.}

\item{We show how to enable efficient oversubscription of multiple GPU resources with dynamic
fine-grained allocation of resources and swapping mechanisms into/out of main
memory. We provide a hardware-software cooperative framework that \One controls the extent of
oversubscription to make an effective tradeoff between higher thread-level
parallelism due to virtualization versus the latency and capacity overheads of
swap space usage, and \two coordinates the virtualization for multiple on-chip
resources, transparently to the programmer.}

\ignore{\item On three generations of real GPU hardware, we demonstrate the
existence and magnitude of performance cliffs for a wide range of workloads. }
\item{We demonstrate that by providing the illusion of having more resources
than physically available, \X \One reduces programmer burden, providing
competitive performance for even suboptimal resource specifications, by reducing
performance variation across different specifications and by alleviating
performance cliffs; \two reduces performance loss when the program with its resource
specification tuned for one GPU platform is ported to a different platform; and
\three retains or enhances performance for highly-tuned code by improving
resource utilization, via dynamic management of resources.}

\squishend

\section{Background}
\ignore{Modern GPUs employ
fine-grained multi-threading to hide the high memory access latencies
with thousands of concurrently running threads~\cite{keckler}. Programmers
exploit parallelism with GPUs using programming languages such as CUDA
and OpenCL. These programming languages allow programmers
to define and invoke parallel functions, called
kernels, on a GPU. Each kernel consists of a number of threads that
execute in parallel on the GPU cores.  
}

\begin{figure}[h!]
\centering
\includegraphics[width=0.23\textwidth]{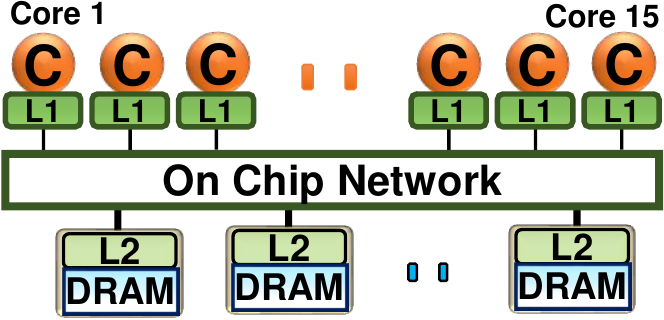}
\caption{Baseline GPU architecture. Reproduced from~\cite{caba-bc}.}
\label{fig:architecture}
\end{figure}

\textbf{The GPU Architecture.} A GPU consists of
multiple simple cores, also called \emph{streaming multiprocessors} (SMs) in NVIDIA
terminology or \emph{compute units} (CUs) in AMD terminology. Each core
\new{contains} a large register file,
programmer-managed shared memory, and an L1 data cache.  Each GPU core time
multiplexes the execution of thousands of threads to hide long latencies due to
memory accesses and ALU operations. The cores and memory controllers
are connected via \new{a} crossbar and every memory controller is
associated with a slice of a shared L2 cache. Every cycle, a set of threads,
referred to as a \emph{warp}, are executed in lockstep. If any warp is
stalled on a long-latency operation, the scheduler swaps in a different
warp for execution. Figure~\ref{fig:architecture}
depicts our baseline architecture with 15 SMs and 6 memory controllers.  
For more details on the internals of modern GPU architectures, we refer the
reader to~\cite{wen-mei-hwu,patterson,largewarp,medic}.

\textbf{The Programming Model - Exploiting Parallelism in GPUs.} 
Programming models like CUDA or OpenCL allow programmers
to define and invoke parallel functions, called
kernels, on a GPU. Each kernel consists of a number of threads that
execute in parallel on the GPU cores. 

Applications running on GPUs require \new{some} on-chip resources for execution. Each
thread, among the hundreds or thousands executing concurrently, requires
\emph{(i)}
registers, \emph{(ii)} scratchpad memory (if used in the application), and
\emph{(iii)} a warpslot
which includes the necessary book-keeping for execution -- a slot in the thread
scheduler, PC, and the SIMT stack (used to track control divergence within a
warp). Programming
languages like CUDA and OpenCL also provide the ability to synchronize execution
of threads as well as exchange data \new{with each other}. These languages provide the
abstraction of a \emph{thread block} or \emph{cooperative thread array (CTA)}, respectively --
which are a group of threads that can synchronize using
barriers or fences, and share data \new{with each other} using scratchpad memory. This form of thread synchronization
requires that \emph{all} the threads within the same thread block make progress in order for
\emph{any} thread to complete execution. As a result, the on-chip resource
partitioning as well as the launch for execution at any streaming multiprocessor
is done at the granularity of a thread block.

The GPU
architecture itself is well provisioned with these on-chip resources to support the
concurrent execution of a large number of threads, and these resources can be
flexibly partitioned across the application threads \new{according to} the application
requirements. This flexible partitioning implies that the \new{amount of} parallelism that the
GPU can support at any time depends on the per-thread block resource
requirement.

The programming models, hence, also require specification of several key
parameters that decide the utilization of these resources. These include
\emph{(i)} the number of thread blocks in the kernel compute grid, \emph{(ii)}
the number of threads within the thread block (which dictates the number of warp
slots required per thread block), \emph{(iii)} the number of registers per
thread and, \emph{(iv)} the scratchpad usage per thread block. These parameters
are typically decided by the programmer and/or compiler. Programmers who aim to
optimize code for high efficiency hand-optimize these parameters or use software
tools such as 
autotuners~\cite{toward-autotuning,atune,maestro,parameter-profiler,autotuner1,autotuner-fft}
and 
optimizing
compilers~\cite{g-adapt,optimizing-compiler1,parameter-selection,porple,optimizing-compiler2,sponge} to find optimized parameter
specifications.

\ignore{A typical GPU application consists of many
kernels. Each kernel is divided into groups of threads referred to as thread
blocks. The thread block is the granularity at which threads are scheduled onto
GPU cores. However, at any point in time, the hardware executes only a small
subset of the thread block, also referred to as a warp. All threads within a
warp are executed in lockstep and every cycle a different warp may be executed.
The number of threads that can run concurrently on each SM depends on the per-thread
register and scratchpad memory requirement. 
}


\section{Motivation} \label{sec:motivation} 
The amount of parallelism that the GPU can provide for any application depends on the
utilization of on-chip resources by threads within the application. As a
result, suboptimal usage of these resources may lead to loss in the parallelism
that can be achieved during program execution. This loss in parallelism 
often leads to significant degradation in performance, as GPUs primarily use
fine-grained multi-threading~\cite{burtonsmith,cdc6600} to hide the long latencies during execution. 

The granularity of synchronization -- i.e., the number of threads in a thread
block -- and the amount of scratchpad memory used per thread block is determined by
the programmer while adapting any algorithm or application for execution on a
GPU. 
\new{This choice involves a complex tradeoff between
minimizing data movement, by using \emph{larger} scratchpad memory sizes, and
reducing the
inefficiency of synchronizing a large number of threads, by using
\emph{smaller} scratchpad memory and thread block sizes. 
A similar tradeoff exists when determining the number of registers
used by the application.
Using \emph{fewer} registers
minimizes hardware register usage and enables higher parallelism during
execution, whereas using \emph{more} registers avoids expensive accesses to
memory.} The resulting application parameters -- the number of
registers, the amount of scratchpad memory, and the number of threads per thread
block -- dictate the on-chip resource requirement and hence, determine the
parallelism that can be obtained for that application on
any GPU. 

In this section, we study the performance implications of different choices of
resource specifications for GPU applications to demonstrate the key issues we aim to alleviate.
\subsection{Performance Variation and Cliffs} \label{sec:perf-cliffs}
To understand the impact of resource specifications and the resulting
utilization of physical resources on GPU performance, we conduct an experiment
on a Maxwell GPU system (GTX 745) with 20 GPGPU workloads from the CUDA
SDK~\cite{sdk}, Rodinia~\cite{rodinia}, GPGPU-Sim benchmarks~\cite{GPGPUSim},
Lonestar~\cite{lonestar}, Parboil~\cite{parboil}, and US DoE application
suites~\cite{DBLP:conf/sc/VillaJOBNLSWMSKD14}. We use the NVIDIA profiling tool
(NVProf)~\cite{sdk} to determine the execution time of each application
kernel\ignore{(detailed methodology is in Section~\ref{sec:methodology})}.
We sweep the three parameters of the specification{\textemdash}number of threads in a
thread block, register usage per thread, and scratchpad memory usage per thread
block{\textemdash}for each workload, and measure their impact on execution time. \ignore{We
ensure that the application performs the same amount of work for each
specification.}

\begin{figure}[!h] \centering
\includegraphics[width=0.48\textwidth]{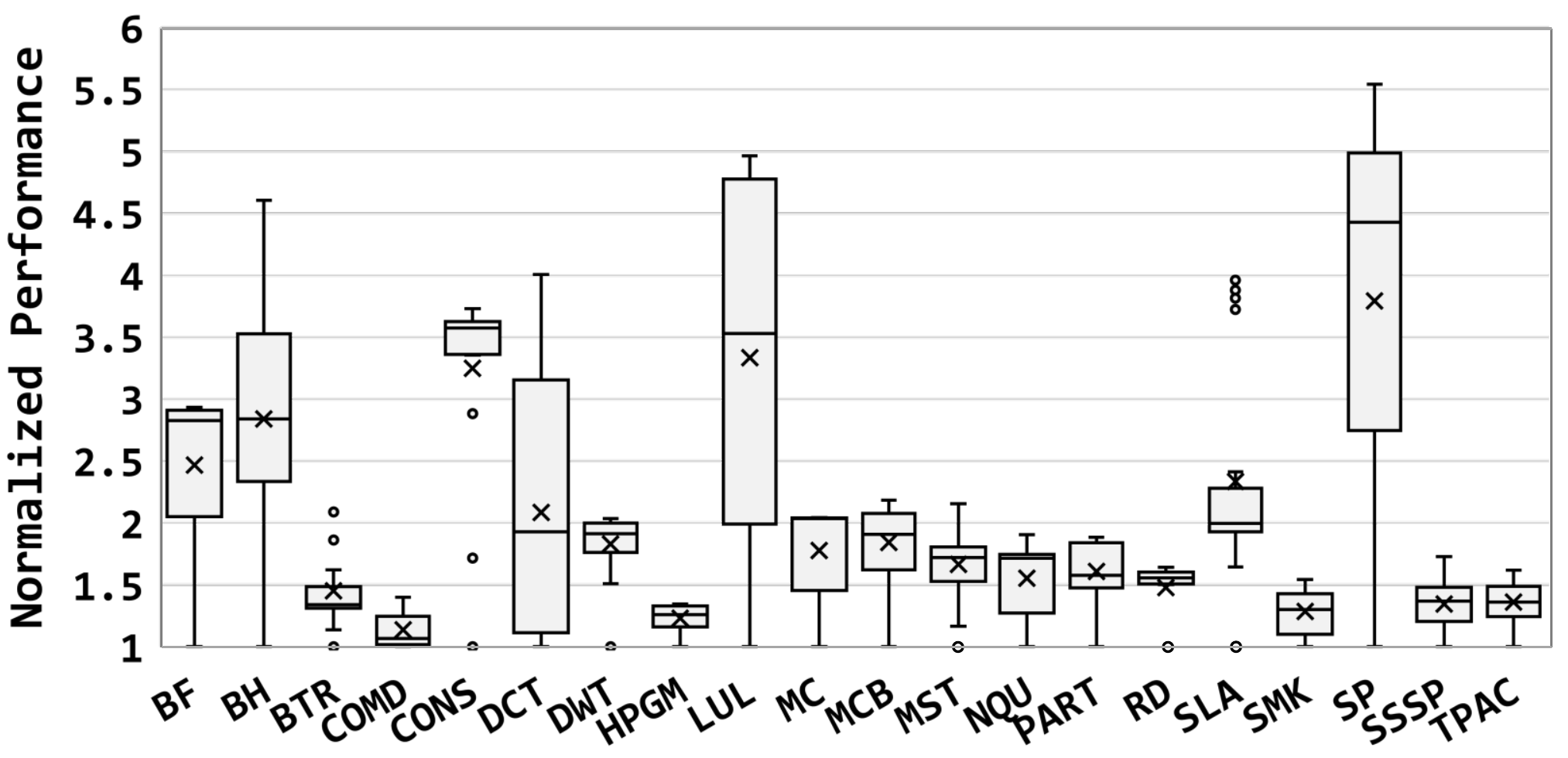}
\caption{\small{Performance variation across specifications. Reproduced
from~\cite{zorua}.}}
\label{fig:summary-variation} \end{figure}

Figure~\ref{fig:summary-variation} shows a summary of variation in performance
(higher is better), normalized to the slowest specification for each application, across all
evaluated specification points for each application\ignore{Our
technical report~\cite{zorua-tr} contains more detail on the evaluated ranges.} in a Tukey box
plot~\cite{mcgill1978variations}. The boxes in the box plot represent the range
between the first quartile (25\%) and the third quartile (75\%). 
The whiskers extending from the
boxes represent the maximum and minimum points of the distribution, or
1.5$\times$ the
length of the box, whichever is smaller. Any points that lie more than
1.5$\times$
the box length beyond the box are considered to be
outliers~\cite{mcgill1978variations}, and are plotted as
individual points. The line in the middle of the box represents the median,
while the ``X'' represents the average. 

\label{sec:motivation:cliffs} 

We can see that there is
significant variation in performance across different specification points (as
much as 5.51$\times$ in \emph{SP}), proving the importance of optimized resource
specifications. In some applications (e.g., \emph{BTR, SLA}), few points
perform well, and these points are significantly better than others, suggesting that it would be
challenging for a programmer to locate these high performing specifications and obtain
the best performance. Many workloads (e.g., \emph{BH, DCT, MST}) also have
higher concentrations of specifications with suboptimal performance in
comparison to the best performing point, implying that, without effort, it is
likely that the programmer will end up with a resource specification that leads
to low performance. 

There are several sources for this performance variation. One important source is the
loss in thread-level parallelism as a result of a suboptimal resource
specification. Suboptimal specifications that are \emph{not} tailored to fit the
available physical resources lead to the underutilization of resources.
This causes a drop in the number of threads that can be executed concurrently,
as there are insufficient resources to support their execution. Hence, better
and more balanced utilization of resources enables higher thread-level
parallelism. Often, this loss in
parallelism from resource underutilization
manifests itself in what we refer to as a \emph{performance cliff}, where a
small deviation from an optimized specification can lead to
significantly worse performance, i.e., there is very high variation in
performance between two specification points that are nearby. To demonstrate the
existence and analyze the behavior of performance cliffs, we examine two representative workloads more closely. 

\begin{figure}[h]
\centering \begin{subfigure}[b]{0.99\linewidth} 
\centering
\includegraphics[width=0.99\textwidth]{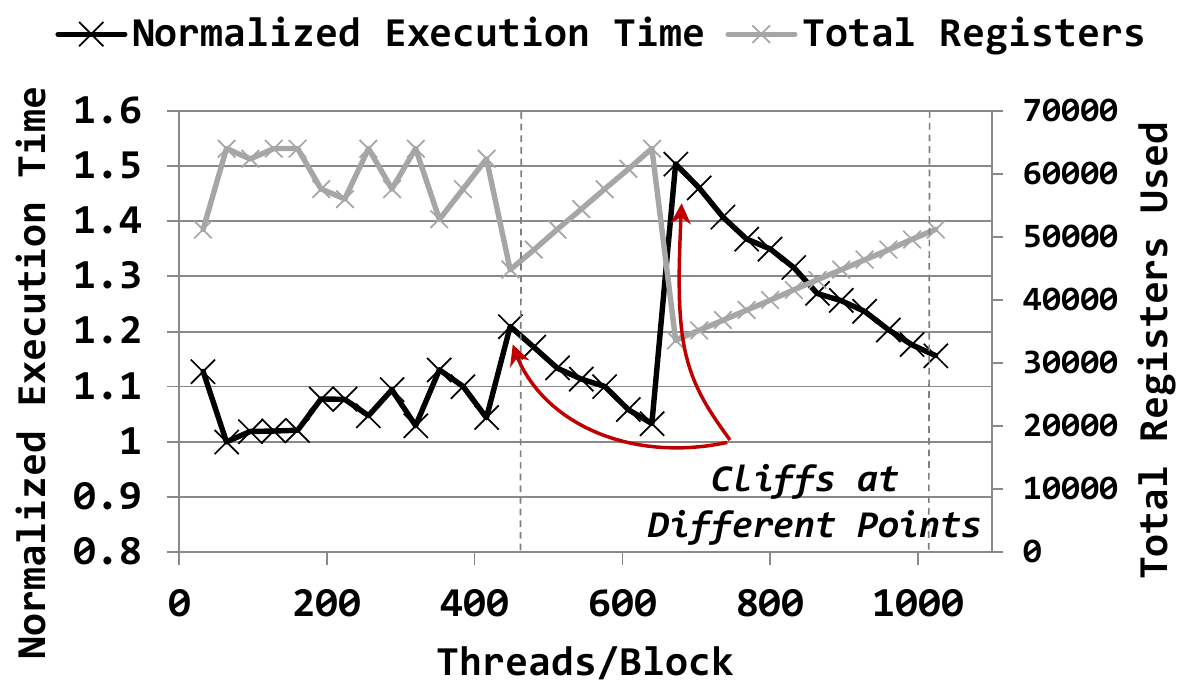}
\caption{Threads/block sweep}
\label{fig:mst-time}
\end{subfigure}
\begin{subfigure}[b]{0.99\linewidth}
\centering
\includegraphics[width=0.99\textwidth]{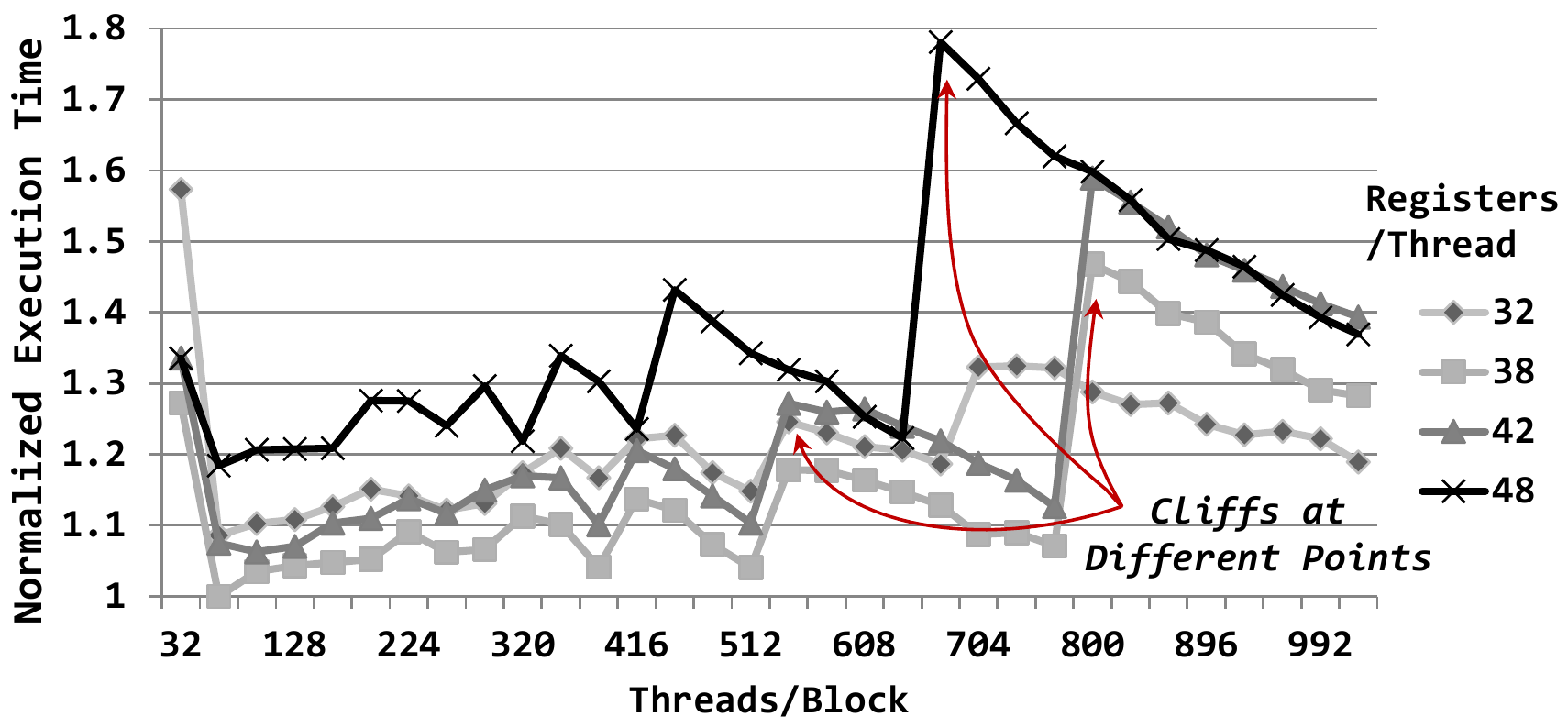} 
\caption{Threads/block \& Registers/thread sweep} 
\label{fig:mst-registers} 
\end{subfigure}
\caption{Performance cliffs in \emph{Minimum Spanning Tree} (\emph{MST}).
Reproduced from~\cite{zorua}.} 
\label{fig:mst-cliff}
\end{figure}

Figure~\ref{fig:mst-time} shows \One how the application execution time changes;
and \two how the corresponding number of registers, statically used, changes when the number of
threads per thread block increases from 32 to 1024 threads, for \emph{Minimum
Spanning Tree (MST)}~\cite{lonestar}. We make two observations.
 
First, let us focus on the execution time between 480 and 1024 threads per
block. As we go from 480 to 640 threads per block, execution time gradually
decreases. Within this window, the GPU can support two thread blocks running
concurrently for \emph{MST}. The execution time falls because the increase in the number of threads per block
improves the overall throughput (the number of thread blocks running
concurrently remains constant at two, but each thread block does more work in
parallel by having more threads per block). However, the corresponding total number of registers used by the blocks
also increases. At 640 threads per block, we reach the point where the total
number of available registers is not large enough to support two blocks. As a
result, the number of blocks executing in parallel drops from two to one,
resulting in a significant increase (50\%) in execution time, i.e., the
\emph{performance cliff}.\footnote{Prior
work~\cite{warp-level-divergence} has studied performing resource allocation at the finer warp
granularity, as opposed to the coarser granularity of a
thread block. As we discuss in
Section~\ref{sec:related} and demonstrate in Section~\ref{sec:eval}, this does
\emph{not} solve the problem of performance cliffs.}
We see many of these cliffs earlier in the graph as
well, albeit not as drastic as the one at 640 threads per block.

Second, Figure~\ref{fig:mst-time} shows the existence of performance cliffs when
we vary \emph{just one} system parameter{\textemdash}the number of threads per block. To make
things more difficult for the programmer, other parameters (i.e., registers per thread or scratchpad memory
per thread block) also need to be decided at the same time\ignore{, which makes the
programmer's task much harder in avoiding such cliffs}. Figure~\ref{fig:mst-registers}
demonstrates \new{that performance cliffs also exist} when the \emph{number of registers per
thread} is varied from 32 to 48.\footnote{We note that the
register usage reported by the compiler may vary from the actual runtime
register usage~\cite{sdk}, hence slightly altering the points at which cliffs
occur.} As this figure shows, performance cliffs now occur at \emph{different points} for \emph{different
registers/thread curves}, which makes optimizing resource specification, so as to avoid
these cliffs, much harder for the programmer.
 

\new{\emph{Barnes-Hut (BH)} is another application that
exhibits very significant performance cliffs depending on the number of threads
per block and registers per thread.  Figure~\ref{fig:bh-motiv}
plots the variation in performance with the number of threads per block
when \emph{BH} is compiled for a range of register sizes (between 24 and 48 registers
per thread). We make two observations from the figure.  First, 
similar to \emph{MST}, we observe a significant variation in
performance that manifests itself in the form of performance cliffs.
Second, we observe that the points at which the performance cliffs occur
change \newII{greatly} depending on the number of registers assigned to each thread
during compilation.}

\begin{figure}[h] \centering
\includegraphics[width=0.99\linewidth]{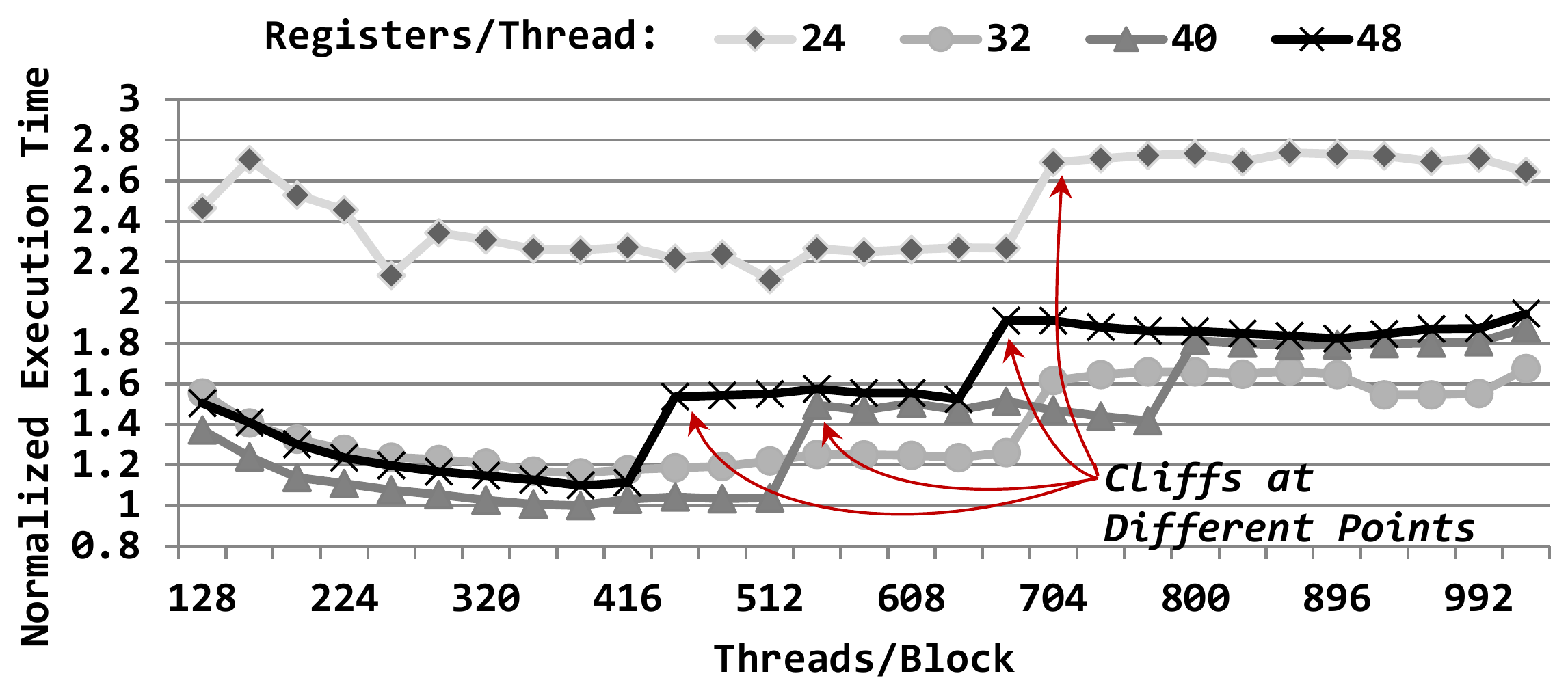}
\label{fig:bh-motiv} \caption{Performance cliffs in \emph{Barnes-Hut}
(\emph{BH}). Reproduced from~\cite{zorua}.}
\label{fig:bh-motiv} \end{figure} 

\new{We conclude that performance cliffs are pervasive across GPU programs,
and occur due to fundamental limitations of existing GPU hardware resource
managers, where resource management is static, coarse-grained, and tightly coupled to the
application resource specification. Avoiding performance cliffs by
determining more optimal resource specifications is a challenging
task, because the occurrence of these cliffs depends on several factors,
including the application characteristics, input data, and the underlying
hardware resources.}

\ignore{This phenomenon is not an artifact
of an application, but rather
a fundamental limitation of the GPU hardware that
distributes existing limited resources in a static and coarse-grain fashion. 
}

\subsection{Portability} 
\label{sec:motivation:port}

As we show in Section~\ref{sec:perf-cliffs}, tuning GPU applications to achieve good
performance on a given GPU is already a challenging task. To make things worse,
even after this tuning is done by
the programmer for one particular GPU architecture, it has to be \emph{redone} for
every new GPU generation (due to changes in the available physical resources
across generations) to ensure that good performance is retained.
We demonstrate this \emph{portability problem} by running sweeps of the three
parameters of the resource specification on various workloads, on three real GPU
generations: Fermi (GTX 480),
Kepler (GTX 760), and Maxwell (GTX 745).
\begin{figure}[h] 
\centering 
\begin{subfigure}[b]{0.90\linewidth} 
\centering
\includegraphics[width=0.99\textwidth]{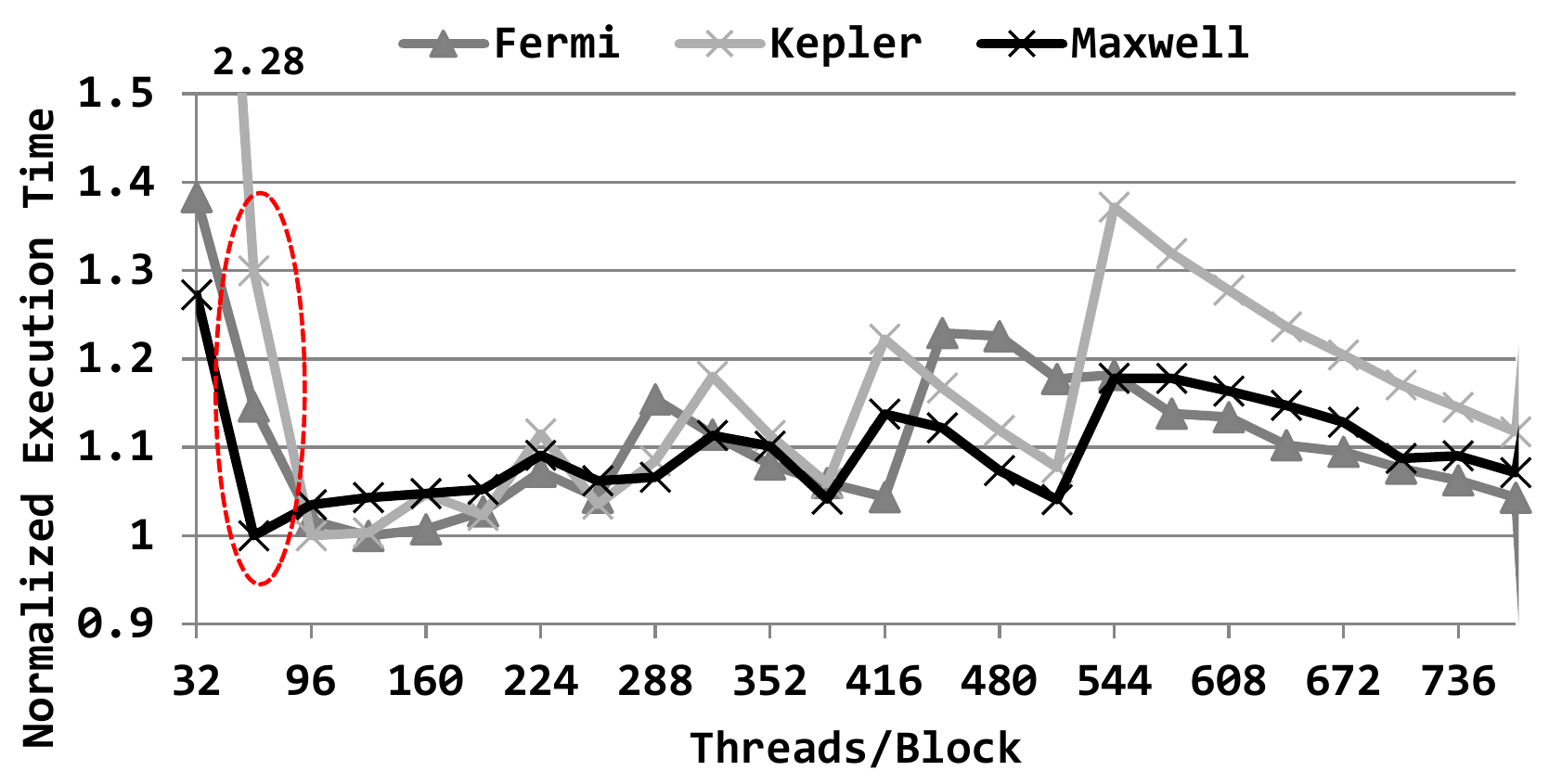} 
\caption{\emph{MST}}
\label{fig:mst-port} 
\end{subfigure} 
\begin{subfigure}[b]{0.90\linewidth}
\centering 
\includegraphics[width=0.99\textwidth]{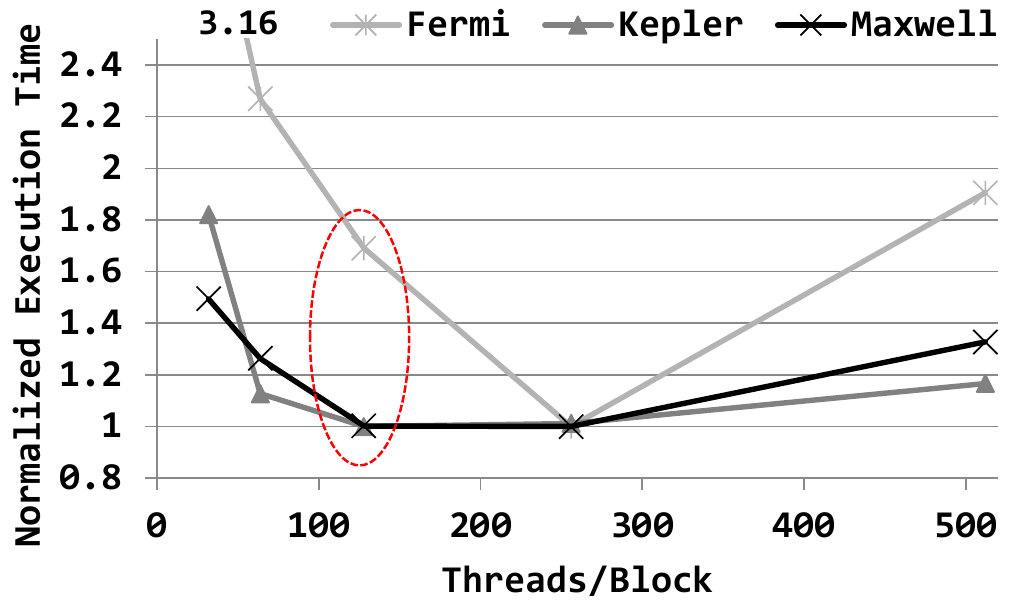}
\caption{\emph{DCT}} 
\label{fig:dct-port} 
\end{subfigure} 
\begin{subfigure}[b]{0.90\linewidth}
\centering 
\includegraphics[width=0.99\textwidth]{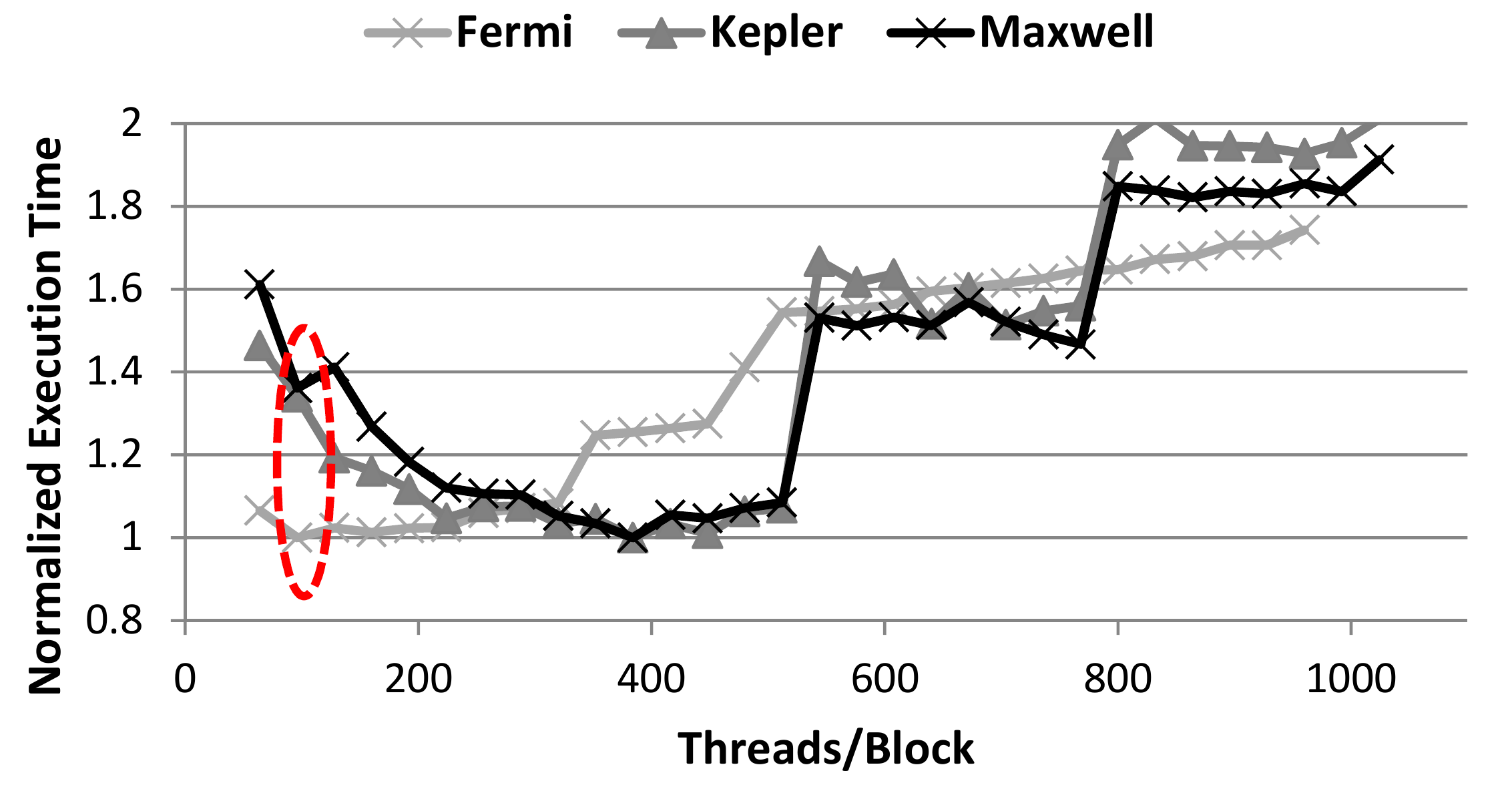}
\caption{\emph{BH}} 
\label{fig:bh-port} 
\end{subfigure} 
\caption{Performance
variation across different GPU generations (Fermi, Kepler, and Maxwell) for
\emph{MST}, \emph{DCT}, and \emph{BH}. Adapted from~\cite{zorua}.}
\label{fig:port} 
\end{figure}

Figure~\ref{fig:port} shows how the optimized performance points change between
different GPU generations for two representative applications (\emph{MST} and
\emph{DCT}). For every generation, results are normalized to the lowest
execution time for that particular generation. As we can see in
Figure~\ref{fig:mst-port}, the best performing points for different generations occur at
\emph{different} specifications because the application behavior changes with the variation in
hardware resources. 
For \emph{MST}, the \emph{Maxwell} architecture performs best at 64
threads per block. However, the same specification point is not efficient
for either of the other generations (\emph{Fermi} and \emph{Kepler}), producing 
15\% and 30\% lower performance, respectively, compared to the best specification
for each generation.
For \emph{DCT} (shown in Figure~\ref{fig:dct-port}),
both \emph{Kepler} and \emph{Maxwell} perform best at 128 threads per
block, but using the same specification for \emph{Fermi} would lead to a 
69\% performance loss. Similarly, for \emph{BH} (Figure~\ref{fig:bh-port}), the
optimal point for 
\emph{Fermi} architecture is at 96 threads per block. \new{However, using the
same configuration for the two later GPU architectures -- \emph{Kepler} and
\emph{Maxwell} could lead to very suboptimal performance results.
Using the same configuration results in as much as a 34\% performance loss
on \emph{Kepler}, and a 36\% performance loss on \emph{Maxwell}.

We conclude that the tight coupling between the programming model and the
underlying resource management in hardware imposes a significant challenge in
performance portability. To avoid suboptimal performance, an application has to
be \emph{retuned} by the programmer to find an optimized resource specification
for \emph{each}
GPU generation.
}

\subsection{Dynamic Resource Underutilization}
\label{sec:underutilized_resources}

Even when a GPU application is \emph{perfectly} tuned for a particular GPU
architecture, the on-chip resources are
typically not fully
utilized~\cite{virtual-register,gebhart-hierarchical,compiler-register,shmem-multiplexing,unified-register,
caba,asplos-sree,spareregister, mask, mosaic, ltrf-sadrosadati-asplos18}.
For example, it is well known that while the compiler conservatively allocates registers to
hold the \emph{maximum number} of live values throughout the execution, the number of
live values at any given time is well below the maximum for large portions of application execution time.
To determine the magnitude of this \emph{dynamic
underutilization},\footnote{\revision{Underutilization of registers occurs in two
major forms{\textemdash}\emph{static}, where registers are unallocated
throughout
execution~\cite{caba,warp-level-divergence,unified-register,spareregister,energy-register,mosaic,mask,ltrf-sadrosadati-asplos18},
and \emph{dynamic}, where utilization of the registers drops during runtime as a
result of early completion of warps~\cite{warp-level-divergence}, short register
lifetimes~\cite{virtual-register,gebhart-hierarchical,compiler-register} and
long-latency operations~\cite{gebhart-hierarchical,compiler-register}. We do not
tackle underutilization from long-latency operations (such as memory
accesses) in this paper, and leave the exploration of alleviating this type of
underutilization to future work.}} we conduct an experiment where we measure
the dynamic usage (per epoch) of both scratchpad memory and registers for
different applications with \emph{optimized} specifications in our workload pool. 

\begin{figure}[h]
  \centering
  \begin{subfigure}[b]{0.99\linewidth}
  \centering
  \includegraphics[width=0.99\textwidth]{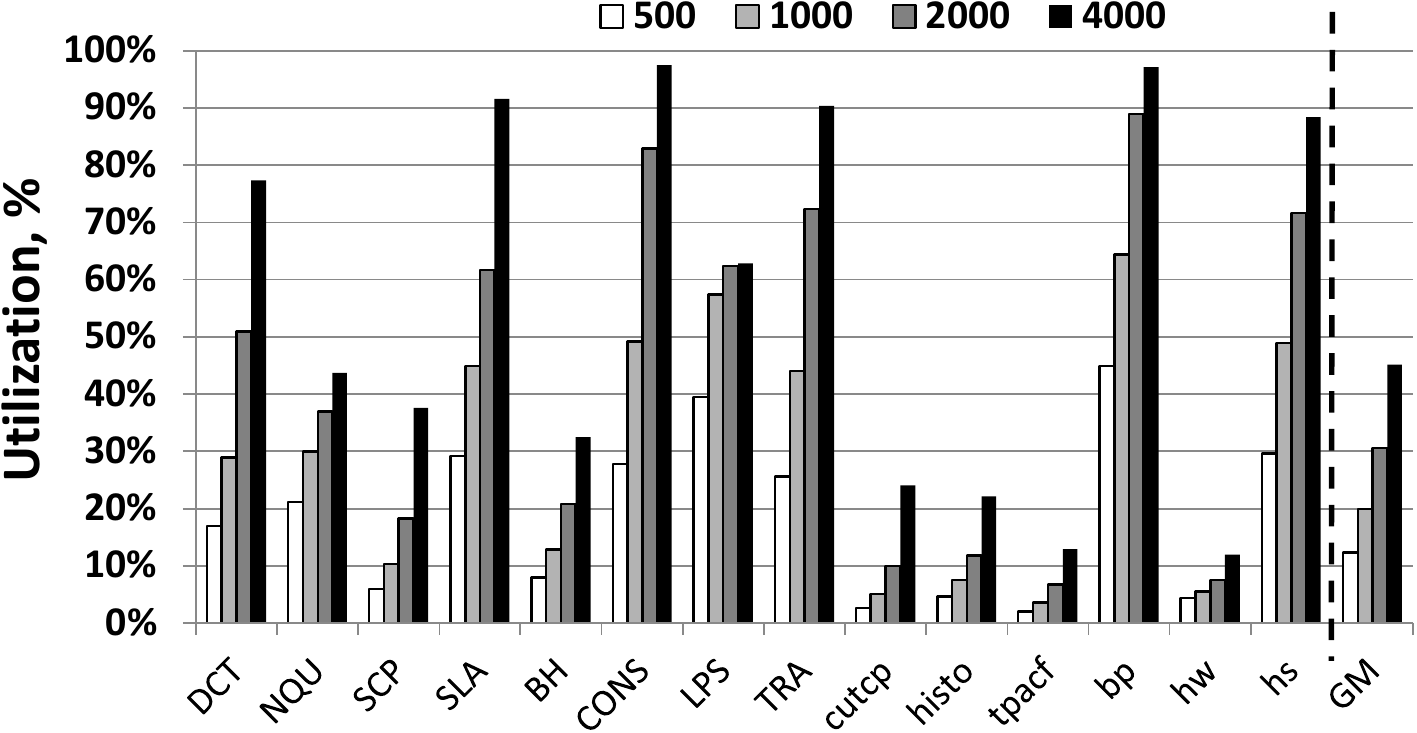}
  \caption{Scratchpad memory}
  \label{fig:scratchpad}
  \end{subfigure}
  \begin{subfigure}[b]{0.99\linewidth}
  \centering
  \includegraphics[width=0.99\textwidth]{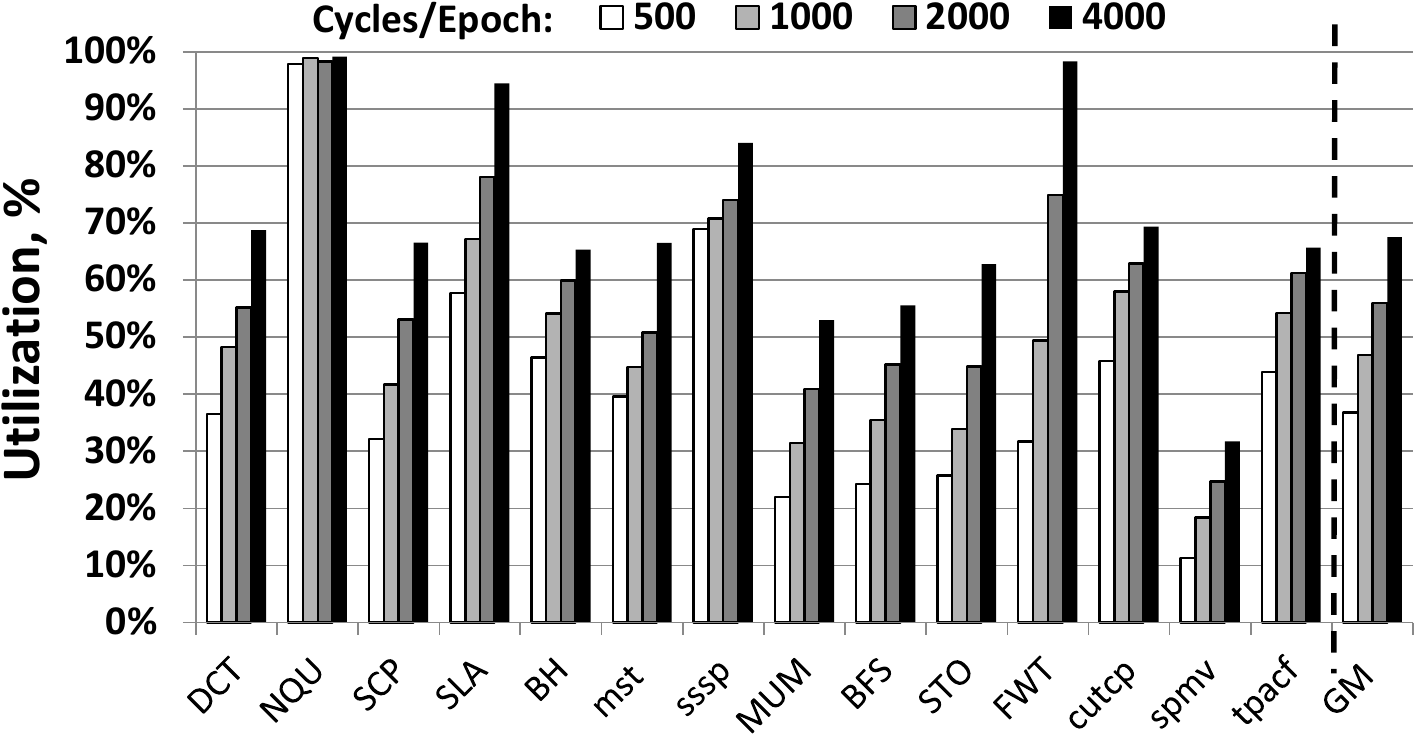}
  \caption{Registers}
  \label{fig:registers}
  \end{subfigure}
  \caption{Dynamic resource utilization for different length epochs.}
  \label{fig:utilization}
\end{figure}

We vary the length of epochs from
500 to 4000 cycles. Figure~\ref{fig:utilization} shows the results of this
experiment for \emph{(i)}~scratchpad memory (Figure~\ref{fig:scratchpad}) and
\emph{(ii)}~on-chip registers (Figure~\ref{fig:registers}).
We make two major observations from these figures.

\new{First, for relatively small epochs (e.g., 500 cycles), the average utilization
of resources is very low (12\% for scratchpad memory and 37\% for registers).
Even for the largest epoch size that we analyze (4000 cycles), the
utilization of scratchpad memory is still less than 50\%, and the utilization of
registers is less than 70\%.} This observation clearly suggests that there is an opportunity
for a better dynamic allocation of these resources that could allow higher
effective GPU parallelism.

Second, there are several noticeable applications, e.g., \emph{cutcp}, \emph{hw},
\emph{tpacf}, where utilization of the scratchpad memory is always lower than
15\%. This dramatic underutilization due to static resource allocation can lead
to significant loss in potential performance benefits for these applications.

In summary, we conclude that existing static on-chip resource allocation in GPUs
can lead to significant resource underutilization that can lead to suboptimal performance
and energy waste.

\subsection{Our Goal}
As we see above, the tight coupling between the resource specification and
hardware resource allocation, and the resulting heavy dependence of performance on
the resource specification, creates a number of challenges.
In this work, our goal is to alleviate these challenges by providing a mechanism
that can \One 
 ease the burden on the programmer by ensuring reasonable
performance, \emph{regardless of the resource specification}, by successfully
avoiding performance cliffs, while retaining performance for code with
optimized specification; \two enhance portability by minimizing the variation
in performance for optimized specifications across different GPU generations; and
\three maximize dynamic resource utilization even in highly optimized code to further
improve performance. We make two key observations from our studies above to help
us achieve this goal.

\textbf{Observation 1: }\emph{Bottleneck Resources.} We find that performance cliffs occur when
the amount of any resource required by an application exceeds the
physically available amount of that resource. This resource becomes a
\emph{bottleneck}, and limits the amount of parallelism that the GPU can
support. If it were possible to provide the application with a \emph{small
additional amount} of the bottleneck resource, the application can see a
significant increase in parallelism and thus avoid the performance cliff. \ignore{For
example, an application may require 18KB of scratchpad memory per thread block.
When run on a system with a 32KB scratchpad memory, the application can only run
one thread block at a time. If we could find a way to effectively provide 4KB
more of scratchpad, the application could run two blocks concurrently.}


\textbf{Observation 2: }\emph{Underutilized Resources.} As discussed in
Section~\ref{sec:underutilized_resources}, there is significant underutilization
of resources at runtime. These underutilized resources could be employed to
support more parallelism at runtime, and thereby alleviate the aforementioned
challenges.

We use these two observations to drive our resource virtualization solution,
which we describe 
next.
\section{\X: Our Approach}
In this work, we design \X, a framework that provides the illusion of more GPU
resources than physically available by decoupling the resource specification
from its allocation in the hardware resources. We introduce a new level of
indirection by virtualizing the on-chip resources to allow the hardware to
manage resources transparently to the programmer.

The virtualization provided by \X builds upon two \emph{key concepts} to
leverage the aforementioned observations. First, when there are insufficient
physical resources, we aim to provide the illusion of the required amount by
\emph{oversubscribing} the required resource. We perform this oversubscription
by leveraging the dynamic underutilization as much as possible, or by spilling to a
swap space in memory. This oversubscription essentially enables the illusion of
more resources than what is available (physically and statically), and supports
the concurrent execution of more threads. Performance cliffs are mitigated by
providing enough additional resources to avoid drastic drops in parallelism.
Second, to enable efficient oversubscription by leveraging underutilization, we
dynamically allocate and deallocate physical resources depending
on the requirements of the application during execution. We manage the
virtualization of each resource \emph{independently} of other resources to maximize
its runtime utilization.

\begin{figure}[h] \centering
\includegraphics[width=0.49\textwidth]{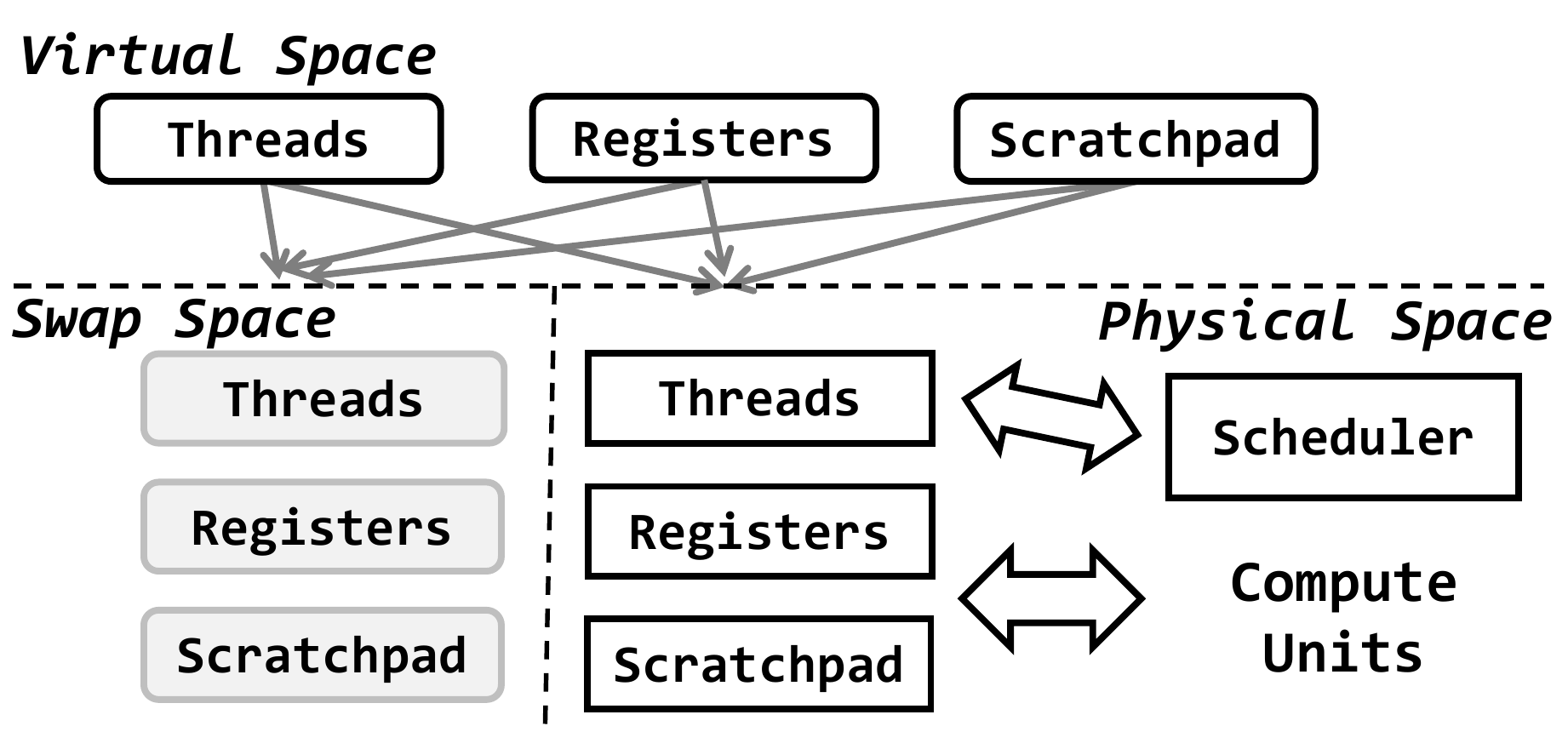}
\caption{High-level overview of \X. Reproduced from~\cite{zorua}.} \label{fig:overview} \end{figure}

Figure~\ref{fig:overview} depicts the high-level overview of the virtualization
provided by \X. 
The \emph{virtual space} refers to the \emph{illusion} of the quantity of available
resources. The \emph{physical space} refers to the \emph{actual} hardware resources
(specific to the GPU architecture), and the \emph{swap space} refers to the resources
that do not fit in the physical space and hence are \emph{spilled} to other
physical locations.
For the register file and scratchpad memory, the swap space is mapped to global
memory space in the memory hierarchy. For threads, only
those that are mapped to the physical space are available for scheduling
and execution at any given time. If a thread is mapped to the swap space, its
state (i.e., the PC and the SIMT stack) is saved in memory. Resources in the
virtual space can be freely re-mapped between the physical and swap
spaces to maintain the illusion of the virtual space resources.
 
In the baseline architecture, the thread-level parallelism that can be
supported,
and hence the throughput obtained from the GPU, depends on the quantity of
\emph{physical resources}. With the virtualization enabled by \X, the
parallelism that can be supported now depends on the quantity of \emph{virtual
resources} (and how their mapping into the physical and swap spaces is managed). Hence, the size of the virtual space for each resource plays the key
role of determining the parallelism that can be exploited. Increasing
the virtual space size enables higher parallelism, but leads to higher swap
space usage. It is critical to minimize accesses to the swap space to avoid the
latency overhead and capacity/bandwidth contention associated with accessing the
memory hierarchy. 

In light of this, there are two key challenges that need to be addressed to
effectively virtualize on-chip resources in GPUs. We now discuss these
challenges and provide an overview of how we address them.

\subsection{Challenges in Virtualization} \label{sec:virt-challenges}
\textbf{Challenge 1:}\emph{ Controlling the Extent of Oversubscription.} A key
challenge is to determine the \emph{extent} of oversubscription, or the size of
the virtual space for each resource. As discussed above, increasing the size of
the virtual space enables more parallelism. Unfortunately, it could
also result in more spilling of resources to the swap space. Finding the tradeoff
between more parallelism and less overhead is challenging, because the dynamic
resource requirements of each thread tend to significantly fluctuate throughout
execution. As a result, the size of the virtual space for each resource
needs to be \emph{continuously} tuned to allow the virtualization to adapt to
the runtime requirements of the program. 

\textbf{Challenge 2:} \emph{Control and Coordination of Multiple Resources.}
\ignore{Another critical challenge is to efficiently map the continuously varying
virtual resource space to the physical and swap spaces. 
This is critical as it is very important to minimize accesses to
the swap space. Accessing the swap space for the register file or scratchpad
involves expensive accesses to global memory, due to the added
latency and contention. Also, only those threads that are mapped to the
physical space are available to the warp scheduler for selection. Furthermore, each
thread requires \emph{multiple} resources for execution, each of which may be mapped to
the physical or swap space. It is critical to \emph{coordinate} the mapping of
these different virtual resources to ensure that a thread has all the
resources required at any given time mapped to the physical space, to enable
execution with minimal overhead. Thus, an effective virtualization framework
must coordinate the allocation of \emph{multiple} physical resources.}
Another critical challenge is to efficiently map the continuously varying
virtual resource space to the physical and swap spaces. 
This is important for two reasons. First, it is critical to minimize accesses to
the swap space. Accessing the swap space for the register file or scratchpad
involves expensive accesses to global memory, due to the added
latency and contention. Also, only those threads that are mapped to the
physical space are available to the warp scheduler for selection. Second, each
thread requires multiple resources for execution. It is critical to
\emph{coordinate} the allocation and mapping of
these different resources to ensure that an executing thread has \emph{all} the
required resources allocated to it, while minimizing accesses
to the swap space. Thus, an effective virtualization framework
must coordinate the allocation of \emph{multiple} on-chip resources.

\subsection{Key Ideas of Our Design}

To solve these challenges, \X employs two key ideas. First, we leverage the software
(the compiler) to provide annotations with information regarding the resource
requirements of each \emph{phase} of the application. This information enables the
framework to make intelligent dynamic decisions, with respect to both the size of
the virtual space and the allocation/deallocation of resources (Section~\ref{sec:key_idea_phases}).

Second, we use an adaptive runtime system to control the allocation of resources
in the virtual space and their mapping to the physical/swap spaces. This allows us to \One
dynamically alter the size of the virtual space to change the extent of
oversubscription; and \two continuously coordinate the allocation of multiple
on-chip resources and the mapping between their virtual and physical/swap
spaces, depending
on the varying runtime requirements of each thread
(Section~\ref{sec:key_idea_coordinator}).

\subsubsection{Leveraging Software Annotations of Phase Characteristics}
\label{sec:key_idea_phases} 
We observe that the runtime variation in resource requirements
(Section~\ref{sec:underutilized_resources}) typically occurs at 
the granularity of \emph{phases} of a few tens of instructions. This variation
occurs because different parts of kernels perform different operations that
require different resources. For example, loops that primarily load/store data
from/to 
scratchpad memory tend to be less register heavy. Sections of
code that perform specific computations (e.g., matrix transformation, graph
manipulation), can either be register heavy or primarily operate out of
scratchpad. Often, scratchpad memory is used for only short
intervals~\cite{shmem-multiplexing}, e.g., when data exchange between threads is
required, such as for a reduction operation. 

\begin{figure}[h] \centering
\includegraphics[width=0.49\textwidth]{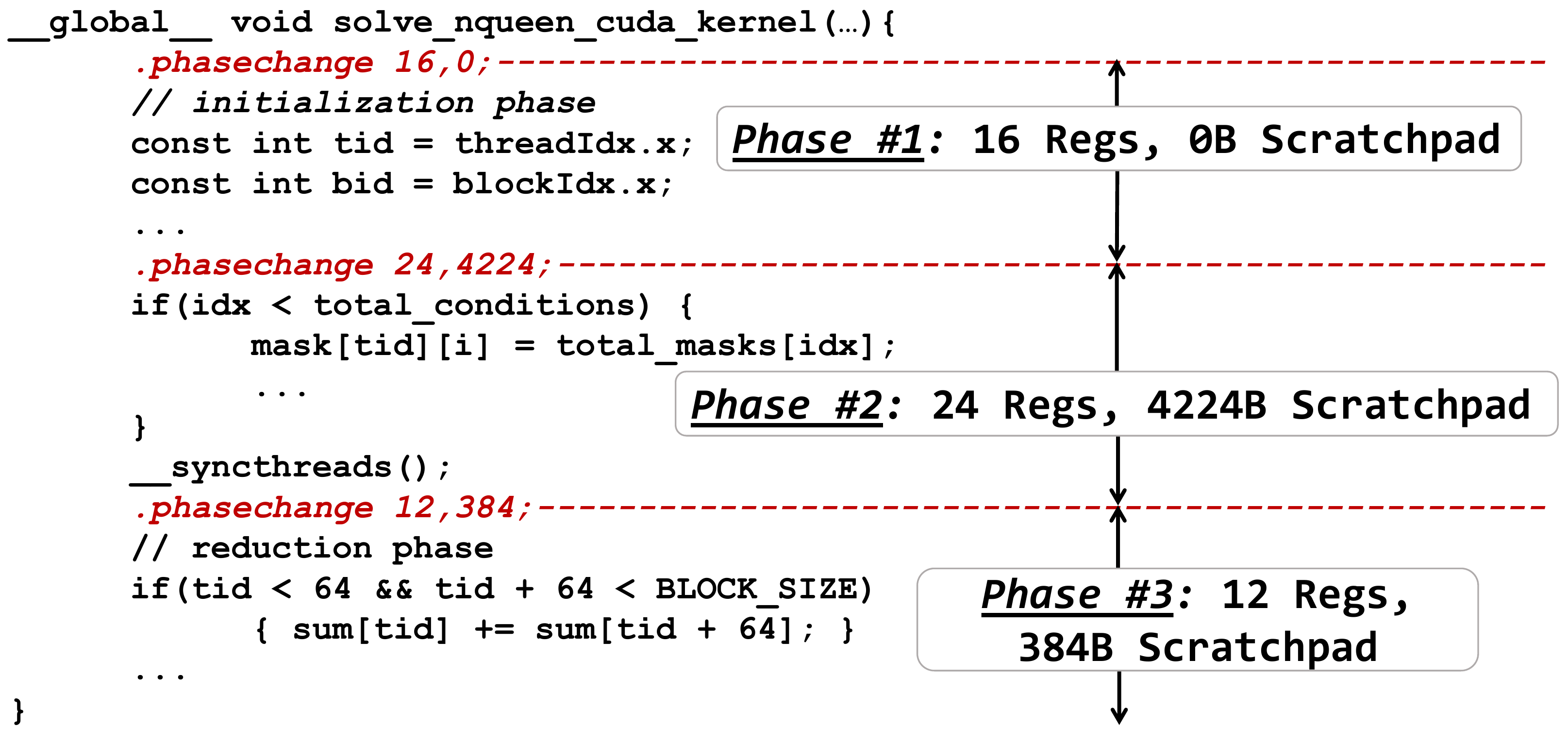}
\caption{Example phases from \emph{NQU}. Reproduced from~\cite{zorua}.} \label{fig:phases} \end{figure}

Figure~\ref{fig:phases} depicts a few example phases from the \emph{NQU}
(\emph{N-Queens
Solver})~\cite{NQU} kernel. \emph{NQU} is a scratchpad-heavy application, but it does
not use the
scratchpad at all during the initial computation phase. During its second phase, it performs
its primary computation out of the scratchpad, using as much as 4224B. During its
last phase, the scratchpad is used only for reducing results, which requires
only 384B. There is also significant variation in the maximum
number of live registers in the different phases.

\begin{figure}[h] \centering
\includegraphics[width=0.49\textwidth]{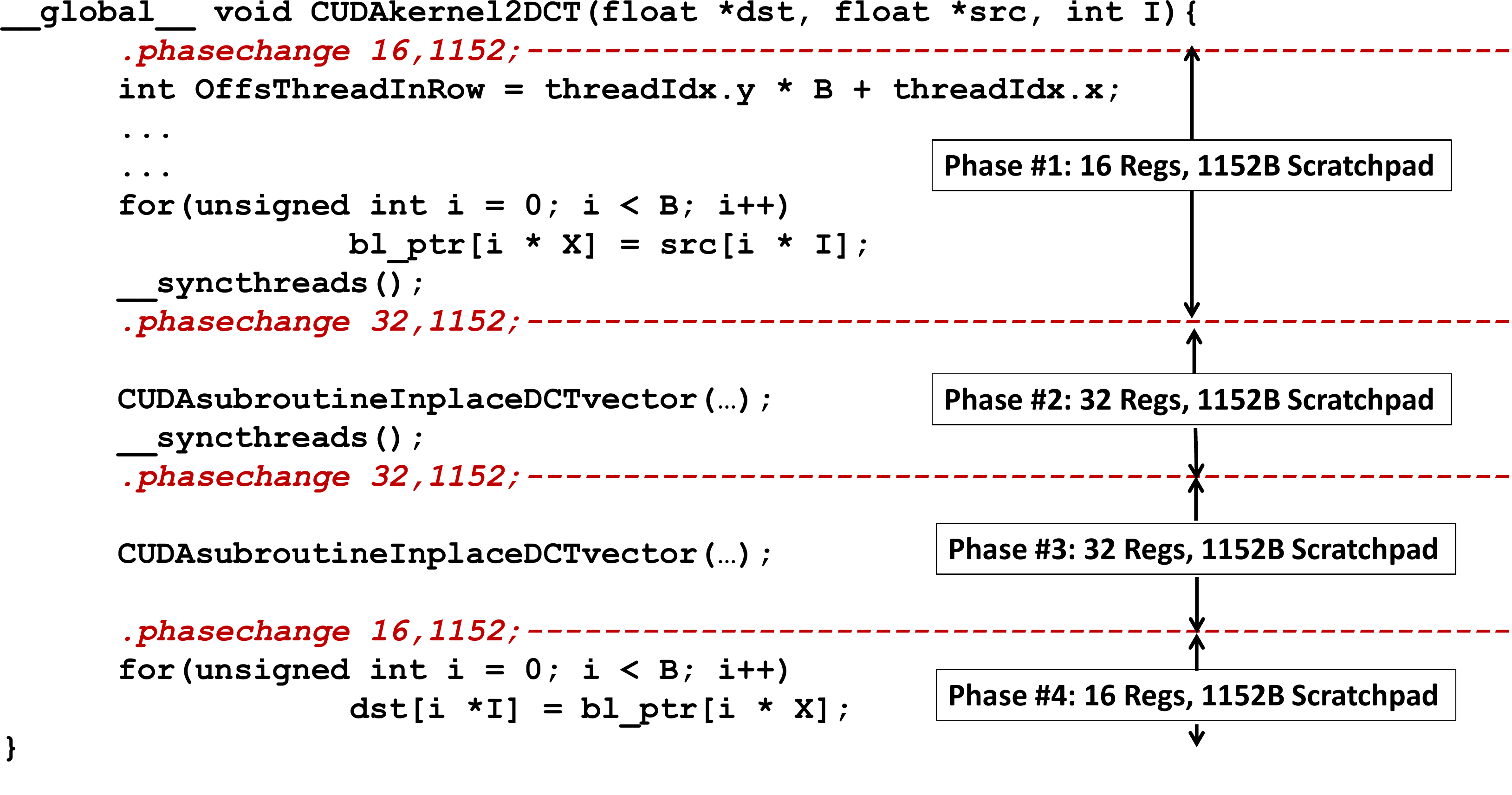}
\caption{Example phases from \emph{DCT}} \label{fig:phases2} \end{figure} 

Another example of phase variation from the \emph{DCT} (\emph{Discrete Fourier
Transform}) kernel is depicted in Figure~\ref{fig:phases2}. \emph{DCT} is both
register and scratchpad-intensive. The scratchpad memory usage \new{does not vary} in
this kernel. However, the register usage significantly
varies -- the register usage increases by 2X in the second and third phase in
comparison with the first and fourth phase.

In order to capture both the resource requirements as well as their variation
over time, we 
partition the program into a number of \emph{phases}. A phase is a
sequence of instructions with sufficiently different resource requirements than
adjacent phases. Barrier or fence operations also indicate a change in
requirements for a different reason{\textemdash}threads that are waiting at a barrier do
not immediately require the thread slot that they are holding. 
We interpret barriers and fences as phase boundaries since they potentially alter
the utilization of their thread slots. The compiler inserts special instructions
called \emph{phase specifiers} to mark the start of a new phase. Each phase
specifier contains information regarding the resource requirements of the next
phase. Section~\ref{sec:phase_specifiers} provides more detail on the semantics
of phases and phase specifiers. 
 
A phase forms the basic unit for resource allocation and
de-allocation, as well as for making oversubscription decisions. It offers
a finer granularity than an \emph{entire thread} to make such decisions.
The phase specifiers provide information on the \emph{future resource usage} of
the thread at a phase boundary. This enables \One preemptively controlling the
extent of oversubscription at runtime, and \two dynamically allocating and
deallocating resources at phase boundaries to maximize utilization of the
physical resources.

\subsubsection{Control with an Adaptive Runtime System}
\label{sec:key_idea_coordinator}
\begin{figure}[h] \centering
\includegraphics[width=0.45\textwidth,scale=0.8]{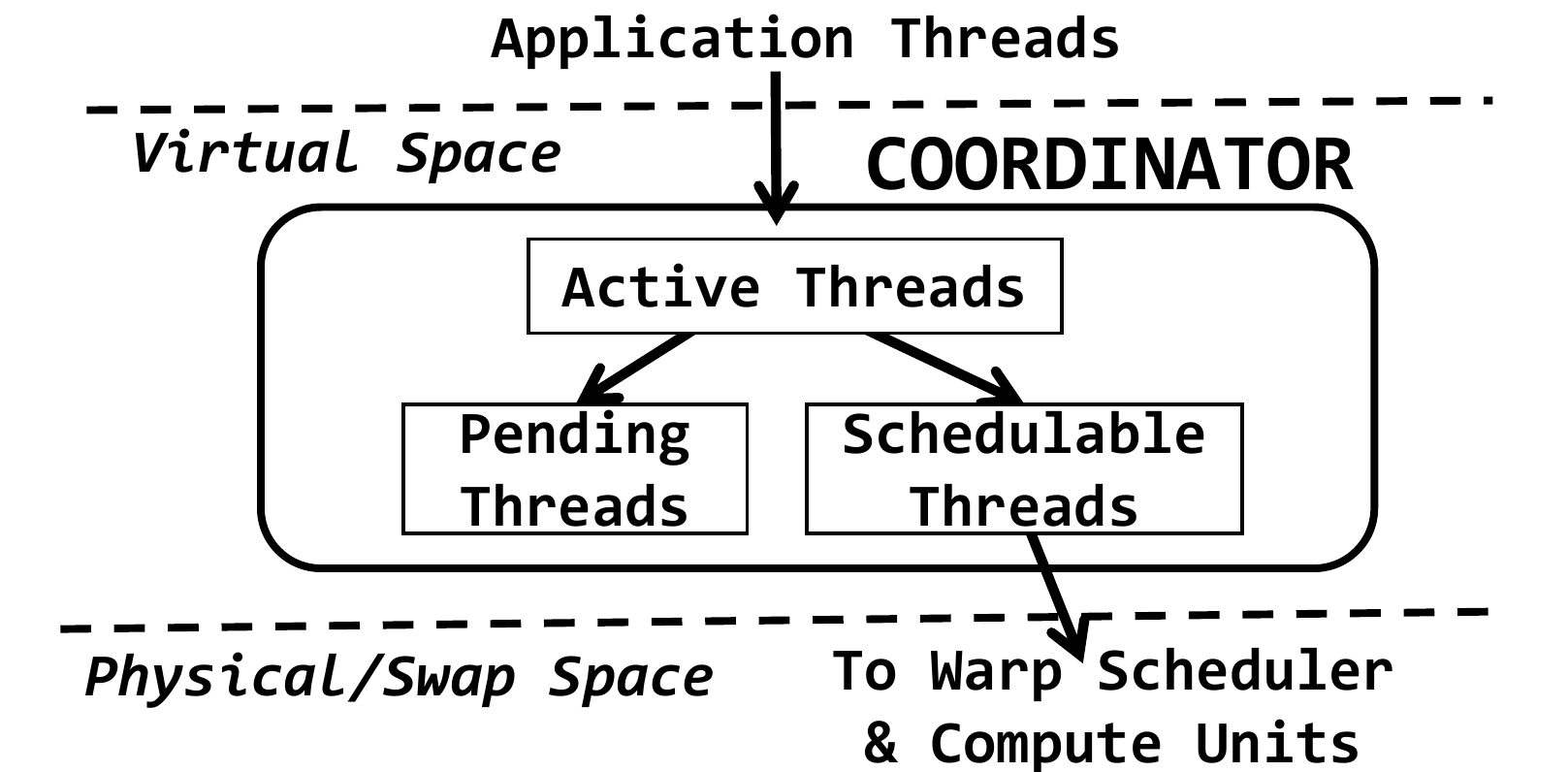}
\caption{Overview of the coordinator. Reproduced from~\cite{zorua}.} \label{fig:coordinator} \end{figure}

Phase specifiers provide information to make oversubscription and allocation/deallocation
decisions. However, we still need a way to make decisions on the
extent of oversubscription and appropriately allocate resources at runtime. To this
end, we use an adaptive runtime system, which we refer to as the
\emph{coordinator}. Figure~\ref{fig:coordinator} presents an overview of the
coordinator.

The virtual space enables the illusion of a larger amount of each of the
resources than what is physically available, to adapt to different application
requirements. This illusion enables higher thread-level parallelism than what
can be achieved with solely the fixed, physically available resources, by
allowing more threads to execute concurrently. The size of the virtual space at a given
time determines this parallelism, and those threads that are effectively
executed
in parallel are referred to as \emph{active threads}. All active threads have
thread slots allocated to them in the virtual space (and hence can be executed),
but some of
them may not be mapped to
the physical space at a given time. As discussed previously,
the resource requirements of each application continuously change during
execution. To adapt to these runtime changes, the coordinator leverages
information from the phase specifiers to make decisions on oversubscription. The
coordinator makes these decisions at
every phase boundary and thereby controls the size of the virtual space for each resource
(see
Section~\ref{sec:walkthrough}).

To enforce the determined extent of oversubscription, the coordinator allocates
all the required resources (in the virtual space) for only a \emph{subset} of
threads from the active threads. Only these dynamically selected threads, referred to
as \emph{schedulable threads}, are available to the warp scheduler and compute
units for execution. The coordinator, hence, dynamically partitions the active threads into
\emph{schedulable threads} and the \emph{pending threads}. 
Each thread is swapped
between \emph{schedulable} and \emph{pending} states, depending on the
availability of resources in the virtual space.
Selecting only a subset of threads to execute at any time ensures that
the determined size of
the virtual space is not exceeded for any resource, and helps coordinate the
allocation and mapping of multiple on-chip resources to minimize expensive data
transfers between the physical and swap spaces (discussed in
Section~\ref{sec:mechanism}). 
%

\subsection{Overview of \X} In summary, to effectively address the challenges
in virtualization by leveraging the above ideas in design, \X employs a
software-hardware codesign that comprises three components: \One
\textbf{\emph{The compiler}} annotates the program by adding special
instructions (\emph{phase specifiers}) to partition it into \emph{phases} and to
specify the resource needs of each phase of the application. \two
\textbf{\emph{The coordinator}}, a hardware-based adaptive runtime system, 
uses the compiler annotations to dynamically allocate/deallocate resources for
each thread at phase boundaries. The coordinator plays the key role of
continuously controlling the extent of the oversubscription (and hence the size
of the virtual space) at each phase boundary.\ignore{by altering the number of threads
that are executing at any given time.} \three \textbf{\emph{Hardware
virtualization support}} includes a mapping table for each resource to
locate each virtual resource in either the physical space or the swap
space in main memory, and the machinery to swap resources between the physical
and swap spaces. 



\section{\X: Detailed Mechanism}
\label{sec:mechanism}
We now detail the operation and implementation of the various components of the \X
framework. 
\subsection{Key Components in Hardware}
\X has two key hardware components: \One the \emph{coordinator}
that contains queues to buffer the \emph{pending threads} and
control logic to make oversubscription and resource management
decisions, and \two \emph{resource mapping tables} to map each of the
resources to their corresponding physical or swap spaces.

\begin{figure}[h]
 \centering
 \includegraphics[width=0.49\textwidth]{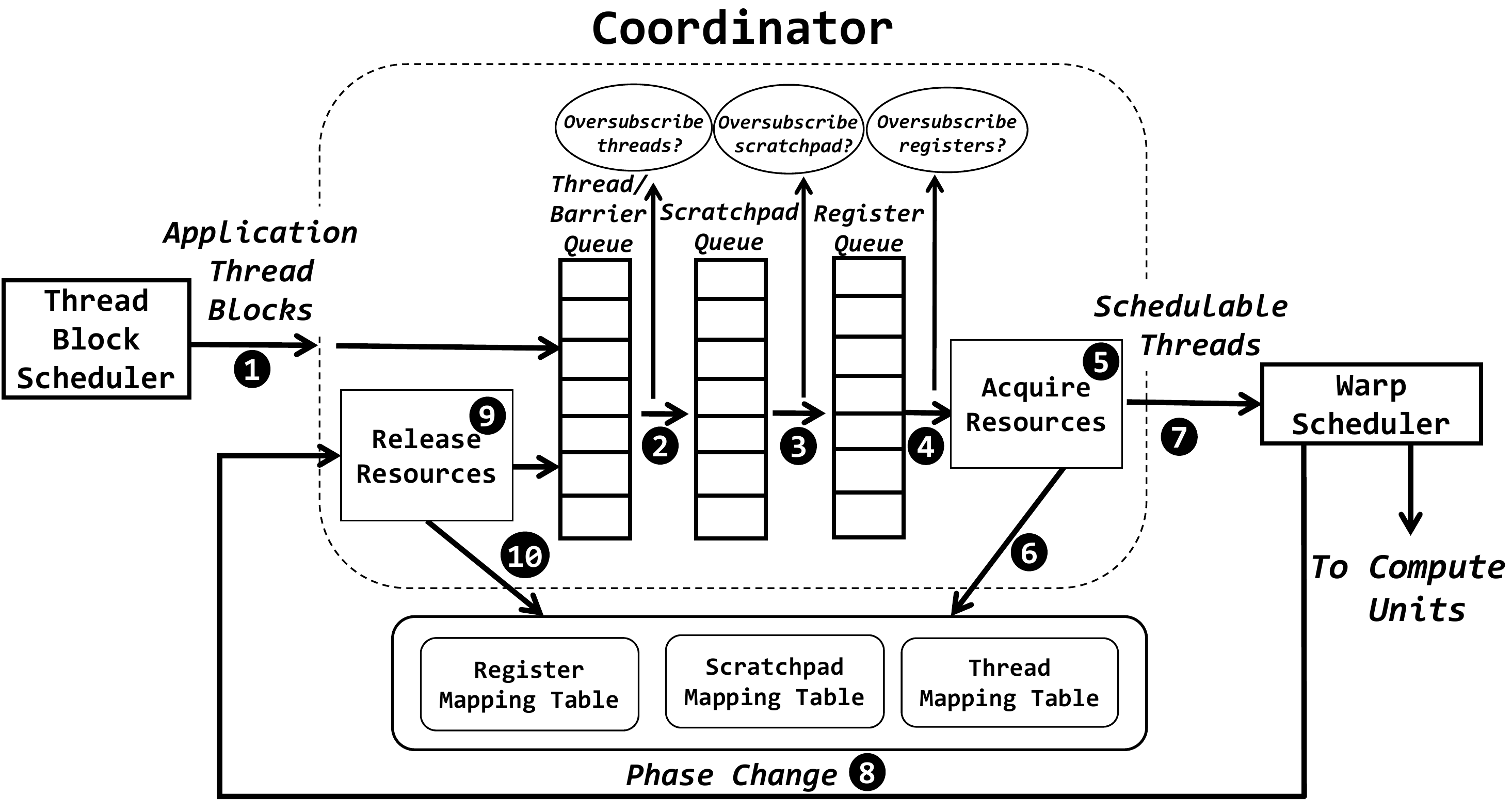}
 \caption{Overview of \X in hardware. Reproduced from~\cite{zorua}.} 
 \label{fig:full_overview}
\end{figure}

Figure~\ref{fig:full_overview} presents an overview of the hardware components
that are added to each SM. The coordinator interfaces with the thread block
scheduler (\ding{182}) to schedule new blocks onto an SM. It also
interfaces with the warp schedulers by providing a list of
\emph{schedulable warps} (\ding{188}).\footnote{We use an additional
 bit in each warp slots to indicate to the
 scheduler whether the warp is schedulable.} The resource mapping
tables are accessible by the coordinator and the compute
units. We present a detailed walkthrough of the operation of \X
and then discuss its individual components in more detail. \ignore{In
 addition, the coordinator interfaces with the on-chip resources
 using per-resource \emph{mapping tables} which are used to control
 the mapping of resources and their oversubscription at runtime
 (described in Section~\ref{sec:virtualizing_resources}). }

\subsection{Detailed Walkthrough}
\label{sec:walkthrough}
The coordinator is called into action by three events: \One
a new thread block is scheduled at the SM for execution, \two a
warp undergoes a phase change, or \three a warp or a thread block reaches the end of execution. 
Between these
events, the coordinator performs no action and execution proceeds as
usual. We now walk through the sequence of actions performed by the
coordinator for each type of event.


\textbf{Thread Block: Execution Start.} When a thread block is
scheduled onto an SM for execution (\ding{182}), the coordinator first
buffers it. 
The primary decision that the coordinator makes is to determine
whether or not to make each thread available to the scheduler for
execution. The granularity at which the coordinator makes decisions is
that of a warp, as threads are scheduled for execution at the
granularity of a warp (hence we use \emph{thread
  slot} and \emph{warp slot} interchangeably). Each warp requires
three resources: a thread slot, registers, and potentially
scratchpad. The amount of resources required is determined by the phase
specifier (Section~\ref{sec:phase_specifiers}) at the start of
execution, which is placed by the compiler into the code. The coordinator must
supply each warp with \emph{all} its required
resources in either the physical or swap space before presenting it to the warp scheduler for execution.

To ensure that each warp is furnished with its resources and
to coordinate potential oversubscription for each resource, the
coordinator has three queues{\textemdash}\emph{thread/barrier, scratchpad,
 and register queues}. The three queues together essentially house
the \emph{pending threads}. Each warp must traverse each queue
(\ding{183}~\ding{184}~\ding{185}), as described next, before becoming eligible to be
scheduled for execution. The coordinator allows a warp to traverse
a queue when \emph{(a)} it has enough of the corresponding resource
available in the physical space, or \emph{(b)} it has an insufficient
resources in the physical space, but has decided to oversubscribe and allocate
the resource in the swap space. The total size of the resource allocated in the physical
and swap spaces cannot exceed the determined virtual space size. The coordinator determines the availability of
resources in the physical space using the mapping tables
(see Section~\ref{sec:virtualizing_resources}). If there is an insufficient
amount of a resource in the physical space, the coordinator needs to decide
whether or not to increase the virtual space size for that particular resource by
oversubscribing and using swap space. We describe the
decision algorithm in Section~\ref{sec:oversubscription_decision}. If
the warp cannot traverse \emph{all} queues, it is left waiting in the first
(\emph{thread/barrier}) queue
until the next coordinator event. Once a warp has traversed \emph{all} the
queues, the coordinator acquires all the resources required for the warp's
execution (\ding{186}). The corresponding mapping tables for each
resource is updated (\ding{187}) to assign resources to the warp, as
described in Section~\ref{sec:virtualizing_resources}.

\textbf{Warp: Phase Change.} At each phase change (\ding{189}), the warp is removed
from the list of schedulable warps and is returned to the coordinator
to acquire/release its resources. Based on the information in its
phase specifier, the coordinator releases the
resources that are no longer live and hence are no longer required (\ding{190}).
The coordinator updates the mapping tables to free these resources
(\ding{191}). The warp is then placed into a specific queue, depending on which
live resources it retained from the previous phase and which new resources it
requires. The warp then
attempts to traverse the remaining queues
(\ding{183}~\ding{184}~\ding{185}), as described above. A warp that
undergoes a phase change as a result of a barrier instruction is
queued in the \emph{thread/barrier queue} (\ding{183}) until all warps
in the same thread block reach the barrier.
  
\textbf{Thread Block/Warp: Execution End.} When a warp completes
execution, it is returned to the coordinator to release any
resources it is holding. Scratchpad is released only when
the entire thread block completes execution. When the coordinator has
free warp slots for a new thread block, it requests the thread block
scheduler (\ding{182}) for a new block.

 
\textbf{Every Coordinator Event.} At any event, the coordinator
attempts to find resources for warps waiting at the queues, to enable them to
execute. Each warp in each queue (starting from the \emph{register queue}) is
checked for the availability of the required resources. If the coordinator is
able to allocate resources in the physical or swap space without exceeding the
determined size of virtual space, the warp is allowed to traverse the queue.
 
\subsection{Benefits of Our Design}
\textbf{Decoupling the Warp Scheduler and Mapping Tables from the Coordinator.} Decoupling the
warp scheduler from the coordinator enables \X to use any scheduling
algorithm over the schedulable warps to enhance performance. One case
when this is useful is when increasing parallelism degrades
performance by increasing cache miss
rate or causing memory
contention~\cite{tor-micro12,nmnl-pact13,Kayiran-micro2014}. Our decoupled
design allows this challenge to be addressed
independently from the coordinator using more intelligent scheduling
algorithms~\cite{tor-micro12,Kayiran-micro2014,largewarp, medic} and cache management
schemes~\cite{DBLP:conf/ics/LiSDSHZ15,DBLP:conf/hpca/XieLWSW15,DBLP:conf/hpca/LiRJOEBFR15,medic}. Furthermore,
decoupling the mapping tables from the coordinator allows
easy integration of any implementation of the mapping tables that may
improve efficiency for each resource.

\ignore{
\textbf{Throttling Warps.} 
Even in cases where the overhead of oversubscription
is outweighed by the increase in parallelism that it offers, increasing
parallelism in some cases can lead to performance degradation by increasing
the cache miss rate in cache-sensitive workloads
~\cite{tor-micro12,nmnl-pact13}. 
The decoupling between the coordinator and the thread scheduler
offers a opportunity to address this challenge independent of the \X framework.
More intelligent scheduling algorithms~\cite{tor-micro12} and cache management schemes~\cite{x,y,z} can be used
along with \X to maximize the caching performance. 
}

\textbf{Coordinating Oversubscription for Multiple Resources.}
The queues help ensure that a warp is allocated
\emph{all} resources in the virtual space before execution. 
They \One ensure an ordering in resource allocation to avoid deadlocks, and
\two enforce priorities between resources. In our evaluated approach, we use
the following order of priorities: threads, scratchpad, and registers. We prioritize scratchpad over
registers, as scratchpad is shared by all warps in a block and hence has a higher
value by enabling more warps to execute. We prioritize threads over scratchpad,
as it is wasteful to allow warps stalled at a barrier to acquire other
resources{\textemdash}other
warps that are still progressing towards the barrier may be starved of the resource they
need. Furthermore, managing each resource independently allows
different oversubscription policies for each resource and enables fine-grained
control over the size of the virtual space for that resource.\ignore{Our
technical report~\cite{zorua-tr} has a more detailed discussion of these
benefits.} 

\textbf{Flexible Oversubscription.} \X's design can flexibly
enable/disable swap space usage, as the dynamic fine-grained
management of resources is independent of the swap space. Hence, in cases where the
application is well-tuned to utilize the available resources, 
swap space usage can be disabled or minimized, and \X can still improve
performance by reducing dynamic underutilization of resources. Furthermore, 
different oversubscription algorithms can be flexibly employed to manage the
size of the virtual space for each resource (independently or cooperatively).
These algorithms can be  
designed for different purposes, e.g., minimizing swap space usage, improving
fairness in a multikernel setting, reducing energy, etc. In
Section~\ref{sec:oversubscription_decision},
we describe an example algorithm to improve performance by making a
good tradeoff between improving parallelism and reducing swap space usage.

\textbf{Avoiding Deadlocks.} 
\new{A resource allocation deadlock could happen if resources are
distributed among too many threads, such that \emph{no} single thread is able to
obtain enough necessary resources for execution.}
Allocating resources using \emph{multiple} ordered queues
helps avoid deadlocks in resource allocation \new{in
three ways}. \new{First, new resources are allocated to a warp only once the warp has
traversed \emph{all} of the queues.} This ensures that resources are not wastefully
allocated to warps that will be stalled anyway. Second, \new{a warp is} allocated
resources based on how many resources it already has, i.e. how many queues it
has already traversed. Warps that \new{already hold} multiple live
resources are prioritized in allocating new resources over warps that \new{do
\emph{not}
hold} any resources. Finally, if there are insufficient resources to maintain a
minimal level of parallelism (e.g., 20\% of SM occupancy in our evaluation), the coordinator
handles this rare case by simply oversubscribing resources to ensure that there
is no deadlock in allocation. 

\textbf{Managing More Resources.} \new{Our} design also allows flexibly adding more
resources to be managed by the virtualization framework, for example, thread
block slots. Virtualizing a new resource with Zorua simply requires adding a new
queue to the coordinator and a new mapping table to manage the virtual to
physical mapping.

\subsection{Oversubscription Decisions}
\label{sec:oversubscription_decision}
\textbf{Leveraging Phase Specifiers.} \X leverages the information
provided by phase specifiers (Section~\ref{sec:phase_specifiers})
to make oversubscription decisions for each
phase. For each resource, the coordinator checks whether allocating
the requested quantity according to the phase specifier would cause
the total swap space to exceed an \emph{oversubscription
 threshold}, or \emph{o\_thresh}. This threshold essentially dynamically sets the size of
 the virtual space for each resource. The coordinator allows
oversubscription for each resource only within its threshold. \emph{o\_thresh}
is dynamically determined to
adapt to the characteristics of the workload, and tp ensure good performance by achieving a good tradeoff between the overhead of
oversubscription and the benefits gained from parallelism.

\textbf{Determining the Oversubscription Threshold.} In order to make
the above tradeoff, we use two architectural statistics: \One idle time at the
cores,
\emph{c\_idle}, as an indicator for potential performance
benefits from parallelism; and \two memory idle time (the idle cycles when
all threads are stalled waiting for data from memory or the memory
pipeline), \emph{c\_mem}, as an indicator of a saturated memory
subsystem that is unlikely to benefit from more parallelism.\footnote{This
is similar to the approach taken by prior work~\cite{nmnl-pact13} to
estimate the performance benefits of increasing parallelism.} We use
Algorithm~\ref{alg:determining_threshold} to determine
\emph{o\_thresh} at runtime. Every \emph{epoch}, the
change in \emph{c\_mem} is compared with the change in
\emph{c\_idle}. If the increase in \emph{c\_mem} is greater, this
indicates an increase in pressure on the memory subsystem, suggesting
both lesser benefit from parallelism and higher overhead from
oversubscription. In this case, we reduce \emph{o\_thresh}. On the
other hand, if the increase in \emph{c\_idle} is higher, this is
indicative of more idleness in the pipelines, and higher potential
performance from parallelism and oversubscription. We increase
\emph{o\_thresh} in this case, to allow more oversubscription and
enable more parallelism. Table~\ref{table:thresholds} describes the variables
used in Algorithm~\ref{alg:determining_threshold}.\ignore{We include more
detail on these variables in our technical report~\cite{zorua-tr}.}

\begin{algorithm} 
{\footnotesize{
\begin{algorithmic}[1]
\State \emph{o\_thresh} $=$ \emph{o\_default} \Comment{Initialize threshold} 
\For{\emph{each epoch}}
\State \emph{c\_idle\_delta $=$ (c\_idle $-$ c\_idle\_prev)} \Comment{Determine the
change in c\_idle and c\_mem from the previous epoch}
\State \emph{c\_mem\_delta $=$ (c\_mem $-$ c\_mem\_prev)}
\If{\emph{(c\_idle\_delta $-$ c\_mem\_delta) > c\_delta\_thresh}}
\Comment{Indicates more idleness and potential for benefits from parallelism}
\State \emph{o\_thresh $+=$ o\_thresh\_step}
\EndIf
\If{\emph{(c\_mem\_delta $-$ c\_idle\_delta) > c\_delta\_thresh}}
\Comment{Traffic in memory is likely to outweigh any parallelism benefit}
\State \emph{o\_thresh $-=$ o\_thresh\_step}
\EndIf
\EndFor
\caption{\small{Determining the oversubscription threshold}}
\label{alg:determining_threshold}
\end{algorithmic} 
}} 
\end{algorithm}

\begin{table}[h]
\begin{scriptsize} \centering 
\begin{tabular}{ll} \toprule
\textbf{Variable} & \textbf{Description} \\ \midrule 
\emph{o\_thresh} & oversubscription threshold (dynamically determined)
\\ \cmidrule(rl){1-2} 
\emph{o\_default} & initial value for \emph{o\_thresh}
, (experimentally determined \\ & to be 10\% of
total physical resource) 
\\ \cmidrule(rl){1-2} 
\emph{c\_idle} & core cycles when no threads are issued to the core \\
& (but the pipeline is not stalled)~\cite{nmnl-pact13} \\ \cmidrule(rl){1-2}
\emph{c\_mem} & core cycles when all warps are waiting for data \\ & from memory or
stalled at the memory pipeline \\ \cmidrule(rl){1-2}
\emph{*\_prev} & the above statistics for the previous epoch \\ \cmidrule(rl){1-2}
\emph{c\_delta\_thresh} & threshold to produce change in \emph{o\_thresh}
\\ & (experimentally determined to be 16) 
\\ \cmidrule(rl){1-2}
\emph{o\_thresh\_step} & increment/decrement to \emph{o\_thresh}
, experimentally
\\ & determined to be 4\% of the total physical resource
\\ \cmidrule(rl){1-2}
\emph{epoch} & interval in core cycles to change \emph{o\_thresh}\\
 & (experimentally determined to be 2048)\\ 
\bottomrule
\end{tabular} 
\caption{Variables for oversubscription}
\label{table:thresholds}	
\end{scriptsize}
\end{table}
 
\subsection{Virtualizing On-chip Resources}
\label{sec:virtualizing_resources}
A resource can be in either the physical space, in
which case it is mapped to the physical on-chip resource, or the swap
space, in which case it can be found in the
memory hierarchy. Thus, a resource is effectively virtualized, and we need
to track the mapping between the virtual and physical/swap spaces. We
use a \emph{mapping table} for each resource to determine \One whether the resource
is in the physical or swap space, and \two the location of the
resource within the physical on-chip hardware. The compute units
access these mapping tables before accessing the real
resources. An access to a resource that is mapped to the swap space is converted
to a global
memory access that is addressed by the logical resource ID and
warp/block ID (and a base register for the swap space of the resource). In addition to the mapping tables, we use two registers
per resource to track the amount of the resource that is \One
free to be used in physical space,\ignore{(indicating availability of
 resources on-chip)} and \two mapped in swap space.\ignore{(indicating the extent of
 oversubscription)} These two counters enable the coordinator to
make oversubscription decisions
(Section~\ref{sec:oversubscription_decision}). We now go into more
detail on virtualized resources in \X.\footnote{Our implementation of
a virtualized resource aims to minimize complexity. This implementation
is largely orthogonal to the
 framework itself, and one can envision other 
 implementations (e.g.,~\cite{virtual-register,shmem-multiplexing,virtual-thread}) for
 different resources.}
\subsubsection{Virtualizing Registers and Scratchpad Memory}
\begin{figure}[h!] \centering
 	\begin{subfigure}[h]{0.49\linewidth} 
		\centering
 		\includegraphics[width=0.9\textwidth]{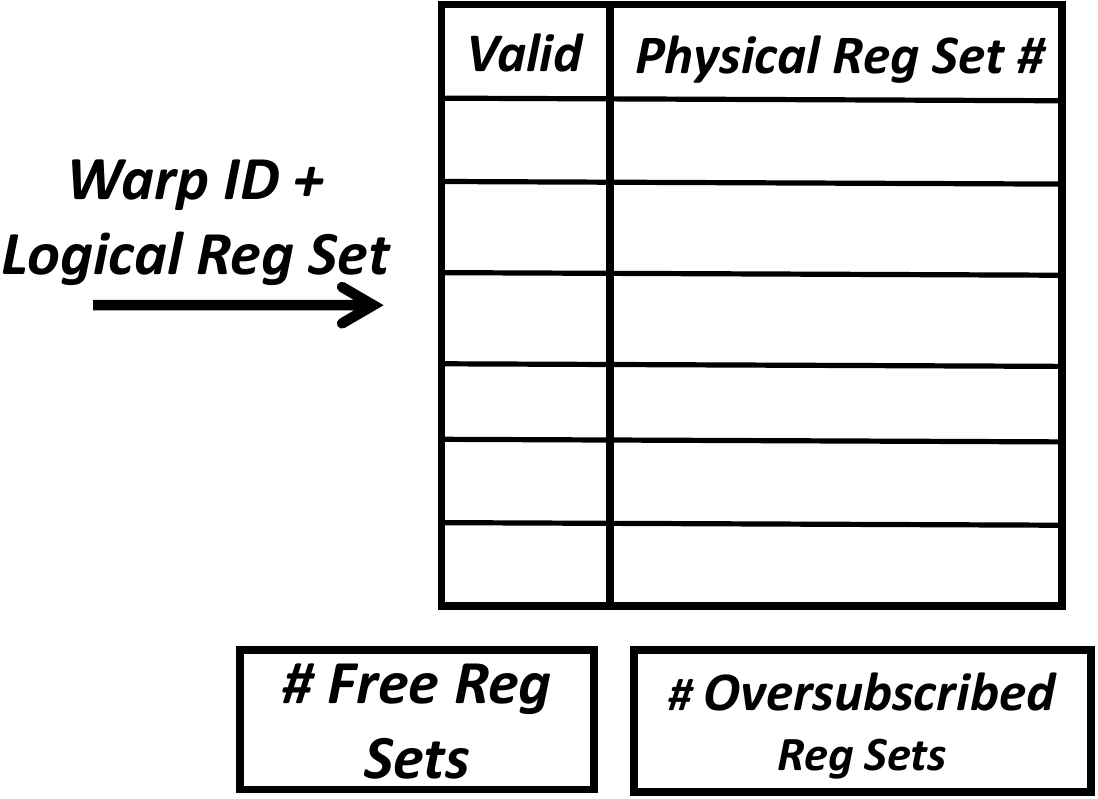}
 		\caption{Register Mapping Table}
	\end{subfigure}%
	\hfill
 	\begin{subfigure}[h]{0.49\linewidth} 
		\centering
 		\includegraphics[width=1\textwidth]{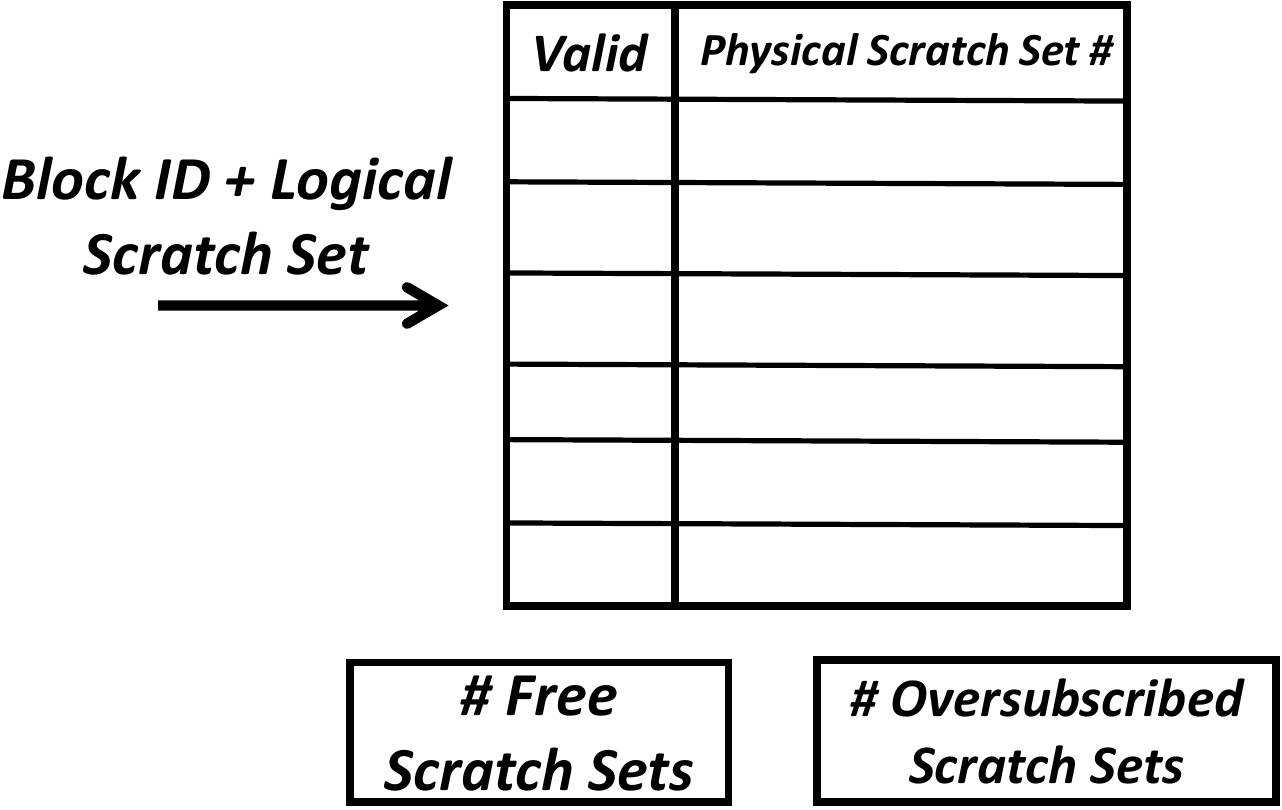}
 		\caption{Scratchpad Mapping Table}
	\end{subfigure}
	\caption{Mapping Tables}
 	\label{fig:mapping_table}
\end{figure}

In order to minimize the overhead of large mapping tables, we map
registers and scratchpad at the granularity of a \emph{set}. The size
of a set is configurable by the architect{\textemdash}we use 4*\emph{warp\_size}\footnote{We
 track registers at the granularity of a warp.} for the register mapping table, and 1KB for
scratchpad\ignore{ in our experiments}. \ignore{The size of a set is a tradeoff
 between the size of the table and the granularity of
 allocation/deallocation and hence, maximizing
 utilization. } Figure~\ref{fig:mapping_table} depicts the
 tables for the registers and scratchpad. The register mapping table
is indexed by the warp ID and the logical register set number
(\emph{logical\_register\_number / register\_set\_size}). The
scratchpad mapping table is indexed by the block ID and the logical
scratchpad set number (\emph{logical\_scratchpad\_address /
 scratchpad\_set\_size}). Each entry in the mapping table contains
the physical address of the register/scratchpad content in the
physical register file or scratchpad. The valid bit indicates whether
the logical entry is mapped to the physical space or the swap
space.\ignore{(in which case it is converted into a global memory
 access).} With 64 logical warps and 16 logical thread blocks (see Section~\ref{sec:meth:model}), the register mapping table
takes 1.125 KB ($64 \times 16 \times 9$ bits, or 0.87\% of the register
file) and the scratchpad mapping table takes 672 B ($16 \times 48 \times 7$
bits, or 1.3\% of the scratchpad).

\ignore{The
physical register file is 128KB in size which implies 256 registers
sets. The physical shared memory size is 48KB which implies 48 scratch
sets. For the register file, the total size of the table is \emph{64
 entries X (log2(256 register sets) + 1 valid bit) X 16 maximum
 register sets per thread = 1.12KB} ~= 0.87\% of the register
file. For the scratchpad table, the total size of the table is
\emph{16 entries X (log2(48 scratch sets) + 1 valid bit) X 48 sets per
 block = 672B} ~= 1.3\% of the scratchpad memory size. The mapping
table is used to determine the location of the register during the
operand collection stage of execution. \todo{Nandita}{Discuss latency
 overhead.}}

\subsubsection{Virtualizing Thread Slots}
Each SM is provisioned with a fixed number of \emph{warp slots}, which
determine the number of warps that are considered for execution every
cycle by the warp scheduler. In order to oversubscribe warp slots, we
need to save the state of each warp in memory before remapping the
physical slot to another warp. This state includes the bookkeeping
required for execution, i.e., the warp's PC (program counter) and
the SIMT stack, which holds divergence information for each executing
warp.\ignore{The SIMT stack includes the next PC, active mask, and
 reconvergent PC which is a total of 384B with a maximum stack depth
 of 32~\cite{dynamic-warp,rhu2013dpe}. } The thread slot mapping
table records whether each warp is mapped to a physical slot or swap space.
The 
table is indexed by the logical warp ID, and stores the address of the
physical warp slot that contains the warp. \ignore{A valid bit
 indicates the whether the slot can be found in the physical space or
 in swap space. }In our baseline design with 64 logical warps, this
mapping table takes 56 B ($64 \times 7$ bits).

\ignore{In our
baseline design, there are 48 physical warp slots per SM, and we
support upto 64 warps in the virtual space. The total size of the
thread slot mapping table is \emph{64 entries X (log2(48) + 1 valid
 bit) = 56 bytes.}}

\subsection{Handling Resource Spills} If the coordinator has oversubscribed any
 resource, it is possible that the resource can be found either 
\new{\emph{(i)}~on-chip (in the
 physical space) or \emph{(ii)}~in the swap space in the memory hierarchy}. As described
 above, the location of any virtual resource is determined by the mapping
 table for each resource. \new{If the resource is found on-chip, the mapping
 table provides the physical location in the register file and scratchpad
 memory.} If the resource is in the swap space, the access to
 that resource is converted to a global memory load that is addressed \new{either} by the
 \new{\emph{(i)}~thread block ID and logical register/scratchpad set, in the case of  
registers or scratchpad memory; or \emph{(ii)} logical warp ID, in the case of warp slots}.
The oversubscribed resource is typically found in the L1/L2 cache but in the
worst case, could be in memory. 
When the coordinator chooses to oversubscribe any resource beyond what is
available on-chip, the least frequently accessed resource set is spilled to the
memory hierarchy \new{using a simple store operation}. 

\subsection{Supporting Phases and Phase Specifiers}
\label{sec:phase_specifiers}
\textbf{Identifying phases.} The compiler partitions each application
into phases based on the liveness of registers and scratchpad
memory. To avoid changing phases too often, the compiler uses
thresholds to determine phase boundaries. In our evaluation, we define
a new phase boundary when there is \One a 25\% change in the number of live
registers or live scratchpad content, and \two a minimum of 10
instructions since the last phase boundary. To simplify hardware
design, the compiler draws phase boundaries only where there is no control
divergence.\footnote{The phase boundaries for the applications in our pool
easily fit this restriction,
but the framework can be extended to support control divergence if needed.}


Once the compiler partitions the application into phases, it inserts
instructions{\textemdash}\emph{phase specifiers}{\textemdash}to specify the
beginning of each new phase and convey information to the
framework on the number of registers and scratchpad memory
required for each phase. As described in Section~\ref{sec:key_idea_phases}, a barrier or
a fence instruction also implies a phase change, but the compiler does
not insert a phase specifier for it as the resource requirement does
not change.

\begin{figure}[h]
 \centering
 \includegraphics[width=0.40\textwidth]{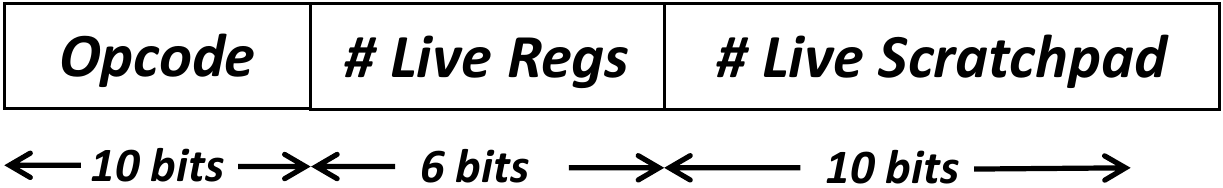}
 \caption{Phase Specifier}
 \label{fig:phase_specifier}
 \end{figure}

\textbf{Phase Specifiers.} The phase specifier instruction contains
fields to specify \One the number of live registers and
\two the amount of scratchpad memory in
bytes, both for the next phase. Figure~\ref{fig:phase_specifier} describes the
fields in the phase specifier instruction. The instruction decoder sends this information to the coordinator along with the phase
change event. The coordinator keeps this information in the
corresponding warp slot.

\subsection{Role of the Compiler and Programmer} 
 The compiler plays an important role, annotating the code with
 phase specifiers to convey information to the coordinator regarding
 the resource requirements of each phase. The compiler, however, does
 {\em not} alter the size of each thread block or the scratchpad memory
 usage of the program. The resource specification provided by the
 programmer (either manually or via auto-tuners) is retained to guarantee correctness. For registers, the
 compiler follows the default policy or uses directives as specified by the user. One could
 envision more powerful, efficient resource allocation with a
 programming model that does {\em not} require \emph{any} resource
 specification and/or compiler policies/auto-tuners that are \emph{cognizant} of the
 virtualized resources. 
 \subsection{Implications to the Programming Model and Software
 Optimization} Zorua offers several new opportunities and implications
 in enhancing the programming model and software optimizations (via libraries,
 autotuners, optimizing compilers, etc.) which we briefly
 describe below. \new{We leave these ideas for exploration in future work.}
\subsubsection{Flexible programming models for GPUs and heterogeneous
systems}
State-of-the-art high-level programming languages and models still
assume a fixed amount of on-chip resources and hence, with the help of
the compiler or the runtime system, are required to find \emph{static} resource
specifications to fit the application to the desired GPU. Zorua, by
itself, also still requires the programmer to specify resource specifications to
ensure correctness{\textemdash}albeit they are not required to be highly optimized for a given
architecture. However, 
by providing a flexible but dynamically-controlled view of the on-chip hardware resources, Zorua changes the abstraction of the on-chip resources
that is offered to the programmer and software. This offers the opportunity to
rethink resource management in GPUs from the ground up. One could envision more powerful resource
allocation and better programmability with programming models that do
\emph{not} require static
resource specification, leaving the compiler/runtime system and the underlying
virtualized framework to completely handle \emph{all} forms of on-chip resource
allocation, unconstrained by the fixed physical resources in a specific
GPU, entirely at runtime. This is especially
significant in future systems that are likely to support a wide range of
compute engines and accelerators, making it important to be able to write
high-level code that can be partitioned easily, efficiently, and at a
fine granularity across any accelerator, \emph{without} statically tuning any code segment to run efficiently on the GPU.

\subsubsection{Virtualization-aware compilation and autotuning} Zorua changes the
contract between the hardware and software to provide a more powerful
resource abstraction (in the software) that is
\emph{flexible and dynamic}, by pushing some more functionality into the hardware,
which can more easily react to the runtime resource requirements of the
\new{running} program. We can re-imagine compilers
 and autotuners to be more intelligent, leveraging this new abstraction and,
 hence the virtualization, to
 deliver more efficient
and high-performing code optimizations 
that are \emph{not} possible with the fixed and static
abstractions of today. They could, for example, \emph{leverage} the
oversubscription and dynamic management that Zorua provides to tune the code to
more aggressively use resources that are underutilized at runtime. As we
demonstrate in this work, static optimizations are limited by the fixed view of
the resources that is available to the program today. Compilation frameworks
that are cognizant of the \emph{dynamic} allocation/deallocation of resources
provided by Zorua could make more efficient use of the available resources. 

\subsubsection{Reduced optimization space} \new{Programs written for applications
in machine learning, computer graphics, computer vision, etc., 
typically follow the \emph{stream} programming paradigm, where the code is
decomposed into
many \emph{stages} in an \emph{execution pipeline.} Each stage processes only a part
of the input data in a pipelined fashion to make better use of the caches. 
A key challenge in writing complex pipelined code is finding \emph{execution schedules}
(i.e., how the work should be partitioned across stages)
and optimizations that perform best for \emph{each} pipeline stage from a
prohibitively large space of potential solutions.} This requires complex tuning
algorithms or profiling runs that are both computationally intensive and
time-consuming. The search for optimized specifications has to be done when
there is a change in input data or in the underlying architecture. By pushing
some of the resource management
functionality to the hardware, Zorua reduces this search space for optimized
specifications by
making it less sensitive to the wide space of resource specifications.


\section{Methodology}
\label{sec:methodology}

\subsection{System Modeling and Configuration} 
\label{sec:meth:model}

We model the \X framework with GPGPU-Sim
3.2.2~\cite{GPGPUSim}. Table~\ref{tab:param} summarizes the major
parameters. Except for the portability
results, all results are obtained using the Fermi configuration. We
use GPUWattch~\cite{gpuwattch} to model the GPU power consumption. We
faithfully model the overheads of the \X framework, including an
additional 2-cycle penalty for accessing each mapping table, and
the overhead of memory accesses for swap space accesses (modeled as a part of
the memory system). We model the energy overhead of mapping table accesses as
SRAM accesses in GPUWattch.

\begin{table}[h] 
	\begin{scriptsize} 
	\centering
	\renewcommand\arraystretch{0.5}
	\begin{tabular}{ll} \toprule
System Overview           &  15 SMs, 32 threads/warp,  6 memory channels\\ \cmidrule(rl){1-2}
Shader Core Config &  1.4 GHz, GTO
scheduler~\cite{tor-micro12}, 2 schedulers per SM\\ \cmidrule(rl){1-2}
Warps/SM    &  Fermi: 48; Kepler/Maxwell: 64 \\ \cmidrule(rl){1-2}
Registers    &  Fermi: 32768; Kepler/Maxwell: 65536 \\ \cmidrule(rl){1-2} 
Scratchpad     &  Fermi/Kepler: 48KB; Maxwell: 64KB \\ \cmidrule(rl){1-2}
On-chip Cache  &  L1: 32KB, 4 ways; L2: 768KB, 16 ways   \\ \cmidrule(rl){1-2}
Interconnect   &  1 crossbar/direction (15 SMs, 6
MCs), 1.4 GHz  \\ \cmidrule(rl){1-2}
Memory Model  &  177.4 GB/s BW, 6 memory controllers (MCs),\\ 
		& FR-FCFS scheduling, 16 banks/MC \\ \bottomrule
 \end{tabular}%
\caption{Major parameters of the simulated systems} 
\label{tab:param}%
 \end{scriptsize}%
\end{table}%

\subsection{Evaluated Applications and Metrics}

We evaluate a number of applications from the Lonestar suite~\cite{lonestar},
GPGPU-Sim benchmarks~\cite{GPGPUSim}, and CUDA SDK~\cite{sdk}, whose
resource specifications
(the number of registers, the amount
of scratchpad memory, and/or the number of threads per thread
block)
 are parameterizable.
Table~\ref{table:applications} shows the applications and the evaluated parameter ranges. For each application, we
make sure the amount of work done is the same for all specifications. The performance metric we use is the execution
time of the GPU kernels in the evaluated applications.

\begin{table}[h]
\begin{scriptsize} \centering 
	\renewcommand\arraystretch{0.5}
	\begin{tabular}{ll} \toprule
\textbf{Name (Abbreviation)} & \textbf{(R: Register, S: Scratchpad,} \\ 
 &\textbf{T: Thread block) Range}  \\ \midrule
Barnes-Hut (BH)~\cite{lonestar} & R:28-44 $\times$ T:128-1024
\\ \cmidrule(rl){1-2} 
Discrete Cosine Transform (DCT)~\cite{sdk}  & R:20-40 $\times$ T:
64-512 \\ \cmidrule(rl){1-2} 
Minimum Spanning Tree (MST)~\cite{lonestar}  & R:28-44 $\times$ T:
256-1024 \\ \cmidrule(rl){1-2} 
Reduction (RD)~\cite{sdk}  & R:16-24 $\times$ T:64-1024 \\ \cmidrule(rl){1-2}  
N-Queens Solver (NQU)~\cite{NQU}~\cite{GPGPUSim}  & S:10496-47232 (T:64-288) \\ \cmidrule(rl){1-2}  
Scan Large Array (SLA)~\cite{sdk}  & R:24-36 $\times$ T:128-1024 \\ \cmidrule(rl){1-2}  
Scalar Product (SP)~\cite{sdk}  & S:2048-8192 $\times$ T:128-512 \\ \cmidrule(rl){1-2}  
Single-Source Shortest Path (SSSP)~\cite{lonestar} & R:16-36 $\times$ T:256-1024
\\ \bottomrule  
\end{tabular} 
\caption{Summary of evaluated applications}
\label{table:applications}	
\end{scriptsize}
\end{table}


\section{Evaluation}
\label{sec:eval}

We evaluate the effectiveness of \X by studying three different mechanisms: (i)~\emph{Baseline}, the
baseline GPU
that schedules kernels and manages resources at the thread block level;
(ii)~\emph{WLM} (Warp Level Management), a state-of-the-art mechanism for GPUs
to schedule kernels and manage registers at the warp level~\cite{warp-level-divergence}; and (iii)~\emph{\X}.
For our evaluations, we run each application on 8{\textendash}65 (36 on average) 
different resource specifications\ignore{Our technical
report~\cite{zorua-tr} provides the specifications.}
(the ranges are in
Table~\ref{table:applications}). 

\subsection{Effect on Performance Variation and Cliffs}
\label{sec:eval:var}

\begin{figure}[h]
 \centering 
 \includegraphics[width=0.47\textwidth]{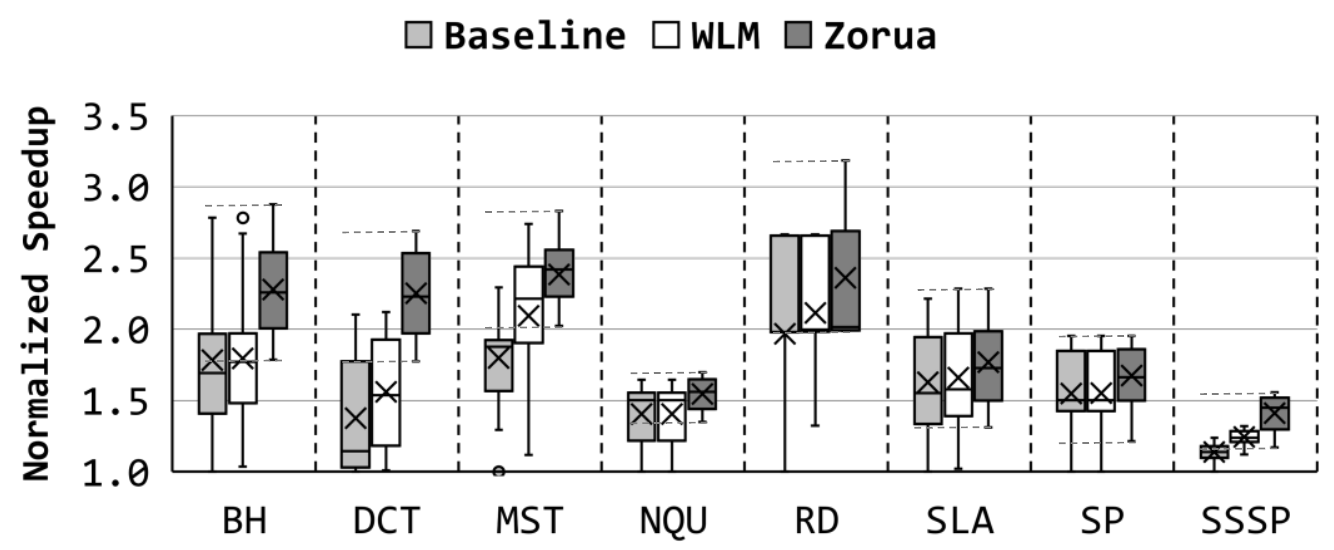}
 \caption{Normalized performance distribution. Reproduced from~\cite{zorua}.}
 \label{fig:performance_range}
\end{figure}

We first examine how \X alleviates the high variation in performance by reducing
the impact of resource specifications on resource utilization.
Figure~\ref{fig:performance_range} presents a Tukey 
box plot~\cite{mcgill1978variations} 
(see Section~\ref{sec:motivation} for a 
description of the presented box plot),
illustrating the performance distribution (higher is better) for each
application (for all different application resource specifications we
evaluated), normalized to the slowest Baseline operating point \emph{for that
application}. 
We make two major
observations.

\ignore{The boxes in the box plot represent the range between the first
quartile (25\%) and the third quartile (75\%). The whiskers extending from the
boxes represent the maximum and minimum points of the distribution, or 1.5X the
length of the box, whichever is smaller. Any points that lie more than 1.5X 
the box length beyond the box are considered to be
outliers~\cite{mcgill1978variations}, and are plotted as
individual points. The line in the middle of the box represents the median,
while the ``X'' represents the average.}

First, we find that \X significantly reduces the \emph{performance range} across
all evaluated resource specifications. Averaged across all of our applications, the 
worst resource specification for Baseline achieves 96.6\% lower performance 
than the best performing resource specification. For WLM~\cite{warp-level-divergence},
this performance range reduces only slightly, to
88.3\%. With \X, the performance range drops significantly, to 48.2\%.
We see
drops in the performance range for \emph{all} applications except \emph{SSSP}. With \emph{SSSP}, the
range is already small to begin with (23.8\% in Baseline), and \X
exploits the dynamic underutilization, which improves performance but also adds a small amount of variation.

Second, while \X reduces the performance range, it also preserves or improves performance
of the best performing points. As we examine in more detail in
Section~\ref{sec:eval:perf}, the reduction in performance range occurs as
a result of improved performance mainly at the lower end of the distribution.

\begin{figure}[h]
 \centering
 \begin{subfigure}[t]{0.335\linewidth}
 \centering
 \includegraphics[width=1.0\textwidth]{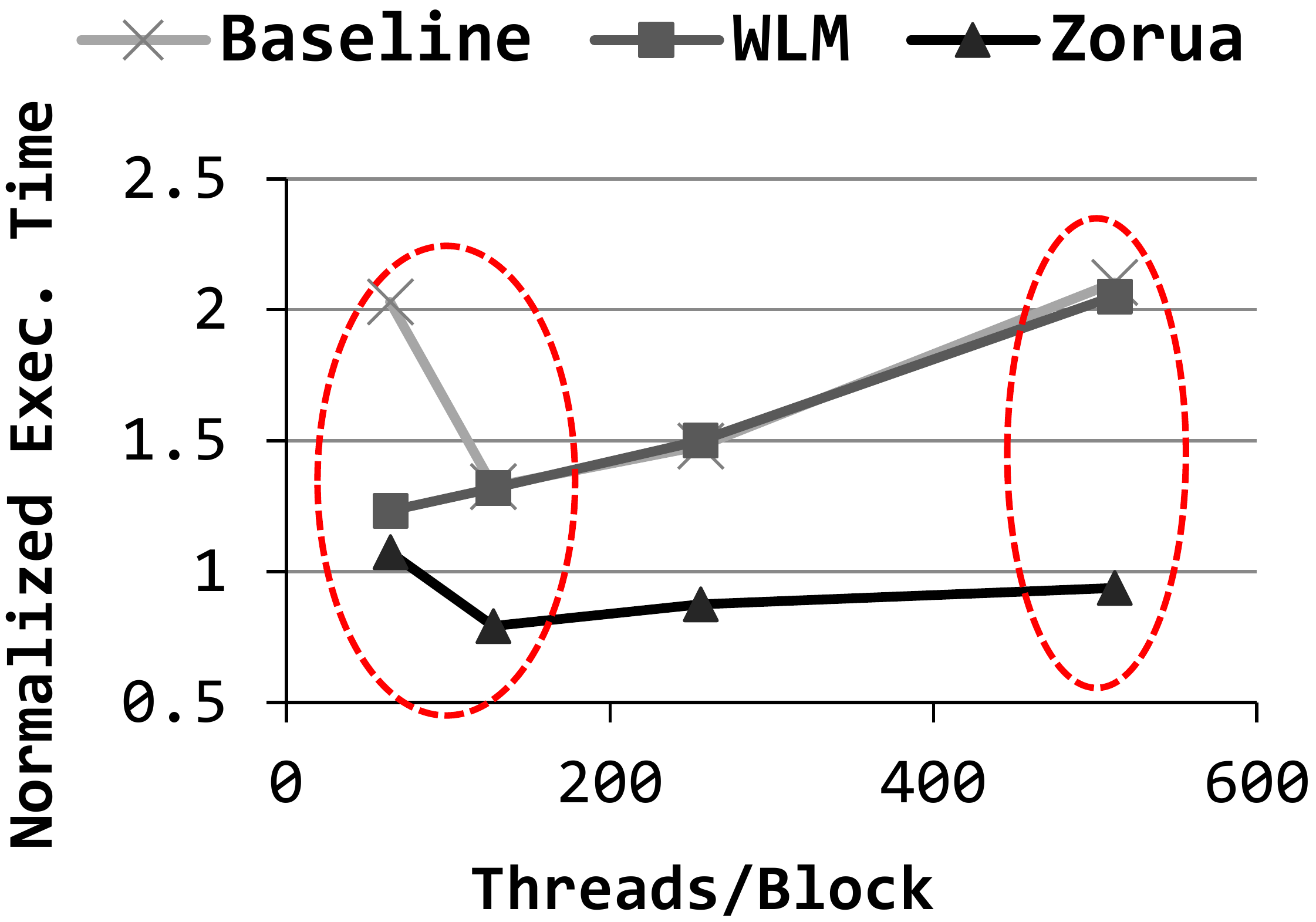}
 \caption{\emph{DCT}} 
 \label{fig:performance_cliff_result_dct}
 \end{subfigure}
 \begin{subfigure}[t]{0.318\linewidth}
 \centering
 \includegraphics[width=1.0\textwidth]{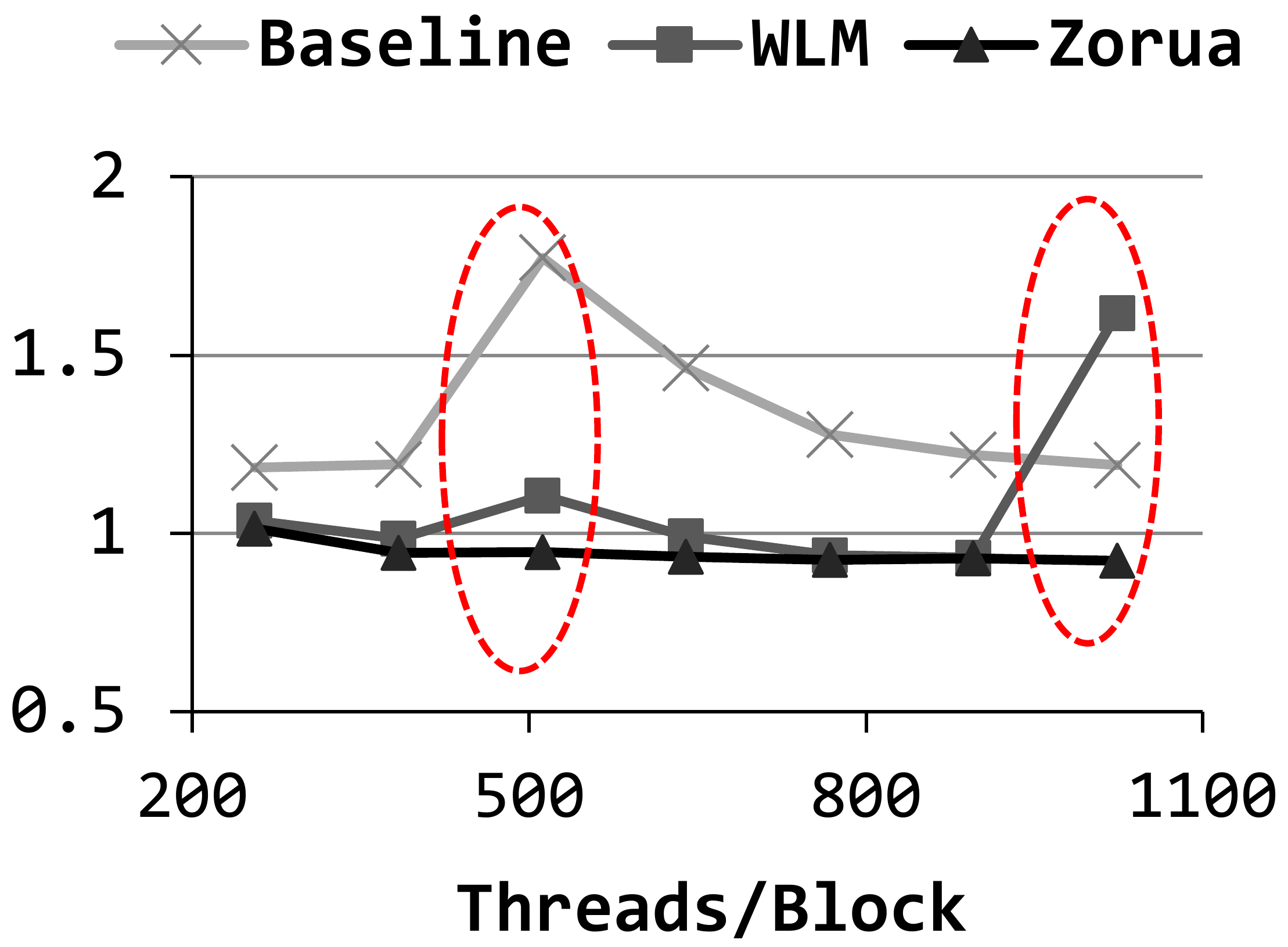} 
 \caption{\emph{MST}}
 \label{fig:performance_cliff_result_mst}
 \end{subfigure}
 \begin{subfigure}[t]{0.316\linewidth}
 \centering
 \includegraphics[width=1.0\textwidth]{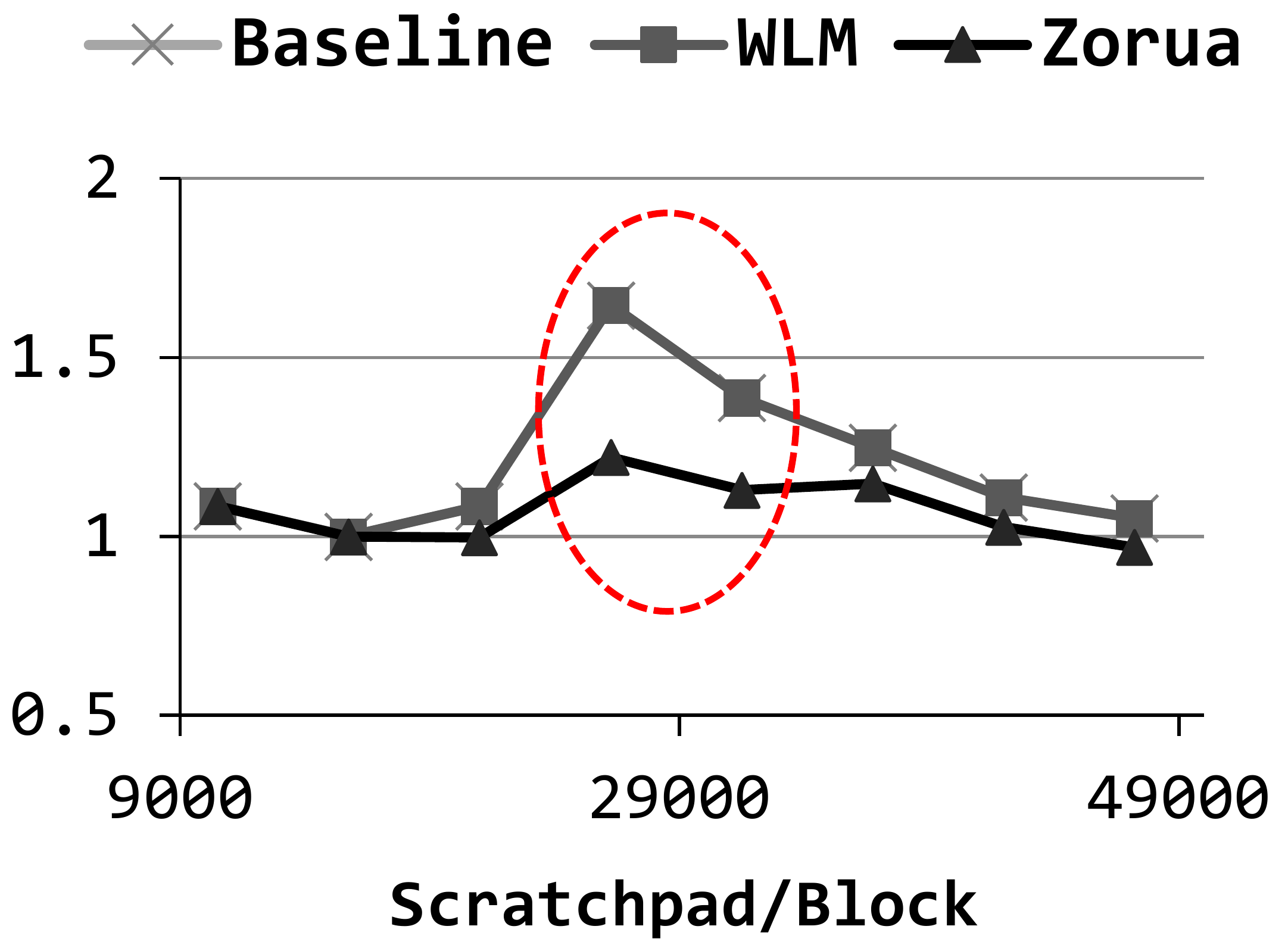}
 \caption{\emph{NQU}}
 \label{fig:performance_cliff_result_nqu}
 \end{subfigure}
 \caption{Effect on performance cliffs. Reproduced from~\cite{zorua}.}
 \label{fig:performance_cliff_result}
\end{figure}

To gain insight into how \X reduces the performance range and improves
performance for the worst performing points, we analyze how it reduces
performance cliffs. With \X, we ideally want to \emph{eliminate} the cliffs we 
observed in Section~\ref{sec:motivation:cliffs}. We study the tradeoff 
between resource specification and execution time for three representative 
applications: 
\emph{DCT} (Figure~\ref{fig:performance_cliff_result_dct}), \emph{MST}
(Figure~\ref{fig:performance_cliff_result_mst}), and
\emph{NQU} (Figure~\ref{fig:performance_cliff_result_nqu}).
For all three figures, we normalize execution time to the \emph{best} execution time
under Baseline. Two observations are in order.


First, \X successfully mitigates the performance cliffs that occur
in Baseline. For example, \emph{DCT} and \emph{MST} are both sensitive to the thread block size, as shown in 
Figures~\ref{fig:performance_cliff_result_dct}
and~\ref{fig:performance_cliff_result_mst}, respectively. We have circled the
locations at which cliffs exist in Baseline. Unlike Baseline, \X
maintains more steady execution times across the  
number of threads per block, 
employing 
oversubscription to overcome the loss in parallelism due to insufficient on-chip
resources. We see
similar results across all of our applications.

Second, we observe that while WLM~\cite{warp-level-divergence} can reduce some of the cliffs by mitigating
the impact of large block sizes, many cliffs still
exist under WLM (e.g., \emph{NQU} in Figure~\ref{fig:performance_cliff_result_nqu}).
This cliff in \emph{NQU} occurs as a result of insufficient scratchpad memory, which
cannot be handled by warp-level management. Similarly, the cliffs for \emph{MST}
(Figure~\ref{fig:performance_cliff_result_mst}) also persist with WLM because \emph{MST}
has a lot of barrier operations, and the additional warps scheduled by WLM
ultimately stall, waiting for other warps within the same block
to acquire resources. We find that, with oversubscription, \X is able to 
smooth out those cliffs that WLM
is unable to eliminate.

Overall, we conclude that \X \One reduces the performance variation across
resource specification points, so that performance depends less on the
specification provided by the programmer; and \two can alleviate the
performance cliffs experienced by GPU applications.

\subsection{Effect on Performance}
\label{sec:eval:perf}

As Figure~\ref{fig:performance_range} shows,
\X either retains or improves the best performing point for each application,
compared to the Baseline. \X improves the best performing point for each
application by 12.8\% on average, and by as much as 27.8\%
(for \emph{DCT}). This improvement comes from the improved parallelism obtained by exploiting the
dynamic underutilization of resources, which exists \emph{even for optimized
specifications}. Applications such as \emph{SP} and \emph{SLA} have little dynamic
underutilization, and hence do not show any performance improvement.
\emph{NQU} \emph{does}
have significant dynamic underutilization, but \X does not improve the best
performing point as the overhead of oversubscription outweighs the benefit,
and \X dynamically chooses not to oversubscribe. We conclude that even for
many specifications that are optimized to fit the hardware resources, 
\X is able to further improve performance. 

We also note that, in addition to reducing performance variation and
improving performance for optimized points, \X improves performance
by 25.2\% on average for all resource specifications across all evaluated
applications.

\ignore{We present a deeper look into
these results in Section~\ref{sec:deeper_look}.}

\subsection{Effect on Portability}
\label{sec:eval:port}

As we describe in Section~\ref{sec:motivation:port},
performance cliffs often behave differently across different GPU architectures, and can
significantly shift the best performing resource specification point.
We study how \X can ease the burden of performance tuning if an application 
has been already tuned for one GPU model, and is later ported to another GPU. 
To understand this, we define a new
metric, \emph{porting performance loss}, that quantifies the performance impact
of porting an application without re-tuning it. To calculate this, we first
normalize the execution time of each specification point to the execution time
of the best performing specification point. We then pick a
source GPU architecture (i.e., the architecture that the GPU was tuned for) and
a target GPU architecture (i.e., the architecture that the code will run on),
and find the point-to-point drop in performance for all points whose
performance on the source GPU comes within 5\% of the performance at the
best performing specification point.\footnote{We include any point within 5\%
of the best performance as there are often multiple points close to the
best point, and the programmer may choose any of them.}

\begin{figure}[h]
 \centering 
 \includegraphics[width=0.48\textwidth]{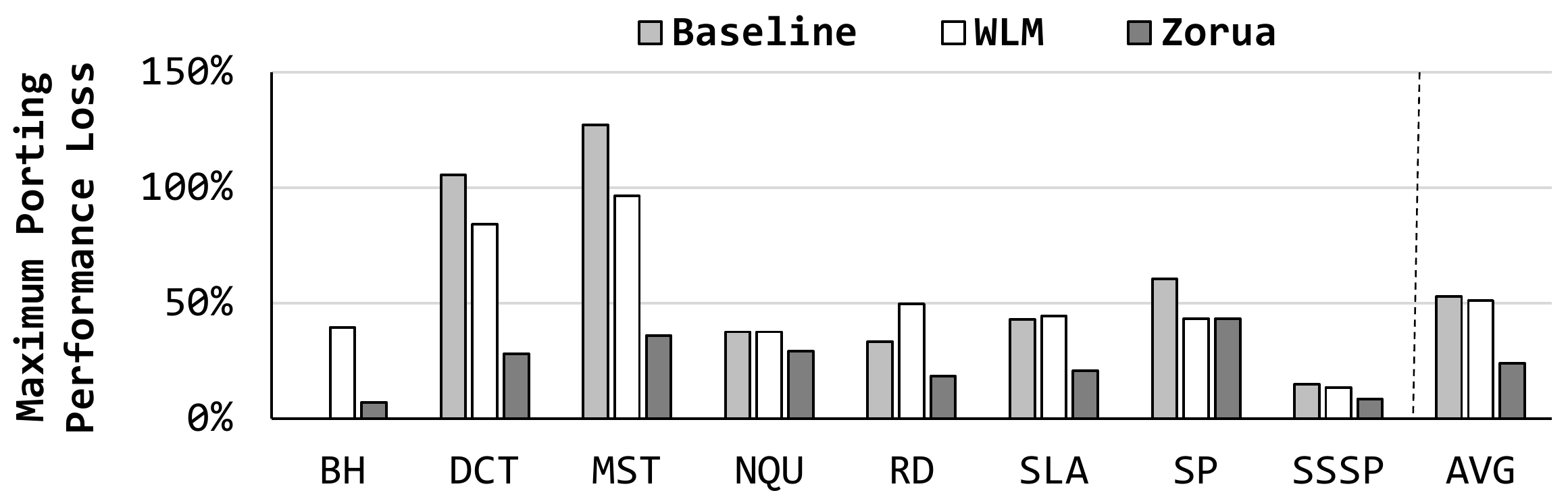}
 \caption{Maximum porting performance loss. Reproduced from~\cite{zorua}.}
 \label{fig:portability_result_overall}
\end{figure}

Figure~\ref{fig:portability_result_overall} shows the \emph{maximum} porting
performance loss for each application, across any two pairings of our three
simulated GPU architectures (Fermi, Kepler, and Maxwell). We find that \X greatly 
reduces the maximum porting performance loss that occurs under both Baseline
and WLM for all but one of our applications. On average, the maximum porting performance 
loss is 52.7\% for Baseline, 51.0\% for WLM, and only 23.9\% for \X.

Notably, \X delivers significant improvements in portability for applications 
that previously suffered greatly when ported to another GPU, such as \emph{DCT} and 
\emph{MST}. For both of these applications, the performance variation differs so much
between GPU architectures that, despite tuning the application on the source
GPU to be within 5\% of the best achievable performance, their performance on the target
GPU is often more than twice as slow as the best achievable performance on the target 
platform. \X significantly lowers this porting performance loss down to 28.1\% for \emph{DCT} and 36.1\% for \emph{MST}. We also observe
that for \emph{BH}, \X
actually increases the porting performance loss slightly with respect to the
Baseline. This is because for Baseline, there are only two points that perform within the 
5\% margin for our metric, whereas with \X, we have five points that fall in
that range. Despite this, the increase in porting performance loss for \emph{BH} is low, deviating
only 7.0\% from the best performance.

\begin{figure}[h]
 \centering 
 \includegraphics[width=0.48\textwidth]{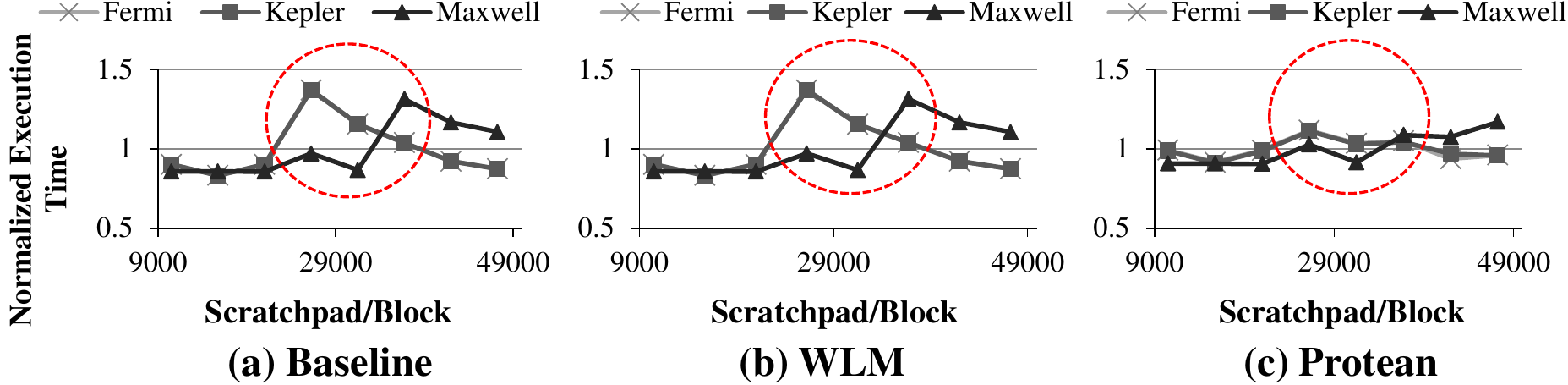}
 \caption{Impact on portability (NQU)}
 \label{fig:portability_nqu}
\end{figure}

\begin{figure}[h]
 \centering 
 \includegraphics[width=0.48\textwidth]{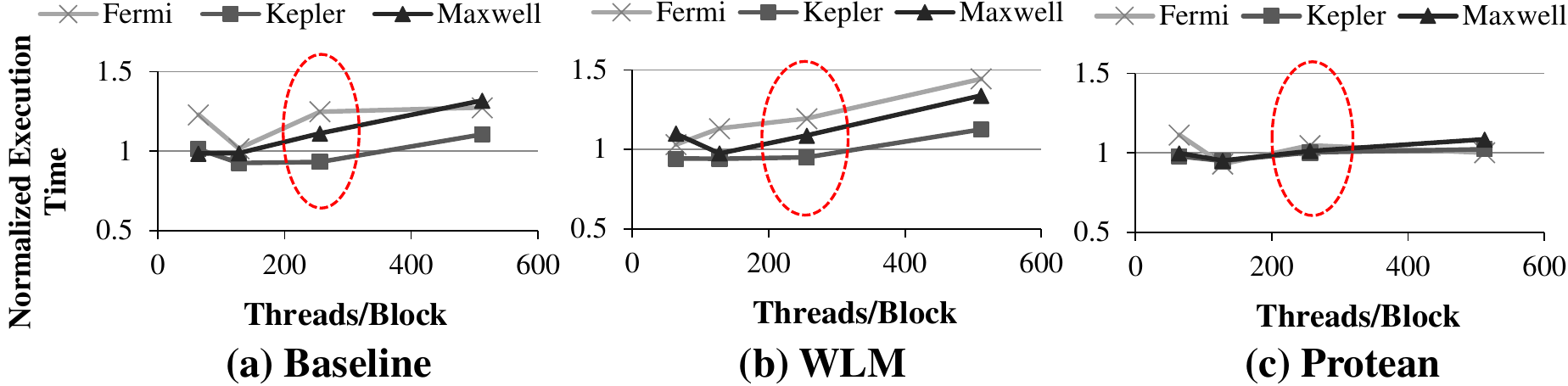}
 \caption{Impact on portability (DCT)}
 \label{fig:portability_dct}
\end{figure}

To take a closer look into the portability benefits of Zorua, we run experiments
to obtain the performance sensitivity curves for each application using
different GPU architectures. Figures~\ref{fig:portability_nqu} and \ref{fig:portability_dct} depict the
execution time curves while sweeping a single resource specification for \emph{NQU} and
\emph{DCT} for the three evaluated GPU architectures -- Fermi, Kepler, and
Maxwell. We make two major observations from the figures. 

First, Zorua significantly alleviates the presence of performance cliffs and
reduces the
performance variation across \emph{all} three evaluated architectures, thereby
reducing the impact of both resource specification and underlying architecture
on the resulting performance curve. In comparison, WLM is unable to make a
significant impact on the performance variations and the cliffs remain for all
the evaluated architectures.  

Second, by reducing the performance variation across all three GPU generations,
Zorua significantly reduces the \new{\emph{porting performance loss}}, i.e., the loss in
performance when code optimized for one GPU generation is run on another (as
highlighted within the figures). 

We conclude that \X enhances portability of applications by reducing the
impact of a change in the hardware resources for a given resource
specification. For applications that have 
already been tuned on one platform, \X significantly lowers the penalty of not 
re-tuning for another platform, allowing programmers to save development time.

\subsection{A Deeper Look: Benefits \& Overheads}
\label{sec:deeper_look}
\begin{figure}[h]
 \centering 
 \includegraphics[width=0.47\textwidth]{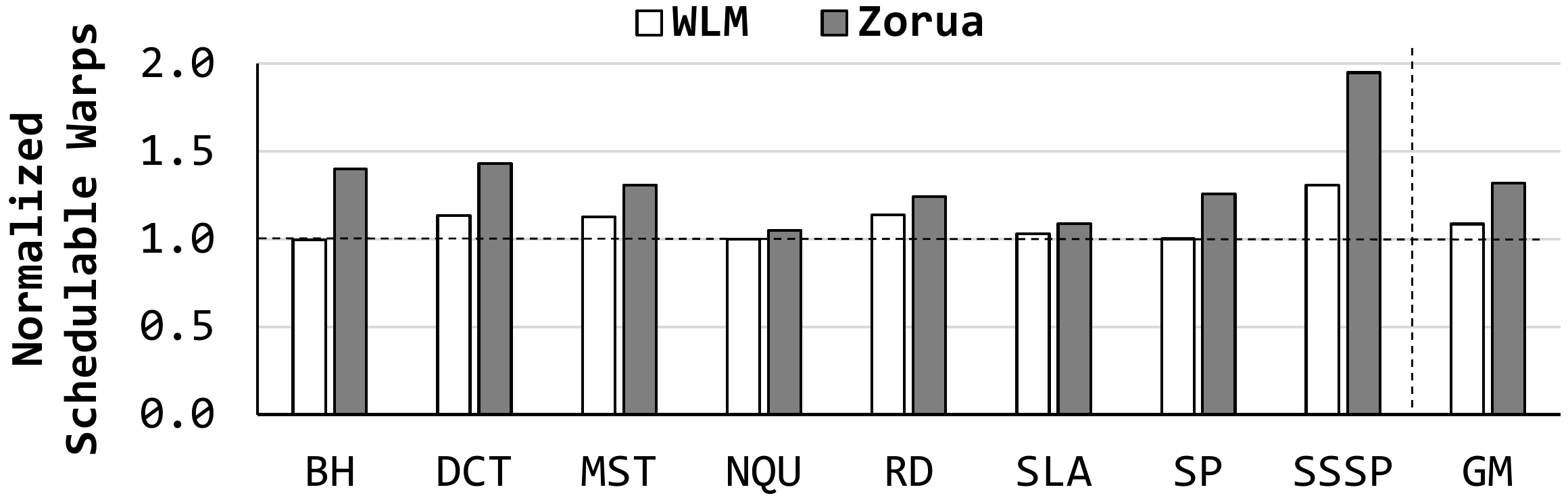}
 \caption{Effect on schedulable warps. Reproduced from~\cite{zorua}.}
 \label{fig:active_warps}
\end{figure}

To take a deeper look into how \X is able to provide the above benefits, 
in Figure~\ref{fig:active_warps}, we show the 
number of \emph{schedulable warps} (i.e., warps that are available to be
scheduled by the warp scheduler at any given time excluding warps
waiting at a barrier), averaged across all of specification points. 
On average, \X increases the number of
schedulable 
warps by 32.8\%, significantly more than WLM (8.1\%), which is constrained by the
fixed amount of available resources. We conclude that by oversubscribing and
dynamically managing resources, \X is able to improve thread-level parallelism, and hence performance. 

\begin{figure}[h]
 \centering 
 \includegraphics[width=0.47\textwidth]{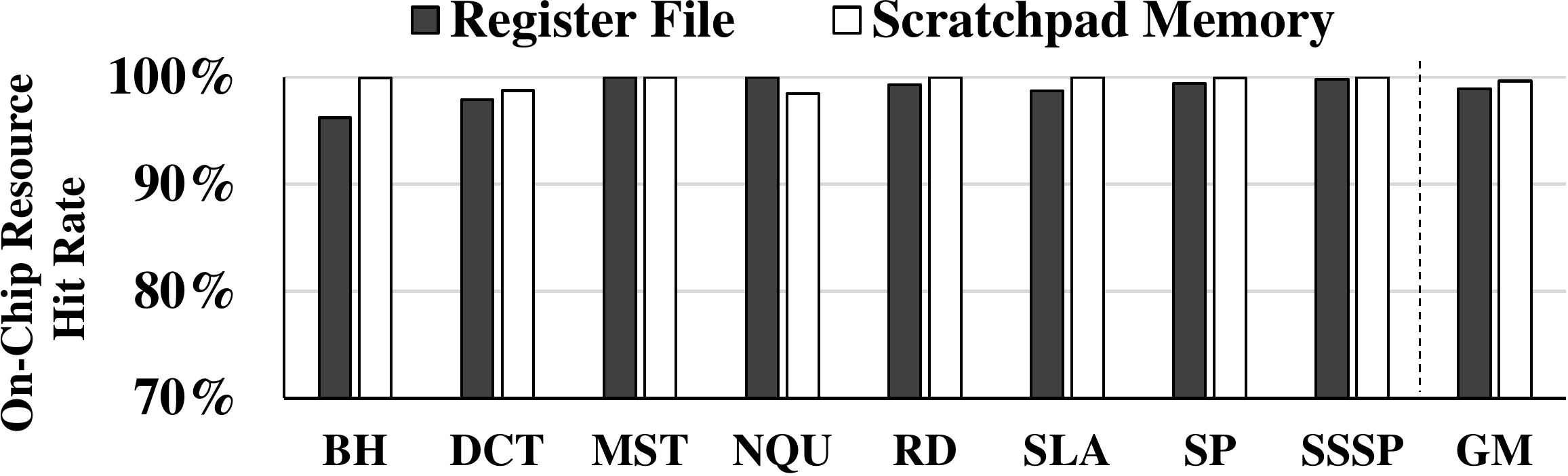}
 \caption{Virtual resource hit rate in \X}
 \label{fig:resource_hit_rate}
\end{figure}

We also find that the overheads due to resource swapping and
contention do not significantly impact the performance of \X.
Figure~\ref{fig:resource_hit_rate} depicts resource hit rates for each
application, i.e., the fraction of all resource accesses that were found on-chip
as opposed to making a potentially expensive off-chip access. The
oversubscription mechanism (directed by the coordinator) is able to keep resource hit rates
very high, with an average hit rate of 98.9\% for the
register file and 99.6\% for scratchpad memory.

\begin{figure}[h]
 \centering 
 \includegraphics[width=0.47\textwidth]{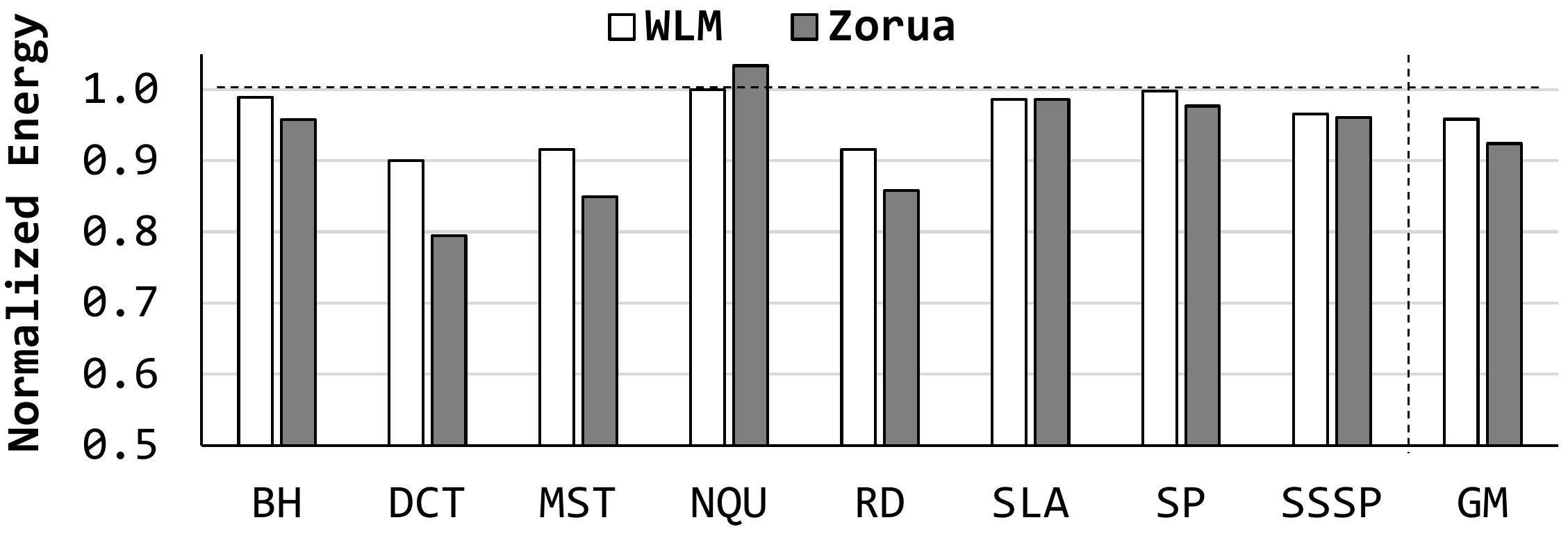}
 \caption{Effect on energy consumption. Reproduced from~\cite{zorua}.}
 \label{fig:energy}
\end{figure}

Figure~\ref{fig:energy} shows the average reduction in total system energy consumption
of WLM and \X over Baseline for each application (averaged across the 
individual energy consumption over Baseline for each evaluated specification
point). We observe that \X
reduces the total energy consumption across all of our applications, except for 
\emph{NQU} (which has a small increase of 3\%). Overall, \X provides a 
mean energy reduction of 7.6\%, up to 20.5\% for \emph{DCT}.\footnote{We note that the energy consumption can be
reduced further by appropriately optimizing the oversubscription algorithm. We
leave this exploration to future work.} We conclude that \X is an energy-efficient virtualization framework for GPUs.


We estimate the die area overhead of \X with CACTI
6.5~\cite{wilton1996cacti}, using the same
40nm process node as the GTX 480 \ignore{(Fermi)}, which our system closely models. We
include all the overheads from the coordinator and the resource
mapping tables (Section \ref{sec:mechanism}). The total
area overhead is 0.735 $mm^2$ for all 15 SMs, which is only
0.134\% of the die area of the GTX 480.

\ignore{We conclude that \One \X improves overall performance, 
even when suboptimal resource specifications are used; \two the overheads of
decoupling resource specifications from the hardware resources are minimal; and
\three \X reduces energy consumption significantly.}


\section{Other Applications}
\label{sec:applications}
By providing the illusion of more resources than physically available, \X provides the
opportunity to help address other important challenges in GPU computing today.
We discuss several such opportunities in this section.
\subsection{Resource Sharing in Multi-Kernel \new{or}
Multi-Programmed Environments}
Executing multiple kernels or applications within the same SM can improve
resource utilization and
efficiency~\cite{asplos-sree,simultaneous-sharing,fine-grain-hotpar,kernelet,app-aware-GPGPU-2014,mosaic,mask,rachata-isca}. 
Hence, providing support to enable fine-grained sharing and partitioning
of resources is critical for future GPU systems. 
This is especially true in environments where multiple different applications
may be consolidated on the same GPU, e.g. in clouds or clusters. 
By providing a flexible view of each of the resources, \X provides a natural way
to enable dynamic and fine-grained control over resource partitioning and
allocation among multiple kernels. Specifically, 
\X provides several key benefits
for enabling better performance and efficiency in multi-kernel/multi-program
environments. First, selecting the optimal resource specification for an
application is challenging in virtualized environments (e.g., clouds), as it is unclear which
other applications may be running alongside it. \X can improve efficiency in resource
utilization \emph{irrespective} of the application specifications and of other kernels
that may be executing on the same SM. Second, \X manages the different
resources independently and at a fine granularity, using a dynamic runtime system
(the coordinator). This enables the maximization of
resource utilization, while providing the ability to control the partitioning of
resources at runtime to provide QoS, fairness, etc., by leveraging the coordinator. 
Third, \X enables oversubscription of the different resources. This obviates the
need to alter the application specifications~\cite{asplos-sree,kernelet} in
order to ensure there are sufficient resources to co-schedule kernels on the
same SM, and hence enables concurrent kernel execution
transparently to the programmer. 

\subsection{Preemptive Multitasking}
A key challenge in enabling true multiprogramming in GPUs is
enabling rapid preemption of
kernels~\cite{isca-2014-preemptive,simultaneous-sharing,chimera}. Context
switching on GPUs incurs a very high latency and overhead, as a result of the large amount of
register file and scratchpad state that needs to be saved 
before a new kernel can be executed. Saving state at a very coarse
granularity (e.g., the entire SM state) leads to very high preemption
latencies. Prior work proposes context minimization~\cite{chimera,igpu}
or context switching at the granularity of a thread
block~\cite{simultaneous-sharing} to improve response time during preemption. \X
enables fine-grained management and oversubscription of on-chip resources. It can be naturally extended to enable quick
preemption of a task via intelligent management of the swap space and the
mapping tables (complementary to approaches taken by prior
work~\cite{chimera,igpu}).
\subsection{Support for Other Parallel Programming Paradigms}
The fixed static resource allocation for each thread in
modern GPU architectures requires statically dictating the
resource usage for the program throughout its execution. Other forms
of parallel execution that are \emph{dynamic} (e.g., \newII{Cilk}~\cite{cilk}\new{,
staged execution\newII{~\cite{marshaling,bis, joao.isca13}}}) require more flexible allocation
of resources at runtime, and are hence more challenging to enable.
Examples of this include \emph{nested parallelism}~\cite{nested}, where a kernel can dynamically spawn new kernels or
thread blocks, and \emph{helper threads}~\cite{caba} to utilize idle resource at runtime to
perform different optimizations or background tasks in parallel. Zorua makes it
easy to enable these paradigms by providing on-demand dynamic allocation of
resources. Irrespective of whether threads in the programming model are created
statically or dynamically, Zorua allows allocation of the required resources on
the fly to support the execution of these threads. The resources are simply
deallocated when they are no longer required. Zorua also enables
\emph{heterogeneous}
allocation of resources -- i.e., allocating different amounts of resources to
different threads. The current resource allocation model, in line with a GPU's
SIMT architecture, treats all threads the
same and allocates the same amount of resources. \new{Zorua makes it easier to
support execution paradigms where each concurrently-running thread
executes different code at the same time, hence requiring different
resources. This includes helper
threads, multiprogrammed execution, nested parallelism, etc.} Hence, with Zorua, applications are no
longer limited by a GPU's fixed SIMT model which only supports a fixed,
statically-determined number of homogeneous threads as a result of the resource
management mechanisms that exist today. 

\subsection{Energy Efficiency and Scalability}
To support massive parallelism, on-chip resources are a precious
and critical resource. However, these resources \emph{cannot} grow arbitrarily large as
GPUs continue to be area-limited and on-chip memory tends to be extremely power
hungry and area
intensive~\cite{energy-register,virtual-register,compiler-register,warped-register,virtual-thread,
ltrf-sadrosadati-asplos18}.
Furthermore, complex thread schedulers that can select a thread for execution
from an increasingly large thread pool are required in order to support an
arbitrarily large number of warp slots. \X enables using smaller
register files, scratchpad memory and less complex or fewer thread schedulers to
save power and area while still retaining or improving parallelism.

\subsection{Error Tolerance and Reliability}
The indirection offered by \X, along with the dynamic
management of resources, could also enable better reliability and simpler
solutions towards error tolerance in the on-chip resources. The
virtualization framework trivially allows remapping resources with hard or soft
faults such that no virtual resource is mapped to a faulty physical resource.
\new{Unlike in the baseline case, faulty resources would not impact the
number of the resources
seen by the thread scheduler while scheduling threads for execution. A few
unavailable faulty registers,
warp slots, etc., could significantly reduce the number of the threads that
are scheduled concurrently (i.e., the runtime parallelism).}

\subsection{Support for System-Level Tasks on GPUs}
As GPUs become increasingly general purpose, a key requirement is better
integration with the CPU operating system, and with complex distributed software
systems such as those employed for large-scale distributed machine
learning~\cite{tensorflow, gaia} or
graph processing~\cite{graphlab, tesseract}. If GPUs are architected to be
first-class compute engines,
rather than the slave devices they are today, they can be programmed and utilized in the
same manner as a modern CPU. 
This integration requires the GPU execution model to support system-level
tasks like interrupts, exceptions, etc. and more
generally provide support for access to distributed file systems,
disk I/O, or network communication. Support for these tasks and
execution models require dynamic provisioning of resources for execution of
system-level code. Zorua provides a building block to enable this.

\subsection{Applicability to General Resource Management in Accelerators} 
Zorua uses a program \emph{phase} as the granularity for managing resources.
This allows handling resources across phases
\emph{dynamically}, while leveraging \emph{static} information regarding resource requirements
from the software by inserting annotations at phase boundaries. Future work could potentially investigate the applicability
of the same approach to manage
resources and parallelism in \emph{other} accelerators (e.g.,
processing-in-memory accelerators\newII{~\cite{pim-enabled, tesseract, tom, impica,
ambit, googlepim-asplos18, shaw1981non, boroumand2016pim, kim.bmc18,
guo-wondp14, stone1970logic, zhang-2014, kogge.iccp94, patterson.ieeemicro97,
pattnaik.pact16, ghose.pim.bookchapter18, seshadri.bookchapter17, akin.isca15,
seshadri.cal15,concurrent-datastructures}}
or direct-memory access engines~\cite{rowclone,decoupled-dma, chang.hpca16}) that require efficient dynamic management of large
amounts of particular critical resources.

\ignore{
\subsection{Better Resource Utilization}
GPUs employ extreme multithreading to provide high throughput computing, and to
support this support massive parallelism, on-chip resources are a precious
and critical resource. However, these resources cannot grow arbitrarily large as
GPUs continue to be area-limited and on-chip memory tends to be extremely power
hungry and area intensive~\cite{energy-register,virtual-register,compiler-register,warped-register}.
Furthermore, complex thread schedulers that can select a thread for execution
from an increasingly large thread pool are required. Similar to prior
work~\cite{virtual-register,energy-register}, \X enables using smaller
register files, scratchpad memory and less complex or fewer thread schedulers to
save power and area while still retaining or improving parallelism. 
 
\subsection{Reliability}
}

\section{Related Work}
\label{sec:related}


To our knowledge, this is the first work to propose a holistic framework to
decouple a GPU application's resource specification from its physical on-chip
resource allocation by virtualizing multiple on-chip resources. This
enables the illusion of more resources than what physically exists to the
programmer, while the hardware resources are managed at runtime by employing a
swap space (in main memory), transparently to the programmer. We design a new
hardware/software cooperative framework to effectively virtualize multiple
on-chip GPU resources in a controlled and coordinated manner, thus enabling many
benefits of virtualization in GPUs. 

We briefly discuss prior work related to different aspects of our
proposal: 
\One virtualization of resources,
\two improving
programming ease and portability,
and
\three more efficient management of
on-chip resources. 
 
\textbf{Virtualization of Resources.}
 \emph{Virtualization}~\cite{virtual-memory1,virtual-memory2,virtualization-1,virtualization-2}
 is a concept designed to provide the illusion, to the software and
 programmer, of more resources than what truly exists in physical
 hardware. It has been applied to the management of hardware
 resources in many different contexts~\cite{virtual-memory1,virtual-memory2,
virtualization-1,virtualization-2,vmware-osdi02,how-to-fake,pdp-10,ibm-360},
 with virtual memory~\cite{virtual-memory1, virtual-memory2, multics}
 being one of the oldest forms of virtualization that is commonly
 used in high-performance processors today. Abstraction of hardware
 resources and use of a level of indirection in their management
 leads to many benefits, including improved utilization,
 programmability, portability, isolation, protection, sharing, and 
 oversubscription.

In this work, we apply the general principle of virtualization to the
management of multiple on-chip resources in modern
GPUs. Virtualization of on-chip resources offers the opportunity to
alleviate many different challenges in modern GPUs. However, in this
context, effectively adding a level of indirection introduces new
challenges, necessitating the design of a new virtualization
strategy. There are two key challenges. First, we need to dynamically determine
the \emph{extent} of the virtualization to reach an effective tradeoff
between improved parallelism due to oversubscription and
the latency/capacity overheads of swap space usage. Second,
we need to coordinate the virtualization of \emph{multiple} latency-critical
on-chip resources. To our knowledge, this is the first work to propose
a holistic software-hardware cooperative approach to virtualizing
multiple on-chip resources in a controlled and coordinated manner that
addresses these challenges, enabling the different benefits provided
by virtualization in modern GPUs.

Prior works propose to virtualize a specific on-chip resource for
specific benefits, mostly in the CPU context. For example, in CPUs,
the concept of virtualized registers was first used in the IBM
360~\cite{ibm-360} and DEC PDP-10~\cite{pdp-10} architectures to allow
logical registers to be mapped to either fast yet expensive physical
registers, or slow and cheap memory. More recent works~\cite{how-to-fake,cpu-virt-regs-1,cpu-virt-regs-2}, propose to virtualize registers to increase the effective
register file size to much larger register counts. This increases the
number of thread contexts that can be supported in a multi-threaded
processor~\cite{how-to-fake}, or reduces register spills and
fills~\cite{cpu-virt-regs-1,cpu-virt-regs-2}.\ignore{ Virtual Local
Stores~\cite{virtual-local-stores} is a scratchpad virtualization
mechanism to map the scratchpad inside the hardware-managed cache and
enable context-switching of the scratchpad state along with the rest
of the process state.} Other works propose to virtualize on-chip
resources in CPUs
(e.g., \cite{virtual-local-stores,spills-fills-kills,hierarchical-scheduling-windows,twolevel-hierarchical-registerfile,virtual-physical-registers-hpca98}). In
GPUs, Jeon et al.~\cite{virtual-register} propose to virtualize the
register file by dynamically allocating and deallocating physical
registers to enable more parallelism with smaller, more
power-efficient physical register files. 
Concurrent to this work, Yoon et al.~\cite{virtual-thread} propose an approach to virtualize thread slots
to increase thread-level parallelism. 
These works propose
specific virtualization mechanisms for a single resource for specific
benefits. None of these works provide a cohesive virtualization
mechanism for \emph{multiple} on-chip GPU resources in a
controlled and coordinated manner, which forms a key contribution of
our MICRO 2016 work. 

\textbf{Enhancing Programming Ease and Portability.}
There is a large body of work that aims to improve programmability and
portability of modern GPU applications using software tools, such as
auto-tuners~\cite{toward-autotuning,atune,maestro,parameter-profiler,autotuner1,autotuner-fft},
optimizing
compilers~\cite{g-adapt,optimizing-compiler1,parameter-selection,porple,optimizing-compiler2,sponge},
and high-level programming languages and
runtimes~\cite{cuda-lite,halide,hmpp,hicuda}. These tools tackle a multitude of
optimization challenges, and have been
demonstrated to be very effective in generating high-performance portable
code. They can also be used to tune the resource specification. 
However, there are several shortcomings in these approaches. First, these tools
often require profiling
runs~\cite{toward-autotuning,atune,maestro,porple,optimizing-compiler1,optimizing-compiler2}
on the GPU to determine the best performing resource specifications. These runs have to
be repeated for each new input set and GPU generation. Second, software-based approaches still require significant
programmer effort to write code in a manner that can be exploited by these
approaches
to optimize the resource specifications.\ignore{ For example, auto-tuners require
\emph{parameterization} of code, where the programmer is required to ensure
correctness of the program for any of the possible specification that an
auto-tuner optimizes. Optimizing compilers require programmers to write
kernels to ensure that each thread block is sized as small as possible for the
algorithm being implemented as the compiler has to conservatively preserve
synchronization primitives within a thread block. Some high-level languages~\cite{cuda-lite,halide,hmpp,hicuda} 
and compilers~\cite{g-adapt}
require annotations from the programmer or require the program to be written in
such a way that the algorithm is decoupled from potential optimization
schedules.} Third, selecting the best performing resource specifications statically using
software tools is a challenging task in virtualized
environments (e.g., cloud computing, data centers), where it is unclear which kernels
may be run together on the same SM or where it is not known, apriori, which GPU
generation the  application 
may execute on. Finally, software tools assume a fixed amount of available
resources. This leads to runtime underutilization due to static allocation of
resources, which cannot be addressed by these tools. 

In contrast, the programmability and portability benefits provided by \X
require no programmer effort in optimizing resource specifications. Furthermore,
these autotuners and compilers can be used in conjunction with \X to
further improve performance.

\textbf{Efficient Resource Management.}
Prior works aim to improve parallelism by increasing resource utilization
using
hardware-based~\cite{warp-level-divergence,shmem-multiplexing,unified-register,virtual-register,alternative-thread-block,register-mapping-patent,owl-asplos13,osp-isca13,largewarp,medic,rachata-isca,toggle-aware,decoupled-dma}, software-based~\cite{shmem-multiplexing,stash,
asplos-sree,automatic-placement,onchip-allocation,enabling_coordinated,fine-grain-hotpar}
and hardware-software cooperative~\cite{mask, mosaic, caba, bis, acs,
marshaling,ltrf-sadrosadati-asplos18}
approaches.
%
Among these works, the closest to ours
are~\cite{virtual-register,virtual-thread}
(discussed earlier),~\cite{shmem-multiplexing} and~\cite{warp-level-divergence}. 
\ignore{In contrast, our goal is to decouple the application and the physically available
hardware resources to enhance programming ease, portability, and performance. 
The important difference is the
key contribution of \X, which 
is to enable this with controlled oversubscription using a swap space and coordinated management of
multiple resources.}
These approaches propose efficient techniques to
dynamically manage a single resource, and can be used along with \X to improve
resource efficiency further. 
Yang et
al.~\cite{shmem-multiplexing} aim to maximize utilization of the scratchpad with software
techniques, and by dynamically allocating/deallocating scratchpad memory. 
Xiang et al.~\cite{warp-level-divergence} propose
to improve resource utilization by scheduling threads at the finer granularity
of a warp rather than a thread block. This approach can help alleviate
performance cliffs, but not 
in the presence of synchronization or scratchpad
memory, nor does it address the dynamic underutilization within a thread during
runtime. We quantitatively compare to this approach in Section~\ref{sec:eval} and demonstrate
\X's benefits over it. 

Other works leverage resource
underutilization to improve energy efficiency~\cite{warped-register,energy-register,gebhart-hierarchical,compiler-register,virtual-register}
or perform other useful
work~\cite{caba,spareregister}. These works are
complementary to \X.


\section{Conclusion}

We propose \X, a new framework that decouples the application resource
specification from the allocation in the physical hardware resources (i.e.,
registers, scratchpad memory, and thread slots) in GPUs. \X
encompasses a holistic virtualization strategy to effectively
virtualize multiple latency-critical on-chip resources in a
controlled and coordinated manner. We demonstrate that by providing
the illusion of more resources than physically available, via dynamic
management of resources and the judicious use of a swap space in
main memory, \X enhances \One \emph{programming ease} (by reducing
the performance penalty of suboptimal resource specification), \two
\emph{portability} (by reducing the impact of different hardware
configurations), and \three \emph{performance} for code with an
optimized resource specification (by leveraging dynamic
underutilization of resources). We conclude that \X is an
effective, holistic virtualization framework for GPUs. We believe
that the indirection provided by {\X}'s virtualization mechanism
makes it a generic framework that can address other challenges in
modern GPUs. For example, \X can enable fine-grained resource
sharing and partitioning among multiple kernels/applications, as well as 
low-latency preemption of GPU programs. We hope that future work
explores these promising directions, building on the insights and
the framework developed in this paper.

\ignore{We conclude that by decoupling the programmer's view of the resources from what
is physically available, \X enhances
programming ease, portability, and performance of GPU applications while paving the
way for many other use cases that can leverage the fluid view of resources.}

\section*{Acknowledgments}

We thank the reviewers and our shepherd for their valuable suggestions. We thank the
members of the SAFARI group for their feedback and the stimulating research
environment they provide. Special thanks to Vivek Seshadri, Kathryn McKinley,
Steve Keckler, Evgeny Bolotin, and Mike O'Connor for their feedback during
various stages of this project. We acknowledge the support of our
industrial partners: Facebook, Google, IBM, Intel, Microsoft, NVIDIA,
Qualcomm, Samsung, and VMware. This research was partially supported
by NSF (grant 1409723), the Intel Science and Technology Center for
Cloud Computing, and the Semiconductor Research Corporation.

\begin{spacing}{0.85}
\begin{footnotesize}
\bibliographystyle{plain}
\bibliography{paper,gpu}

\begin{thebibliography}{100}

\bibitem{tensorflow}
M.~Abadi et~al.
\newblock {TensorFlow: A System for Large-Scale Machine Learning}.
\newblock In {\em OSDI}, 2016.

\bibitem{warped-register}
M.~Abdel-Majeed et~al.
\newblock {Warped Register File: A Power Efficient Register File for GPGPUs}.
\newblock In {\em HPCA}, 2013.

\bibitem{opencl}
{Advanced Micro Devices, Inc.}
\newblock {AMD Accelerated Parallel Processing OpenCL Programming Guide}, 2011.

\bibitem{pim-enabled}
J.~Ahn et~al.
\newblock {PIM-Enabled Instructions: A Low-Overhead, Locality-Aware
  Processing-in-Memory Architecture}.
\newblock In {\em ISCA}, 2015.

\bibitem{tesseract}
J.~Ahn et~al.
\newblock A scalable processing-in-memory accelerator for parallel graph
  processing.
\newblock In {\em ISCA}, 2015.

\bibitem{akin.isca15}
B.~Akin et~al.
\newblock {Data Reorganization in Memory Using 3D-Stacked DRAM}.
\newblock In {\em ISCA}, 2015.

\bibitem{ibm-360}
G.~M. Amdahl et~al.
\newblock {Architecture of the IBM System/360}.
\newblock {\em IBM JRD}, 1964.

\bibitem{medic}
R.~Ausavarangnirun et~al.
\newblock {Exploiting Inter-Warp Heterogeneity to Improve GPGPU Performance}.
\newblock {\em PACT}, 2015.

\bibitem{rachata-isca}
R.~Ausavarungnirun et~al.
\newblock {Staged Memory Scheduling: Achieving High Prformance and Scalability
  in Heterogeneous Systems}.
\newblock In {\em ISCA}, 2012.

\bibitem{mosaic}
R.~Ausavarungnirun et~al.
\newblock {Mosaic: A GPU Memory Manager with Application-Transparent Support
  for Multiple Page Sizes.}
\newblock {\em MICRO}, 2017.

\bibitem{mask}
R.~Ausavarungnirun et~al.
\newblock {MASK: Redesigning the GPU Memory Hierarchy to Support
  Multi-Application Concurrency}.
\newblock {\em ASPLOS}, 2018.

\bibitem{GPGPUSim}
A.~Bakhoda et~al.
\newblock {Analyzing CUDA Workloads Using a Detailed GPU Simulator}.
\newblock In {\em {ISPASS}}, 2009.

\bibitem{pdp-10}
C.~G. Bell et~al.
\newblock {The Evolution of the DEC System 10}.
\newblock {\em CACM}, 1978.

\bibitem{multics}
A.~Bensoussan et~al.
\newblock {The Multics Virtual Memory}.
\newblock In {\em SOSP}, 1969.

\bibitem{cilk}
R.~D. Blumofe et~al.
\newblock {Cilk: An Efficient Multithreaded Runtime System}.
\newblock In {\em ASPLOS}, 1995.

\bibitem{boroumand2016pim}
A.~Boroumand et~al.
\newblock {LazyPIM: An Efficient Cache Coherence Mechanism for
  Processing-in-Memory}.
\newblock {\em CAL}, 2016.

\bibitem{googlepim-asplos18}
A.~Boroumand et~al.
\newblock {Google Workloads for Consumer Devices: Mitigating Data Movement
  Bottlenecks}.
\newblock In {\em ASPLOS}, 2018.

\bibitem{hierarchical-scheduling-windows}
E.~Brekelbaum et~al.
\newblock {Hierarchical Scheduling Windows}.
\newblock In {\em MICRO}, 2002.

\bibitem{lonestar}
M.~Burtscher et~al.
\newblock {A quantitative study of irregular programs on GPUs}.
\newblock In {\em IISWC}, 2012.

\bibitem{chang.hpca16}
K.~K. Chang et~al.
\newblock {Low-Cost Inter-Linked Subarrays (LISA): Enabling Fast Inter-Subarray
  Data Movement in DRAM}.
\newblock In {\em HPCA}, 2016.

\bibitem{rodinia}
S.~Che et~al.
\newblock {Rodinia: A Benchmark Suite for Heterogeneous Computing}.
\newblock In {\em IISWC}, 2009.

\bibitem{porple}
G.~Chen et~al.
\newblock {PORPLE: An extensible optimizer for portable data placement on GPU}.
\newblock In {\em MICRO}, 2014.

\bibitem{NQU}
P.~Chen.
\newblock {N-Queens Solver}.
\newblock {\em \url{http://forums.nvidia.com/index.php?showtopic=76893}}, 2008.

\bibitem{virtual-local-stores}
H.~Cook et~al.
\newblock Virtual local stores: Enabling software-managed memory hierarchies in
  mainstream computing environments.
\newblock Technical Report UCB/EECS-2009-131, University of California,
  Berkeley, EECS Dept., 2009.

\bibitem{virtualization-2}
R.~J. Creasy.
\newblock {The Origin of the VM/370 Time-sharing System}.
\newblock {\em IBM JRD}, 1981.

\bibitem{toward-autotuning}
A.~Davidson et~al.
\newblock {Toward Techniques for Auto-Tuning GPU Algorithms}.
\newblock In {\em Applied Parallel and Scientific Computing}. Springer, 2010.

\bibitem{virtual-memory1}
P.~J. Denning.
\newblock Virtual memory.
\newblock {\em ACM Comput. Surv.}, 1970.

\bibitem{hmpp}
R.~Dolbeau et~al.
\newblock {HMPP: A hybrid multi-core parallel programming environment}.
\newblock In {\em GPGPU}, 2007.

\bibitem{autotuner-fft}
Y.~Dotsenko et~al.
\newblock {Auto-tuning of Fast Fourier Transform on Graphics Processors}.
\newblock {\em PPoPP}, 2011.

\bibitem{spills-fills-kills}
M.~Erez et~al.
\newblock {Spills, Fills, and Kills - An Architecture for Reducing
  Register-Memory Traffic}.
\newblock Technical Report TR-23, Stanford Univ., Concurrent VLSI Architecture
  Group, 2000.

\bibitem{compiler-register}
M.~Gebhart et~al.
\newblock {A Compile-time Managed Multi-level Register File Hierarchy}.
\newblock In {\em MICRO}, 2011.

\bibitem{energy-register}
M.~Gebhart et~al.
\newblock {Energy-efficient Mechanisms for Managing Thread Context in
  Throughput Processors}.
\newblock In {\em ISCA}, 2011.

\bibitem{gebhart-hierarchical}
M.~Gebhart et~al.
\newblock A hierarchical thread scheduler and register file for
  energy-efficient throughput processors.
\newblock {\em TOCS}, 2012.

\bibitem{unified-register}
M.~Gebhart et~al.
\newblock {Unifying Primary Cache, Scratch, and Register File Memories in a
  Throughput Processor}.
\newblock In {\em MICRO}, 2012.

\bibitem{ghose.pim.bookchapter18}
S.~Ghose et~al.
\newblock {Enabling the Adoption of Processing-in-Memory: Challenges,
  Mechanisms, Future Research Directions}.
\newblock arxiv:1802.00320 [cs.AR], 2018.

\bibitem{virtual-physical-registers-hpca98}
A.~Gonzalez et~al.
\newblock Virtual-physical registers.
\newblock In {\em HPCA}, 1998.

\bibitem{fine-grain-hotpar}
C.~Gregg et~al.
\newblock {Fine-grained resource sharing for concurrent GPGPU kernels}.
\newblock In {\em HotPar}, 2012.

\bibitem{virtualization-1}
P.~H. Gum.
\newblock {System/370 Extended Architecture: Facilities for Virtual Machines}.
\newblock {\em IBM JRD}, 1983.

\bibitem{guo-wondp14}
Q.~Guo et~al.
\newblock {3D-Stacked Memory-Side Acceleration: Accelerator and System Design}.
\newblock In {\em WoNDP}, 2014.

\bibitem{hicuda}
T.~D. Han et~al.
\newblock {hiCUDA: High-Level GPGPU Programming}.
\newblock {\em TPDS}, 2011.

\bibitem{onchip-allocation}
A.~B. Hayes et~al.
\newblock {Unified On-chip Memory Allocation for SIMT Architecture}.
\newblock In {\em ICS}, 2014.

\bibitem{sponge}
A.~H. Hormati et~al.
\newblock {Sponge: Portable Stream Programming on Graphics Engines}.
\newblock In {\em ASPLOS}, 2011.

\bibitem{impica}
K.~Hsieh et~al.
\newblock {Accelerating Pointer Chasing in 3D-Stacked Memory: Challenges,
  Mechanisms, Evaluation}.
\newblock In {\em ICCD}, 2016.

\bibitem{gaia}
K.~Hsieh et~al.
\newblock {Gaia: Geo-Distributed Machine Learning Approaching LAN Speeds}.
\newblock In {\em NSDI}, 2016.

\bibitem{tom}
K.~Hsieh et~al.
\newblock {Transparent Offloading and Mapping (TOM): Enabling
  Programmer-Transparent Near-Data Processing in GPU Systems}.
\newblock In {\em ISCA}, 2016.

\bibitem{patterson}
{J. L. Hennessey and D. A. Patterson}.
\newblock {\em {Computer Architecture, A Quantitaive Approach}}.
\newblock {Morgan Kaufmann}, 2010.

\bibitem{virtual-memory2}
B.~Jacob et~al.
\newblock Virtual memory in contemporary microprocessors.
\newblock {\em IEEE Micro}, 1998.

\bibitem{virtual-register}
H.~Jeon et~al.
\newblock {GPU register file virtualization}.
\newblock In {\em MICRO}, 2015.

\bibitem{bis}
J.~A. Joao et~al.
\newblock {Bottleneck Identification and Scheduling in Multithreaded
  Applications}.
\newblock In {\em ASPLOS}, 2012.

\bibitem{joao.isca13}
J.~A. Joao et~al.
\newblock {Utility-based Acceleration of Multithreaded Applications on
  Asymmetric CMPs}.
\newblock In {\em ISCA}, 2013.

\bibitem{osp-isca13}
A.~Jog et~al.
\newblock {Orchestrated Scheduling and Prefetching for GPGPUs}.
\newblock In {\em ISCA}, 2013.

\bibitem{owl-asplos13}
A.~Jog et~al.
\newblock {OWL: Cooperative Thread Array Aware Scheduling Techniques for
  Improving GPGPU Performance}.
\newblock In {\em ASPLOS}, 2013.

\bibitem{app-aware-GPGPU-2014}
A.~Jog et~al.
\newblock {Application-aware Memory System for Fair and Efficient Execution of
  Concurrent GPGPU Applications}.
\newblock In {\em GPGPU}, 2014.

\bibitem{parameter-selection}
J.~C. Juega et~al.
\newblock {Adaptive Mapping and Parameter Selection Scheme to Improve Automatic
  Code Generation for GPUs}.
\newblock In {\em CGO}, 2014.

\bibitem{nmnl-pact13}
O.~Kayiran et~al.
\newblock {Neither More Nor Less: Optimizing Thread-level Parallelism for
  GPGPUs}.
\newblock In {\em PACT}, 2013.

\bibitem{Kayiran-micro2014}
O.~Kayiran et~al.
\newblock {Managing GPU Concurrency in Heterogeneous Architectures}.
\newblock In {\em MICRO}, 2014.

\bibitem{kayiran-pact16}
O.~Kayiran et~al.
\newblock {${\mu}$C-States: Fine-grained GPU Datapath Power Management}.
\newblock In {\em PACT}, 2016.

\bibitem{autotuner1}
M.~Khan et~al.
\newblock {A Script-based Autotuning Compiler System to Generate
  High-performance CUDA Code}.
\newblock {\em TAC0}, 2013.

\bibitem{kim.bmc18}
J.~S. Kim et~al.
\newblock {GRIM-Filter: Fast Seed Location Filtering in DNA Read Mapping Using
  Processing-in-Memory Technologies}.
\newblock {\em BMC Genomics}, 2018.

\bibitem{wen-mei-hwu}
D.~B. Kirk and W.~W. Hwu.
\newblock {\em {Programming Massively Parallel Processors: A Hands-on
  Approach}}.
\newblock Morgan Kaufmann, 2010.

\bibitem{kogge.iccp94}
P.~M. Kogge.
\newblock {EXECUBE---A New Architecture for Scaleable MPPs}.
\newblock In {\em ICPP}, 1994.

\bibitem{stash}
R.~Komuravelli et~al.
\newblock Stash: Have your scratchpad and cache it too.
\newblock In {\em ISCA}, 2015.

\bibitem{spareregister}
N.~B. Lakshminarayana et~al.
\newblock {Spare register aware prefetching for graph algorithms on GPUs}.
\newblock In {\em HPCA}, 2014.

\bibitem{decoupled-dma}
D.~Lee et~al.
\newblock {Decoupled Direct Memory Access: Isolating CPU and IO Traffic by
  Leveraging a Dual-data-port DRAM}.
\newblock In {\em PACT}, 2015.

\bibitem{nested}
H.~Lee et~al.
\newblock {Locality-aware Mapping of Nested Parallel Patterns on GPUs}.
\newblock In {\em MICRO}, 2014.

\bibitem{alternative-thread-block}
M.~Lee et~al.
\newblock {Improving GPGPU resource utilization through alternative thread
  block scheduling}.
\newblock In {\em HPCA}, 2014.

\bibitem{gpuwattch}
J.~Leng et~al.
\newblock {GPUWattch: Enabling Energy Optimizations in GPGPUs}.
\newblock In {\em ISCA}, 2013.

\bibitem{automatic-placement}
C.~Li et~al.
\newblock {Automatic data placement into GPU on-chip memory resources}.
\newblock In {\em CGO}, 2015.

\bibitem{DBLP:conf/ics/LiSDSHZ15}
C.~Li et~al.
\newblock Locality-driven dynamic {GPU} cache bypassing.
\newblock In {\em ICS}, 2015.

\bibitem{DBLP:conf/hpca/LiRJOEBFR15}
D.~Li et~al.
\newblock Priority-based cache allocation in throughput processors.
\newblock In {\em HPCA}, 2015.

\bibitem{g-adapt}
Y.~Liu et~al.
\newblock {A cross-input adaptive framework for GPU program optimizations}.
\newblock In {\em IPDPS}, 2009.

\bibitem{concurrent-datastructures}
Z.~Liu et~al.
\newblock {Concurrent Data Structures for Near-Memory Computing}.
\newblock {\em SPAA}, 2017.

\bibitem{graphlab}
Y.~Low et~al.
\newblock {Distributed GraphLab: A Framework for Machine Learning and Data
  Mining in the Cloud}.
\newblock {\em Proc. VLDB Endow.}, April 2012.

\bibitem{jpeg-occ}
J.~Matela et~al.
\newblock {Low GPU occupancy approach to fast arithmetic coding in JPEG2000}.
\newblock In {\em MEMICS}, 2011.

\bibitem{mcgill1978variations}
R.~McGill et~al.
\newblock Variations of box plots.
\newblock {\em The American Statistician}, 1978.

\bibitem{igpu}
J.~Menon et~al.
\newblock {iGPU: Exception Support and Speculative Execution on GPUs}.
\newblock {\em ISCA}, 2012.

\bibitem{largewarp}
V.~Narasiman et~al.
\newblock {Improving GPU Performance via Large Warps and Two-level Warp
  Scheduling}.
\newblock In {\em MICRO}, 2011.

\bibitem{pokemon}
{Nintendo/Creatures Inc./GAME FREAK inc.}
\newblock {Pok\'{e}mon}.
\newblock \url{http://www.pokemon.com/us/}.

\bibitem{cuda}
{NVIDIA Corp.}
\newblock {CUDA}.

\bibitem{sdk}
{NVIDIA Corp.}
\newblock {CUDA C/C++ SDK Code Samples}, 2011.

\bibitem{how-to-fake}
D.~W. Oehmke et~al.
\newblock {How to Fake 1000 Registers}.
\newblock In {\em MICRO}, 2005.

\bibitem{asplos-sree}
S.~Pai et~al.
\newblock {Improving GPGPU concurrency with elastic kernels}.
\newblock In {\em ASPLOS}, 2013.

\bibitem{chimera}
J.~Park et~al.
\newblock {Chimera: Collaborative Preemption for Multitasking on a Shared GPU}.
\newblock In {\em ASPLOS}, 2015.

\bibitem{patterson.ieeemicro97}
D.~Patterson et~al.
\newblock {A Case for Intelligent RAM}.
\newblock {\em IEEE Micro}, 1997.

\bibitem{pattnaik.pact16}
A.~Pattnaik et~al.
\newblock {Scheduling Techniques for GPU Architectures with
  Processing-in-Memory Capabilities}.
\newblock In {\em PACT}, 2016.

\bibitem{toggle-aware}
G.~Pekhimenko et~al.
\newblock {Toggle-Aware Compression for GPUs}.
\newblock In {\em HPCA}, 2016.

\bibitem{halide}
J.~Ragan-Kelley et~al.
\newblock Halide: a language and compiler for optimizing parallelism, locality,
  and recomputation in image processing pipelines.
\newblock In {\em PLDI}, 2013.

\bibitem{tor-micro12}
T.~Rogers et~al.
\newblock {Cache-Conscious Wavefront Scheduling}.
\newblock In {\em MICRO}, 2012.

\bibitem{OptGPU4}
S.~Ryoo et~al.
\newblock {Optimization Principles and Application Performance Evaluation of a
  Multithreaded GPU Using CUDA}.
\newblock In {\em PPoPP}, 2008.

\bibitem{OptGPU3}
S.~Ryoo et~al.
\newblock {Program optimization carving for GPU computing}.
\newblock {\em JPDC}, 2008.

\bibitem{OptGPU1}
S.~Ryoo et~al.
\newblock {Program Optimization Space Pruning for a Multithreaded GPU}.
\newblock In {\em CGO}, 2008.

\bibitem{ltrf-sadrosadati-asplos18}
M.~Sadrosadati et~al.
\newblock {LTRF: Enabling High-Capacity Register Files for GPUs via
  Hardware/Software Cooperative Register Prefetching}.
\newblock In {\em ASPLOS}, 2018.

\bibitem{parameter-profiler}
K.~Sato et~al.
\newblock {Automatic Tuning of CUDA Execution Parameters for Stencil
  Processing}.
\newblock In {\em Software Automatic Tuning: From Concepts to State-of-the-Art
  Results}. Springer-Verlag, 2010.

\bibitem{atune}
C.~A. Schaefer et~al.
\newblock {Atune-IL: An instrumentation language for auto-tuning parallel
  applications}.
\newblock In {\em Euro-Par}, 2009.

\bibitem{rowclone}
V.~Seshadri et~al.
\newblock {RowClone: Fast and Energy-efficient In-DRAM Bulk Data Copy and
  Initialization}.
\newblock In {\em MICRO}, 2013.

\bibitem{seshadri.cal15}
V.~Seshadri et~al.
\newblock {Fast Bulk Bitwise AND and OR in DRAM}.
\newblock {\em CAL}, 2015.

\bibitem{ambit}
V.~Seshadri et~al.
\newblock {Ambit: In-memory Accelerator for Bulk Bitwise Operations Using
  Commodity DRAM Technology}.
\newblock In {\em MICRO}, 2017.

\bibitem{seshadri.bookchapter17}
V.~Seshadri and O.~Mutlu.
\newblock {Simple Operations in Memory to Reduce Data Movement}.
\newblock In {\em Advances in Computers, Volume 106}. Academic Press, 2017.

\bibitem{shaw1981non}
D.~E. Shaw et~al.
\newblock {The NON-VON Database Machine: A Brief Overview}.
\newblock {\em IEEE Database Eng. Bull.}, 1981.

\bibitem{burtonsmith}
B.~Smith.
\newblock {A Pipelined, Shared Resource MIMD Computer}.
\newblock In {\em ICPP}, 1978.

\bibitem{maestro}
K.~Spafford et~al.
\newblock {Maestro: Data Orchestration and Tuning for OpenCL Devices}.
\newblock In {\em Euro-Par}, 2010.

\bibitem{stone1970logic}
H.~S Stone.
\newblock {A Logic-in-Memory Computer}.
\newblock {\em IEEE TC}, 1970.

\bibitem{OptGPU2}
J.~A. Stratton et~al.
\newblock Algorithm and data optimization techniques for scaling to massively
  threaded systems.
\newblock {\em IEEE Computer}, 2012.

\bibitem{parboil}
J.~A. Stratton et~al.
\newblock {Parboil: A Revised Benchmark Suite for Scientific and Commercial
  Throughput Computing}.
\newblock Technical Report {IMPACT-12-01}, Univ.\ of Illinois at
  Urbana-Champaign, IMPACT Research Group, 2012.

\bibitem{acs}
M.~A. Suleman et~al.
\newblock {Accelerating Critical Section Execution with Asymmetric Multi-core
  Architectures}.
\newblock In {\em ASPLOS}, 2009.

\bibitem{marshaling}
M.~A. Suleman et~al.
\newblock {Data Marshaling for Multi-core Architectures}.
\newblock In {\em ISCA}, 2010.

\bibitem{isca-2014-preemptive}
I.~Tanasic et~al.
\newblock {Enabling Preemptive Multiprogramming on GPUs}.
\newblock In {\em ISCA}, 2014.

\bibitem{register-mapping-patent}
D.~Tarjan et~al.
\newblock {On Demand Register Allocation and Deallocation for a Multithreaded
  Processor}, 2011.
\newblock U.S.\ Patent Application 20110161616.

\bibitem{cdc6600}
J.~E. Thornton.
\newblock {Parallel Operation in the Control Data 6600}.
\newblock In {\em AFIPS FJCC}, 1964.

\bibitem{cuda-lite}
Sain-Zee Ueng et~al.
\newblock {CUDA-Lite: Reducing GPU Programming Complexity}.
\newblock In {\em LCPC}, 2008.

\bibitem{caba}
N.~Vijaykumar et~al.
\newblock { A Case for Core-Assisted Bottleneck Acceleration in GPUs: Enabling
  Flexible Data Compression with Assist Warps}.
\newblock In {\em ISCA}, 2015.

\bibitem{caba-bc}
N.~Vijaykumar et~al.
\newblock {A Framework for Accelerating Bottlenecks in GPU Execution with
  Assist Warps}.
\newblock {\em Advances in GPU Research and Practices, Elsevier}, 2016.

\bibitem{zorua}
N.~Vijaykumar et~al.
\newblock {Zorua: A Holistic Approach to Resource Virtualization in GPUs}.
\newblock In {\em MICRO}, 2016.

\bibitem{DBLP:conf/sc/VillaJOBNLSWMSKD14}
O.~Villa et~al.
\newblock {Scaling the Power Wall: {A} Path to Exascale}.
\newblock In {\em SC}, 2014.

\bibitem{vmware-osdi02}
C.~A. Waldspurger.
\newblock {Memory Resource Management in VMware ESX Server}.
\newblock In {\em OSDI}, 2002.

\bibitem{simultaneous-sharing}
Z.~Wang et~al.
\newblock {Simultaneous Multikernel GPU: Multi-tasking Throughput Processors
  via Fine-Grained Sharing}.
\newblock In {\em HPCA}, 2016.

\bibitem{wilton1996cacti}
S.~Wilton et~al.
\newblock {{CACTI: An enhanced cache access and cycle time model}}.
\newblock {\em JSSC}, 1996.

\bibitem{warp-level-divergence}
P.~Xiang et~al.
\newblock {Warp-level divergence in GPUs: Characterization, impact, and
  mitigation}.
\newblock In {\em HPCA}, 2014.

\bibitem{DBLP:conf/hpca/XieLWSW15}
X.~Xie et~al.
\newblock {{Coordinated static and dynamic cache bypassing for GPUs}}.
\newblock In {\em HPCA}, 2015.

\bibitem{enabling_coordinated}
X.~Xie et~al.
\newblock {Enabling coordinated register allocation and thread-level
  parallelism optimization for GPUs}.
\newblock In {\em MICRO}, 2015.

\bibitem{cpu-virt-regs-1}
J.~Yan et~al.
\newblock {Virtual Registers: Reducing Register Pressure Without Enlarging the
  Register File}.
\newblock In {\em HIPEAC}, 2007.

\bibitem{cpu-virt-regs-2}
J.~Yan et~al.
\newblock {Exploiting Virtual Registers to Reduce Pressure on Real Registers}.
\newblock {\em TACO}, 2008.

\bibitem{optimizing-compiler1}
Y.~Yang et~al.
\newblock {A GPGPU Compiler for Memory Optimization and Parallelism
  Management}.
\newblock In {\em PLDI}, 2010.

\bibitem{optimizing-compiler2}
Y.~Yang et~al.
\newblock {A Unified Optimizing Compiler Framework for Different GPGPU
  Architectures}.
\newblock {\em TACO}, 2012.

\bibitem{shmem-multiplexing}
Y.~Yang et~al.
\newblock {Shared memory multiplexing: a novel way to improve GPGPU
  throughput}.
\newblock In {\em PACT}, 2012.

\bibitem{virtual-thread}
M.~Yoon et~al.
\newblock {Virtual Thread: Maximizing Thread-Level Parallelism beyond GPU
  Scheduling Limit}.
\newblock In {\em ISCA}, 2016.

\bibitem{twolevel-hierarchical-registerfile}
J.~Zalamea et~al.
\newblock {Two-level Hierarchical Register File Organization for VLIW
  Processors}.
\newblock In {\em MICRO}, 2000.

\bibitem{zhang-2014}
D.~Zhang et~al.
\newblock {TOP-PIM: Throughput-oriented Programmable Processing in Memory}.
\newblock In {\em HPDC}, 2014.

\bibitem{kernelet}
J.~Zhong et~al.
\newblock {Kernelet: High-throughput GPU kernel executions with dynamic slicing
  and scheduling}.
\newblock {\em TPDS}, 2014.

\end{thebibliography}
\end{footnotesize}
\end{spacing}

\end{document}